\documentclass[%
reprint,
superscriptaddress,
showpacs,
nofootinbib,
nobibnotes,
 amsmath,amssymb,
 aps,
pra
]{revtex4-1}

\usepackage{graphicx}
\usepackage{dcolumn}
\usepackage{dsfont}
\usepackage[english]{babel}
\usepackage[T1]{fontenc}
\newcommand{\ad}    [1] { \hat{a} ^\dagger _{#1 } }
\renewcommand{\a}   [1] { \hat{a}   _{#1 } }

\newcommand{\tr}{ {\rm tr}}
\renewcommand{\vec} [1] { {\bf {#1}}}
\newcommand{\uvec} [1] {\vec{e}_{#1}}
\begin{document}

\title{The BBGKY hierarchy for ultracold bosonic systems}

\author{Sven Kr\"onke}
	\email{skroenke@physnet.uni-hamburg.de}
	\affiliation{Zentrum f\"ur Optische Quantentechnologien, Universit\"at
Hamburg, Luruper Chaussee 149, 22761 Hamburg, Germany}
	\affiliation{The Hamburg Centre for Ultrafast Imaging, Universit\"at 
Hamburg, Luruper Chaussee 149, 22761 Hamburg, Germany}
\author{Peter Schmelcher}
	\email{pschmelc@physnet.uni-hamburg.de}
	\affiliation{Zentrum f\"ur Optische Quantentechnologien, Universit\"at 
Hamburg, Luruper Chaussee 149, 22761 Hamburg, Germany}
	\affiliation{The Hamburg Centre for Ultrafast Imaging, Universit\"at 
Hamburg, Luruper Chaussee 149, 22761 Hamburg, Germany}

\date{\today}

\begin{abstract}
We establish a theoretical framework for 
exploring
the quantum dynamics of
finite ultracold bosonic ensembles based on the Born-Bogoliubov-Green-Kirkwood-Yvon (BBGKY)
hierarchy of equations of motion for few-particle reduced density matrices (RDMs).
The theory applies to zero as well as low temperatures and is formulated 
in a highly efficient way by utilizing dynamically optimized
single-particle basis states and representing the RDMs
in terms of permanents with respect to those. 
An energy, RDM compatibility  and symmetry conserving
closure approximation is developed on the basis of a recursively formulated
cluster expansion for these finite systems. 
In order to enforce necessary representability conditions,
two novel, minimal-invasive and energy-conserving correction algorithms are proposed, 
involving the dynamical purification of the solution of the truncated BBGKY hierarchy 
and the correction of the equations of motion themselves, respectively.
For gaining conceptual insights, 
the impact of two-particle correlations on the dynamical
quantum depletion is studied analytically.
We apply this theoretical framework to both a tunneling and an
interaction-quench scenario. Due to our efficient formulation of the theory, we
can reach truncation orders as large as twelve and thereby systematically study
the impact of the truncation order on the results.
While the short-time dynamics is
 found to be excellently described with controllable accuracy, significant
 deviations occur on a longer time-scale in sufficiently far off-equilibrium 
 situations. Theses deviations are
accompanied by exponential-like instabilities leading to unphysical
results.  The phenomenology
of these instabilities is investigated in detail and we show that
the minimal-invasive correction algorithm of the equation of motion
can indeed stabilize the BBGKY hierarchy truncated 
at the second order.
\end{abstract}

\maketitle
\section{Introduction}\label{sec:intro}

Solving the full stationary or time-dependent Schr\"odinger equation for an interacting many-body system
is an intriguing task, which is why 
various theoretical approaches rely on a description based on much fewer,
effective degrees of freedom in order to avoid the exponential scaling 
of complexity with respect to the number of particles. 
These effective degrees of freedom involve a fictitious single-particle system in the density functional theory \cite{hohenberg_kohn_1964,kohn_self-consistent_1965},
a subsystem consisting of few modes (Wannier functions in a lattice problem) in
the large coordinate-number expansion
\cite{schuetzhold_2008,navez_emergence_2010,navez_epj2014,navez_quasi-particle_2014,queisser_equilibration_2014}
or a subsystem consisting of few particles in Green's function \cite{fetter,green_bonitz13} as well as 
reduced density matrix approaches \cite{qua_kin_theo_bonitz,rdm_then_and_now01,many_electr_and_red_damats,maziotti_rdm_book07,Ruggenthaler17}.
Having solved the problem for the effective degrees of freedom, predictions
for certain classes of observables can be made without
the knowledge of the full many-body wavefunction. Expectation values
of arbitrary $o$-particle operators can be computed from the $o$-body reduced 
density operator for instance, implying that e.g.\ the energy expectation value of the full
many-body system can be determined from the reduced two-body density operator alone 
if only binary interactions are involved \cite{bopp59}. Besides the computational
advantage of being potentially size-intensive, these subsystem-based methods
constitute natural approaches for investigating e.g.\ whether and how 
certain subsystem properties of a closed many-body system 
thermalize when starting from a non-equilibrium initial state \cite{Polkovnikov_RevModPhys2011}.

On the other hand, the lowest order reduced density matrices constitute a
comprehensive analysis tool for characterizing many-body states
\cite{loewdin_norb55,Onsager_Penrose_BEC_liquid_He_PR_1956,concept_of_ODLRO_YangRevModPhys1962,glauber_quantum_1963,coleman_order_indices92,rdm_Sakmann_PRA2008,kronke_two-body_2015}. This holds in particular for bosonic ultracold 
quantum gases where intriguing states of quantum matter such as a Bose-Einstein
condensate or fragmented condensates \cite{Onsager_Penrose_BEC_liquid_He_PR_1956,concept_of_ODLRO_YangRevModPhys1962,nozieres82,coleman_order_indices92,mueller06,rdm_Sakmann_PRA2008,kang_revealing_2014} can be diagnosed by analyzing the one-body reduced density matrix.
Due to the immense flexibility and the controllability of essentially all relevant parameters, these systems serve as 
an ideal platform for systematically studying the impact of correlations on
the many-body quantum dynamics in a unprecedented manner \cite{Pethick_Smith2008,many_body_physics_bloch,atoms_lattice_quantum_sim__book2012}.
For these systems, an efficient description of the quantum dynamics dealing only with a few effective
degrees of freedom is highly desirable since experiments on ultracold ensembles can easily involve several hundred thousands 
or even millions of atoms. 
Because few-particle reduced density matrices are very handy for characterizing 
correlated many-body states, we aim here at a closed theory for the
dynamics of these entities in the context of
ultracold bosonic systems, i.e.\ the appropriately truncated quantum version of
the
Born-Bogoliubov-Green-Kirkwood-Yvon (BBGKY) hierarchy of equations of motion
\cite{bogoliubov65,born_green47,kirkwood46,yvon57,qua_kin_theo_bonitz,gessner_reduced_2016}.

While exactly solvable systems with analytically known reduced density operators
are rare
\cite{cohen_exact_1985,helbig_physical_2010,schilling_natural_2013,when_np_vanish,rdm_calogero15,Klaiman2017362,alon_solvable_2017},
truncating the BBGKY hierarchy usually involves a closure approximation (see \cite{rapp_equations_2014}
for an exception). 
For ultracold quantum gases being extremely dilute, a binary-collision closure approximation, neglecting three-particle
correlations, is expected to be very suitable \cite{boercker_degenerate_1979,SNIDER1995155}. 
The latter can be extended to higher-order correlations by means of a cluster 
expansion \cite{N_body_corr_Cassing85,cassing_self-consistent_1992,hauser_two-body_1995,kira_cluster-expansion_2008,semicond_qo_Kira2011,kira_excitation_2014,kira_coherent_2015,kira_hyperbolic_2015,Leymann_2014,Leymann_2017}, and by using the particle-hole duality or a Green's function method \cite{valdemoro_approximating_1992,valdemoroII,yasuda_direct_1997,mazziotti_contracted_1998}.
The Bogoliubov backaction method 
\cite{vardi_bose-einstein_2001,anglin_dynamics_2001,tikhonenkov_quantum_2007,trimborn_decay_2011,witthaut_beyond_2011,kordas_dissipative_2015-1}
as well as non-commuting cumulants \cite{kohler_microscopic_2002,kohler_microscopic_2003}
constitute alternative but conceptually similar approaches. 
Recently, novel approaches using semi-classical correlations \cite{elliott_density-matrix_2016} or solving a time-dependent variational optimization
problem \cite{jeffcoat_n-representability-driven_2014} have been pursued.

At this point, it shall be noted that while there are numerous theoretical
works on the BBGKY hierarchy and its truncation, the literature on the accuracy 
and numerical stability of this approach in dependence on the truncation order
by explicit simulations is limited to the best of our knowledge 
\cite{akbari_challenges_2012,jeffcoat_n-representability-driven_2014,schuck14,schuck17,lackner_propagating_2015,lackner_high-harmonic_2017,elliott_density-matrix_2016}.
The comprehensive study \cite{akbari_challenges_2012} 
unravels that instabilities as a consequence of the non-linear closure approximation 
can occur and lead to unphysical states, i.e.\ reduced density matrices that are 
not representable. In the context of electronic dynamics in atomic and molecular
systems subjected to strong laser pulses, significant progress has been made
by enforcing compatibility to lower-order reduced density matrices and stabilizing the
truncated BBGKY equations of motion by a dynamical purification of their solution
\cite{lackner_propagating_2015,lackner_high-harmonic_2017}.
 
Since most studies deal with fermions (for bosons, see \cite{kira_excitation_2014,kira_coherent_2015,kira_hyperbolic_2015}
 as well the BBGKY-related approaches 
\cite{kohler_microscopic_2002,kohler_microscopic_2003,vardi_bose-einstein_2001,anglin_dynamics_2001,tikhonenkov_quantum_2007,trimborn_decay_2011,witthaut_beyond_2011,kordas_dissipative_2015-1})
and are based on the truncation of the BBGKY hierarchy after the second order,
this work aims at a highly efficient formulation of the BBGKY such
that it can be truncated at high orders by
a closure approximation tailored to ultracold bosonic systems featuring
a fixed number of atoms.

Our starting-point here is an efficient representation
of the few-particle reduced density operators by tracing out particles from
the 
variational ansatz for the full many-body wavefunction of the established Multi-Configuration Time-Dependent Hartree
method for Bosons (MCTDHB) \cite{MCTDHB_PRA08} (Section \ref{sec:notation}). Similarly to the fermionic
case \cite{lackner_propagating_2015,lackner_high-harmonic_2017}, we thereby 
employ an efficient, dynamically optimized single-particle basis. Representing
the reduced density operators in terms of bosonic number states with respect to this
time-dependent single-particle basis, we derive the truncated BBGKY hierarchy of equations of motion (EOM) from the MCTDHB EOM and provide a compact formulation of the result
in the second-quantization picture (Section \ref{sec:eom}). The properties 
of theses equations are carefully discussed, their validity at also 
low but finite temperatures proven and a technically as well as conceptually useful
spectral representation is provided. Thereafter, we discuss requirements
on the truncation approximation for fulfilling certain conservation laws
and introduce a compatible cluster expansion for bosonic systems with a fixed number
of atoms, where an appropriate normalization of the reduced density operators and 
symmetrization operators is essential
(Section \ref{sec:truncation}). This cluster expansion is formulated in a recursive 
way, which allows for going to truncation orders as 
large as twelve in our numerical simulations. In addition, we provide conceptual insights
into the role of two-body correlations for dynamical quantum depletion and fragmentation.
Since the truncation scheme does not ensure that important necessary representability
conditions such as the positive semi-definiteness of the reduced density matrices are 
fulfilled in the course of the time-evolution, two novel, minimal invasive and 
energy-conserving
correction schemes are developed, aiming at a dynamical purification of the
solution of the truncated BBGKY EOM and at a correction of the EOM themselves, respectively
(Section \ref{sec:rep_corr}). 

Thereafter, we apply this methodological framework to two 
examples. The first scenario is concerned with the tunneling dynamics
of bosonic atoms in a double-well potential. Treating
the system in the tight-binding approximation allows us to 
go to large truncation orders without the need of dynamically optimizing the 
single-particle basis via the corresponding MCTDHB EOM. 
Thereby, we probe
solely effects stemming from the truncation of the BBGKY hierarchy. Here, we
find the short-time dynamics to be excellently described by the truncated BBGKY hierarchy
and the accuracy to increase monotonously with increasing truncation order. 
For longer times, strong deviations are observed, which are linked to high-order
correlations becoming dominant as well as exponential instabilities of the 
truncated BBGKY EOM resulting in unphysical solutions. 
The phenomenology of these instabilities is analyzed in detail and we show that the
minimal-invasive correction scheme for the BBGKY EOM truncated at the second
order can stabilize these EOM indeed.

In the second scenario,
we consider a harmonically trapped bosonic ensemble subjected to an interaction
quench. Here, we solve the full system of coupled EOM for the reduced density matrices
and the dynamically optimized single-particle basis.
For low excitation energies, we find the system to be highly accurately described by the
truncated BBGKY approach. For higher excitation energies, however, exponential instabilities
again occur. Also in this case, we can stabilize the BBGKY EOM truncated at the second order by
our EOM correction scheme and obtain reasonably accurate results for longer times.
Finally, we conclude and provide our perspectives
in Section \ref{sec:concl}.

\section{Setting and formal framework}\label{sec:notation}
In the following, we first specify the general physical setting for which
we aim to develop a theoretical description. Thereafter, we describe how 
the state of the whole many-body system is efficiently represented 
by means of a dynamically adapted, truncated single-particle basis.
Our ultimate goal, however, is not to theoretically describe the 
dynamics of the complete many-body system but to find an effective
description for the dynamics of few-particle subsystems. Here, an efficient
representation of such subsystem states is crucial, which we derive from the 
efficient representation of the total system state.

\subsection{Physical setting}\label{sec:setting}
In this work, we are interested in effectively describing the non-equilibrium quantum
dynamics of $N$ identical bosons governed by the Hamiltonian $\hat H=\sum_{\kappa=1}^N\hat h_\kappa
+\sum_{\kappa<\kappa'}\hat v_{\kappa\kappa'}$. Here, $\hat h_\kappa$ denotes the
one-body Hamiltonian acting on the particle $\kappa$,
which typically consists of kinetic and external trapping potential
contributions, and $\hat v_{\kappa\kappa'}$ refers to the binary interaction potential 
between the particle $\kappa$ and $\kappa'$, an example of which is the contact potential
$\hat v_{12}\propto\delta(\hat x^{(1)}-\hat x^{(2)})$ 
in the context of $s$-wave scattering ultracold atoms \cite{Pethick_Smith2008}.
In
what follows, all terms of the Hamiltonian may be explicitly time-dependent and
for simplicity we mainly focus on the zero-temperature case while commenting on the 
validity of the resulting theory for low but finite temperatures in 
Section \ref{sec:BBGKY-prop}. Although
we are in the end interested in effectively describing the dynamics of few-particle
sub-systems, we nevertheless have to first describe how the total system state 
is represented, i.e.\ the many-body wavefunction in the mainly considered zero-temperature case. 

\subsection{Representation of the many-body wavefunction}\label{sec:wfn_rep}
Instead of relying on some fixed {\it a-priori} basis as commonly
pursued, we employ the wavefunction representation of the
Multi-Configuration Time-Dependent Hartree method for Bosons (MCTDHB)
\cite{MCTDHB_PRA08}.
Here, the central idea is to use $m$ time-dependent, dynamically
optimized single-particle functions (SPFs), $|\varphi_i(t)\rangle$ with $i=1,...,m$,
as a truncated single-particle basis. 
By considering all bosonic number-states $|n_1,...,n_m\rangle_t$ with
the SPFs as the underlying single-particle states and 
its occupation numbers summing up to the total number of particles, $\sum_{r=1}^m
n_r=N$, a time-dependent many-body basis is constructed, with respect to which
the many-body wavefunction is expanded
\begin{equation}
 \label{eq:mctdhb_ansatz}
 |\Psi_t\rangle = \sum_{{\vec n}|N}A_{\vec n}(t)\,|{\vec n}\rangle_t.
\end{equation}
Here, $\vec{n}=(n_1,...,n_m)$ abbreviates a vector of occupation numbers and 
the summation is restricted to all $\vec{n}$ with $\sum_{r=1}^m
n_r=N$,
which we indicate by the symbol $|N$. Using a variational principle, equations of 
motion (EOM) can be derived for both the expansion coefficients
$A_\vec{n}(t)$ and the SPFs $|\varphi_i(t)\rangle$ \cite{MCTDHB_PRA08}.
These EOM, which we explicate in Section 
\ref{sec:MCTDHB_EOM},
ensure that the SPFs move in an optimal manner such that the number 
of SPFs $m$, i.e.\ the numerical control parameter by increasing of which convergence can be achieved, 
can be drastically reduced compared to the case of a time-independent 
single-particle basis. In particular, $m$ may often be chosen to be much smaller 
than the number of time-independent basis states (i.e.\ grid points) with respect to
which the SPFs  $|\varphi_i(t)\rangle$ are represented.
Nevertheless, the number of terms in the full configuration-interaction expansion \eqref{eq:mctdhb_ansatz}
equals $C_m^N\equiv\binom{N+m-1}{m-1}$, which increases drastically with 
an increasing number of bosons $N$. Even if convergence can be achieved with $m\ll N$, 
which is often the operating regime of MCTDHB, 
we have the scaling
$C_m^N\sim N^{m-1}/(m-1)!$, which is not exponential in $N$ but prevents going to huge
systems of $N=\mathcal{O}(10^6)$ particles (unless $m=2$). In the following,
we omit the time-dependence of all entities in our notation and stress that 
all number states $|\vec n\rangle$ are always given with respect to the
time-dependent SPF basis $|\varphi_i\rangle$ unless stated otherwise.

\subsection{Representation of reduced density operators}\label{sec:rdm_rep}
Instead of describing the complete $N$-body system in terms of the wavefunction
$|\Psi_t\rangle$ being expanded according to \eqref{eq:mctdhb_ansatz}, 
we are concerned with the state of an $o$-particle subsystem, $o<N$, 
given by the $o$-body reduced density matrix\footnote{For simplicity, we employ the same acronym ``RDM''
for referring to both the abstract reduced density operator and its representation as
a matrix with respect to a given basis.} ($o$-RDM) of the wavefunction
$|\Psi_t\rangle$
\begin{equation}
\label{eq:D_matrix}
 D^o_{(i_1,...,i_o),(j_1,...,j_o)}(t)=\langle\Psi_t|\ad{j_1}\dots\ad{j_o}\a{i_o}
 \dots\a{i_1}|\Psi_t\rangle.
\end{equation}
Here $\a{i}$ ($\ad{i}$) denotes the time-dependent bosonic annihilation (creation) operator
corresponding to the $i$th time-dependent SPF, $|\varphi_i\rangle$, and obeying
the canonical commutation relations $[\a{i},\ad{j}]=\delta_{ij}$ and $[\a{i},\a{j}]=0$.
Since we aim at a closed theory for the states of few-particle subsystems
taking $o$-particle correlations systematically into account up to high orders $o$, an efficient representation of RDMs is vital. 
Starting with the abstract density operator of $o$-th order $\hat D_o=\sum_{i_1,...,j_o}
D^o_{(i_1,...,i_o),(j_1,...,j_o)}\,|\varphi_{i_1}...\varphi_{i_o}\rangle\!\langle
\varphi_{j_1}...\varphi_{j_o}|$
and using the
bosonic symmetry, manifesting itself in an invariance of \eqref{eq:D_matrix} under
permutations of the first (last) $o$ indices, we may expand the $o$-RDM 
with respect to SPF-based, bosonic $o$-particle number-states
\begin{align}\label{eq:rdm_rep}
\hat\rho_o&=\sum_{{\vec n},{\vec m}|o}
\rho^o_{{\vec n},{\vec m}}\,|{\vec n}\rangle\!\langle{\vec m}|\quad\text{with}\\\nonumber
 \rho^o_{{\vec n},{\vec m}}
&=\binom{N}{o}^{-1} 
\sum_{{\vec l}|N-o} A^*_{{\vec l}+{\vec m}}A_{{\vec l}+{\vec n}}\\\nonumber
&\phantom{=\binom{N}{o}^{-1}}\prod_{r=1}^m\,\binom{l_r+m_r}{m_r}^{\frac{1}{2}}
\binom{l_r+n_r}{n_r}^{\frac{1}{2}}.
\end{align}
We use the probabilistic normalization $\tr(\hat\rho_o)=1$
in this work, meaning $\hat \rho_o=\hat D_o\,(N-o)!/N!$, which
turns out to be crucial for the definition of
few-particle correlations for finite bosonic systems in Section \ref{sec:truncation}. 

\begin{figure}[t]
\includegraphics[width=0.4\textwidth]{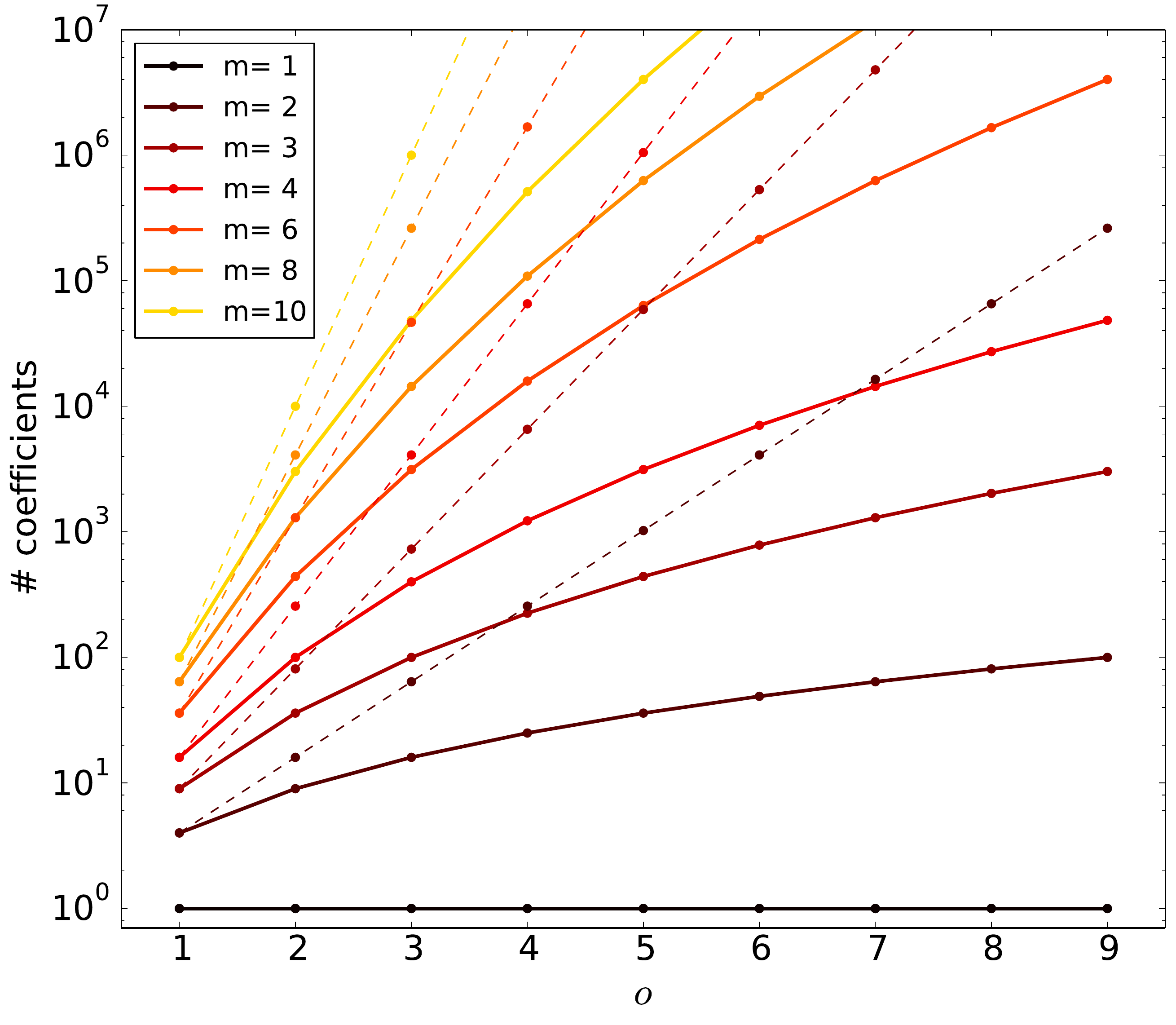}
\caption[]{(color online) Complexity of representing the $o$-RDM, being
measured by the number of complex-valued coefficients,
in dependence on the order $o$ for various numbers of SPFs $m$. Solid lines
refer to bosonic number states as underlying basis functions, see 
Eq.\ \eqref{eq:rdm_rep}, 
dashed ones to Hartree products, see Eq.\ \eqref{eq:D_matrix}. For $m=1$, the solid
and dashed line lie on top of each other.
The hermiticity
of RDMs is not taken into account. Making use of it would roughly reduce the
number of coefficients by a factor of two.
}
\label{fig:num_coeff}
\end{figure}

As a matter of fact, the representation \eqref{eq:rdm_rep} of RDMs is beneficial
in a three-fold manner: (i) Employing $m$ dynamically adapted SPFs as the underlying 
single-particle basis can drastically reduce the necessary number of basis states for
convergence
\cite{MCTDH_BJMW2000}.
(ii) Exploiting the bosonic symmetry strongly reduces the number of
complex coefficients needed for representing an $o$-RDM, namely from $m^{2o}$ for \eqref{eq:D_matrix} to
$(C^o_m)^2$ (if one does not make use of the hermiticity). 
In Fig.\ \ref{fig:num_coeff}, we show the number of coefficients in dependence on $o$ and $m$,
showing clearly that we may effectively represent RDMs of relatively high
order with Eq.\ \eqref{eq:rdm_rep} in contrast to Eq.\ \eqref{eq:D_matrix}.
We note that the depicted range of $m$ is highly relevant for practical
applications since for not too strong correlations in the system, few (optimized) SPFs are often enough
to properly capture the relevant physical processes 
due to the bosonic
bunching-tendency \cite{exact_quantum_dynamics_Cederbaum_PRL2009,lode_numerically_2012,kronke_many-body_2015}.
(iii) Explicitly 
using bosonic number states as the many-body basis for expanding RDMs
is very convenient for analytical manipulations and leads to
equations of motion in a compact second-quantization representation, which is
highly suitable for programming.

Having discussed an efficient representation of RDMs, we also have to
consider how to efficiently perform operations on them. The super-operators that
are crucial for this work cover
the partial trace $\tr_1(\cdot)$, which maps a bosonic $o$-body  
operator to an $(o-1)$-body operator, a raising operation $\hat R_1(\cdot)$, which
maps an $o$-body to an $(o+1)$-body operator, and a joining operation 
$\hat J_{o_1}^{o_2}(\cdot,\cdot)$, which maps an $o_1$- and $o_2$-body operator
to an $(o_1+o_2)$-body operator. In  Appendices \ref{app:superoperators} and \ref{app:1st2ndquant},
we introduce these operations and discuss their efficient
application to e.g.\ RDMs being represented as \eqref{eq:rdm_rep}. Using the formulas provided
in these Appendices, one can easily see
that the $o$-RDM \eqref{eq:rdm_rep} stems from integrating out $(N-o)$ 
degrees of freedom from the $N$-RDM, i.e.\ the total system state $\hat\rho_N=|\Psi_t\rangle\!\langle\Psi_t|$,
meaning
$\hat\rho_o=\tr_{N-o}(\hat\rho_N)$, which implies
the compatibility $\tr_1(\hat\rho_{o+1})=\hat\rho_o$ of the RDMs.

\section{Equations of motion}\label{sec:eom}
Instead of starting with the time-dependent Schr\"odinger equation for 
deriving the EOM 
for the RDMs, we take the MCTDHB EOM of \cite{MCTDHB_PRA08} for 
the wavefunction ansatz \eqref{eq:mctdhb_ansatz}
as our starting point (see also \cite{lackner_propagating_2015}
for the fermionic case). 
Thereby, we tacitly assume that the considered number $m$ of dynamically
optimized SPFs in the MCTDHB ansatz \eqref{eq:mctdhb_ansatz} is sufficient 
for obtaining converged results of desired accuracy up to some time $t_f$ of interest.
At the same time, our starting-point also covers the case of the Schr\"odinger equation
 in the limiting case $m\rightarrow\infty$. After briefly
reviewing the MCTDHB EOM
and their properties in Section \ref{sec:MCTDHB_EOM},
we present the corresponding hierarchy of EOM for the RDMs derived
from the wavefunction ansatz \eqref{eq:mctdhb_ansatz} in Section \ref{sec:BBGKY_EOM}
and discuss equivalent representations in Section \ref{sec:BBGKY_EOM_rep}. By means
of the latter, we draw important conclusions about the role of few-particle
correlations for the dynamics in Section \ref{sec:truncation}, where
truncation approximations for the hierarchy of EOM are discussed.

\subsection{MCTDHB equations of motion}
\label{sec:MCTDHB_EOM}
The EOM of the MCTDHB theory \cite{MCTDHB_PRA08} can be derived by applying the Lagrangian
variational principle to the wavefunction ansatz \eqref{eq:mctdhb_ansatz}, which
ensures that the SPFs are dynamically adapted in a variationally optimal manner.
As a result, one obtains a family of equivalent EOM, whose members 
are specified by fixing the gauge $\langle\varphi_j(t)|i\partial_t|\varphi_k(t)\rangle = g_{jk}(t)$
with an arbitrary, possibly time-dependent hermitian $m\times m$ matrix $g_{jk}$, the so-called
constraint operator \cite{MCTDH_BJMW2000}. For a given gauge $g_{jk}$,
the expansion coefficients obey a Schr\"odinger equation with a time-dependent
Hamiltonian matrix because of the employed time-dependent
many-body basis (setting $\hbar=1$)
\begin{align}
\label{eq:mctdhb_top_eom}
 i\partial_t A_{\vec n} = \sum_{{\vec m}|N}\langle {\vec n}|\,[
  \hat H - \sum_{j,k=1}^mg_{jk}\,\ad{j}\a{k}]\,|{\vec m}\rangle\,
  A_{\vec m}.
\end{align}
We stress here again that all time dependencies are suppressed in the notation.
Correspondingly, the dynamics of the SPFs is governed by the following non-linear
integro-differential equations
\begin{align}
\label{eq:mctdhb_part_eom}
 &i\partial_t |\varphi_l\rangle = \hat g |\varphi_l\rangle +[\mathds{1}-\hat{\mathbb{P}}]
 \Big(\hat h |\varphi_l\rangle +\\\nonumber
 &+(N-1)\sum_{q,p,r,s=1}^m f_{qp}f_{rs}[\hat\rho_1^{-1}]_{rl}\,
 \rho^2_{\uvec{q}+\uvec{p},\uvec{r}+\uvec{s}}\,[\hat v]_{sp}\,
 |\varphi_q\rangle 
 \Big),
\end{align}
where $\hat g\equiv\sum_{i,j=1}^m g_{ij}\,|\varphi_i\rangle\!\langle\varphi_j|$ and
$\hat{\mathbb{P}}= \sum_{i=1}^m |\varphi_i\rangle\!\langle\varphi_i|$ projects onto
the subspace spanned by the instantaneous SPFs. Besides the constraint operator, both
the single-particle Hamiltonian $\hat h\equiv\hat h_1$ and the coupling to the ``other''
$(N-1)$ bosons via the mean-field operator matrix 
$[\hat v]_{sp}\equiv{}^{(2)\!}\langle\varphi_s|\hat v_{12}|\hat\varphi_p\rangle^{\!(2)}$ 
(the super-script $(2)$ denotes a particle label)
drive the time-evolution of the SPFs. We remark that despite of the naming
``mean-field operator matrix'' no mean-field approximation is involved (except for the
limiting case $m=1$).
The matrix $[\hat\rho_1^{-1}]_{rl}$ refers to
the inverse of the regularized\footnote{\label{foot:regularization}See \cite{MCTDH_BJMW2000}
for a regularization recipe and \cite{manthe_multi-configurational_2015,Meyer18} for two recent
alternatives to this regularization.} 1-RDM in SPF representation. As we expand also the 2-RDM with respect to 
two-particle number states, we attain the additional factors $f_{ij}\equiv\sqrt{(1+\delta_{ij})/2}$
as compared to \cite{MCTDHB_PRA08}. Moreover, the occupation-number vectors like $\uvec{q}$, which occur in the indices of the 2-RDM,
describe a state where one boson resides in the $q$th SPFs and all other SPFs are unoccupied.
Thereby, $\uvec{q}+\uvec{p}$ refers to a two-particle state with one boson residing 
in the $q$th and one boson residing in the $p$th SPF.
We note that the density matrices entering the MCTDH(B) EOM in the literature
\cite{MCTDH_BJMW2000,MCTDHB_PRA08}
are the transposed of the RDM definition in this work, for which we have decided since
it allows for evaluating expectation values of few-body operators in the usual manner, namely as the trace
over the observable times the corresponding RDM, e.g.\ 
$\langle\sum_{\kappa=1}^N\hat h_\kappa\rangle=N\,\tr(\hat h\hat\rho_1)$.

The solutions to Eq.\ \eqref{eq:mctdhb_top_eom}, \eqref{eq:mctdhb_part_eom}
for different gauges actually correspond to the same solution for the total wavefunction \eqref{eq:mctdhb_ansatz}
since $\hat g$ in Eq.\ \eqref{eq:mctdhb_part_eom} only induces a unitary 
transformation within the subspace spanned by the instantaneous SPFs, which 
is compensated by a correspond unitary transformation of the coefficients
$A_\vec{n}$ induced by the $g_{jk}$ term in Eq.\ \eqref{eq:mctdhb_top_eom} \cite{MCTDH_BJMW2000,MCTDHB_PRA08}.
For any number of SPFs $m$, the MCTDHB EOM were shown to obey norm as well as
energy\footnote{Higher order moments of a time-independent Hamiltonian are, however, not conserved in general \cite{MCTDH_BJMW2000}.}
conservation \cite{MCTDH_BJMW2000,MCTDHB_PRA08} and to respect
single-particle symmetries such as a parity symmetry if existent \cite{cao_multi-layer_2013}.
Finally, we remark that MCTDHB covers both Gross-Pitaevskii mean-field
theory for fully Bose-Einstein condensed systems (given contact interaction) in the case of $m=1$ and
the exact Schr\"odinger equation\footnote{Setting furthermore $g_{jk}=0$
leads to the Schr\"odinger equation in a time-independent many-body basis.} 
in the formal limit $m\rightarrow\infty$
where $\hat{\mathbb{P}}$ becomes the identity.

\subsection{BBGKY equations of motion}
\label{sec:BBGKY_EOM}
There are at least three ways how the EOM of the RDMs can be derived from the MCTDHB theory.
(i) Since the elements of the $o$-RDM $\hat\rho^o_{\vec n,\vec m}$
only depend explicitly on the expansion coefficients $A_\vec{n}$ (and not on the SPFs),
one may differentiate Eq.\  \eqref{eq:rdm_rep} with respect to time and use the EOM \eqref{eq:mctdhb_top_eom}.
(ii) One could also take the time-derivative of $|\Psi_t\rangle\!\langle\Psi_t|$ using
both Eq.\ \eqref{eq:mctdhb_top_eom} and  \eqref{eq:mctdhb_part_eom}, then trace out $(N-o)$ particles 
via Eq.\ \eqref{eq:part_tr_k} and finally project onto 
$o$-particle number-states from the left and right.
(iii) Instead of MCTDHB, one can equivalently start with the corresponding MCTDH EOM \cite{MCTDH_BJMW2000} such that
individual particles are addressable via artificial labels. Thereby, one can directly apply the standard derivation of the
BBGKY EOM (see e.g.\ \cite{qua_kin_theo_bonitz}) and 
successively trace out particles in the von-Neumann like EOM for the $N$-RDM elements. After
translating this first-quantization result to the second-quantization picture, one obtains the
following EOM for the $o$-RDM
\begin{equation}
\label{eq:BBGKY_arb_gauge}
 i\partial_t \rho^o_{{\vec m},{\vec n}} -
 \langle {\vec m}|\,[\hat {\tilde H},\hat \rho_o]\,|{\vec
n}\rangle
 =\langle {\vec m}|\,\hat I_o\big(\hat\rho_{o+1}\big)\,
 |{\vec n}\rangle.
\end{equation}
Here, the von-Neumann like term $[\hat {\tilde H},\hat \rho_o]$
with the Hamiltonian
\begin{equation}
\label{eq:H_tild}
 \hat {\tilde H} = \sum_{i,j=1}^m (h_{ij} -g_{ij})\,
    \ad{i} \a{j}+\frac{1}{2}\sum_{i,j,q,p=1}^m v_{ijqp}\,
    \ad{i}\ad{j} \a{q} \a{p},
\end{equation}
and $h_{ij}\equiv\langle\varphi_i|\hat h|\varphi_j\rangle$,
$v_{ijqp}\equiv\langle\varphi_i\varphi_j|\hat v_{12}|\varphi_q\varphi_p\rangle$,
describes the unitary evolution of the $o$-particle subsystem in the 
state $\hat\rho_o$ and accounts for all interactions within this subsystem.
Interactions with the sub-system's environment consisting of $(N-o)$ 
particles, however, render the overall dynamics non-unitary in general, which becomes
manifest in the inhomogeneity of the EOM \eqref{eq:BBGKY_arb_gauge}, the
so-called collision integral
\begin{equation}
\label{eq:coll_int}
\hat I_o\big(\hat\rho_{o+1}\big) = \frac{N-o}{o+1}\,\sum_{i,j,q,p=1}^m
  v_{qjpi}\,\big[\hat a^\dagger_q \hat a_p, \hat a_i\,\hat \rho_{o+1}\,\hat
a^\dagger_j\big],\,
\end{equation}
coupling the dynamics of $\hat\rho_o$ to $\hat\rho_{o+1}$. Since 
this second-quantization formulation of the collision integral might appear
less familiar compared to e.g.\ \cite{qua_kin_theo_bonitz}, let us reformulate
the above expression. Using the mixed first and second quantization representation
of Appendix \ref{app:1st2ndquant} [see formula \eqref{eq:ns_1part_sep}] and the 
representation \eqref{eq:rdm_rep}, one easily verifies $\hat a_i\,\hat \rho_{o+1}\,\hat
a^\dagger_j/(o+1)={}^{(o+1)\!}\langle\varphi_i|\hat\rho_{o+1}|\varphi_j\rangle^{\!(o+1)}$,
which may loosely be interpreted as the ``state'' of the $o$-particle subsystem 
conditioned on the $|\varphi_i\rangle\leftrightarrow|\varphi_j\rangle$ 
transition of a further particle. 
By employing the mean-field operator matrix $[\hat v]_{ji}^{(\kappa)}
\equiv {}^{(o+1)}\!\langle\varphi_j|\hat v_{\kappa,o+1}|\varphi_i\rangle^{\!(o+1)}$
with $\kappa=1,...,o$, we can translate 
$\sum_{q,p=1}^mv_{qjpi}\hat a^\dagger_q \hat a_p$
into the first-quantization picture in the $o$-particle sector, namely to
$\sum_{\kappa=1}^o \hat{\mathbb{P}}^{(\kappa)}[\hat v]_{ji}^{(\kappa)}\hat{\mathbb{P}}^{(\kappa)}$
where the mean-field operator matrix and the projector act on the particle $\kappa$ as
indicated by the superscript index. Putting both ingredients together,
we obtain a more familiar representation
\begin{align}
\nonumber
\frac{\hat I_o\big(\hat\rho_{o+1}\big)}{(N-o)}&=\sum_{\kappa=1}^o\sum_{i,j=1}^m
\big[
\hat{\mathbb{P}}^{(\kappa)}[\hat v]_{ji}^{(\kappa)}\hat{\mathbb{P}}^{(\kappa)},
 {}^{(o+1)}\!\langle\varphi_i|\hat\rho_{o+1}|\varphi_j\rangle^{\!(o+1)}
\big],\\
\label{eq:coll_int_1st_quant}
&=\sum_{\kappa=1}^o
\hat{\mathbb{P}}^{(\kappa)}\,\tr_1\left(
\big[
\hat {v}_{\kappa,o+1},
\hat\rho_{o+1}
\big]
\right)\,\hat{\mathbb{P}}^{(\kappa)},
\end{align}
where the partial trace effectively runs over the SPFs only.
The latter representation directly shows that the collision integral 
describes the interaction of any particle of the considered $o$-particle
subsystem with one particle of its environment. Next, we briefly comment on
some properties of the EOM.

\subsubsection{Properties}\label{sec:BBGKY-prop}
 Solved together with the EOM for the SPFs
\eqref{eq:mctdhb_part_eom}, the complete hierarchy of RDM EOM \eqref{eq:BBGKY_arb_gauge}
(with $o=1,...,N$)
is equivalent to MCTDHB, of course, and thereby inherits all properties such
as gauge invariance, norm, energy and, if existent, single-particle symmetry conservation. 
In particular, the solution of the complete hierarchy corresponds to an exact solution
of the many-body Schr\"odinger equation for $m\rightarrow\infty$.
Trivially, the RDM EOM respect the compatibility of the RDMs by construction, meaning
$\partial_t\rho^o_{\vec n,\vec m}=\langle\vec n|\tr_1(\partial_t\hat\rho_{o+1})|\vec m\rangle$.
Although the above EOM are derived from MCTDHB and the RDMs are represented with respect to a dynamically
optimized basis, the EOM for the matrix elements $\rho_{\vec n,\vec m}^o$
are formally identical to the BBGKY EOM derived from the time-dependent 
Schr\"odinger equation \cite{qua_kin_theo_bonitz}, which one can see by using
the representation \eqref{eq:coll_int_1st_quant} for the collision integral
and translating the Hamiltonian \eqref{eq:H_tild} into the first-quantization picture
for the $o$-particle sector. This is due to the fact that
the elements $\rho_{\vec n,\vec m}^o$ depend only on the expansion coefficients
$A_\vec n$ (see Eq.\ \eqref{eq:rdm_rep}), which obey a linear Schr\"odinger-like equation
\eqref{eq:mctdhb_top_eom}. Moreover, in the limit $m\rightarrow\infty$ and the gauge
$g_{ij}=0$, the above equations exactly coincide with the BBGKY hierarchy of EOM 
represented in some time-independent basis (see e.g.\ \cite{qua_kin_theo_bonitz}).
In the opposite limit $m=1$, where all bosons are forced to reside in the same SPF,
the time-derivative of the RDM elements vanishes and the dynamics is solely governed
by Eq.\ \eqref{eq:mctdhb_part_eom}, which becomes equivalent to the Gross-Pitaevskii mean-field
equation for the case of contact interaction
\cite{MCTDHB_PRA08}.

Since the BBGKY EOM derived from MCTDHB are formally identical to
the BBGKY EOM stemming from the von-Neumann equation, the question of their
validity at finite temperatures boils down to the question of whether 
Eq.\ \eqref{eq:mctdhb_part_eom} results in an optimal dynamics for the SPFs
in this case. By purifying the $N$-RDM (see also \cite{raab_multiconfigurational_2000}), 
we show in Appendix \ref{app:finiteT} in which sense Eq.\ \eqref{eq:mctdhb_part_eom}
ensures optimality of the SPFs also for arbitrary mixed initial states $\hat\rho_N(0)$
of the total $N$-particle system, given that the $N$-particle dynamics is unitary.
Thereby, the above equations can safely be applied also to the case of low temperatures
as long as one can computationally account for sufficiently many SPFs to resolve
all significantly populated single-particle states. Otherwise, one would have to
combine the BBGKY approach with some suitable Monte-Carlo sampling technique, which, however, goes
far beyond the scope and aims of this work.

In order to use the above BBGKY EOM for simulating the quantum dynamics of systems 
which are too large for a
MCTDHB calculation, one needs to truncate the hierarchy of EOM at a certain
order $\bar{o}$ and approximate the unknown collision integral $\hat I_{\bar o}[\hat\rho_{\bar o+1}]$. 
This closure approximation
may therefore be regarded as an additional approximation to the MCTDHB theory in the
case of a finite number of SPFs $m$. If successful,
the total particle-number $N$ would directly\footnote{In some situations, however, the number of particles 
does affect the minimal number of SPFs required for convergence, cf.\ the Mott insulating 
state of ultracold bosons in an optical lattice at unit filling where one needs $m= N$.} enter 
the resulting theory only as a prefactor of the collision integral \eqref{eq:coll_int}.
While truncation schemes are discussed in Section \ref{sec:truncation}, 
we provide in the following comments on (i) how to find an appropriate initial state 
for the $o$-RDM with $o=1,...,\bar o$ (Section \ref{sec:init_state}) and (ii) different representations of the
BBGKY hierarchy (Section \ref{sec:BBGKY_EOM_rep}).

\subsubsection{Initial-state calculation}\label{sec:init_state}
In the following, let us assume that we have already truncated the BBGKY
at the order $\bar o$ by means of an appropriate closure approximation (see 
Section \ref{sec:truncation}) and discuss different approaches 
for determining the initial RDMs $[\hat\rho_1(t=0),..., \hat\rho_{\bar o}(t=0)]$.

First, if the system is initially fully condensed or in a non-interacting 
thermal state, 
the initial $o$-RDM can be stated (semi-)analytically for arbitrary orders,
given that the occupied single particle state(s) are known.

Second, if, however, correlations do play a role initially, e.g.\ if the system
is initially in the correlated ground-state of some reference Hamiltonian $\hat H_0$,
numerical methods such as
MCTDHB with imaginary time propagation or improved relaxation
\cite{DensityOperator_MCTDH_Meyer_TCA_2003}
can be employed. Due to the closure approximation, however, the resulting $o$-RDM cease to
be an exact stationary point of the EOM
\eqref{eq:BBGKY_arb_gauge} (with $\hat H$ replaced by the reference Hamiltonian $\hat H_0$). In such a situation, the initial RDMs can
be improved by propagating the RDM EOM \eqref{eq:BBGKY_arb_gauge} with
fixed SPFs and $\hat H$ still replaced by the reference Hamiltonian $\hat H_0$ 
for some time and performing a time-average over the solution
as done in \cite{lackner_high-harmonic_2017}. 

Third, one may alternatively take some initial guess for $\hat\rho_o(t=0)$ with $o=1,...,\bar o$, obtained e.g.\ from
an accurate MCTDHB calculation or, if infeasible, a rough one taking too few SPFs 
into account\footnote{In order to perform the subsequent calculations accurately,
one has to add further, e.g.\ randomly chosen SPFs and
embed the given $o$-RDM with smaller $m$ into 
an $o$-RDM with larger $m$ such that those additional SPFs are unoccupied.}, 
and aim at finding a fixed point of the EOM \eqref{eq:mctdhb_part_eom}, \eqref{eq:BBGKY_arb_gauge},
where $\hat H$ is again replaced by the reference Hamiltonian $\hat H_0$. The MCTDHB EOM \eqref{eq:mctdhb_top_eom}
in negative imaginary time leads to the following trace-conserving
EOM for the $N$-RDM $\partial_\tau\rho^N_{\vec n,\vec m}=
\langle\vec n|\{\tr(\hat H_0\hat\rho_N)-\hat H_0,\,\hat\rho_N\}|\vec m\rangle$ with $\{\cdot,\cdot\}$ 
denoting the anti-commutator. Given a gapped reference Hamiltonian $\hat H_0$, 
the latter EOM exponentially contracts all initial states
with $\langle E_0|\hat\rho_N(0)|E_0\rangle\neq0$ to a state
proportional to the projector onto the (approximate) ground state\footnote{For a degenerate ground state, 
the asymptotic solution is proportional to $\hat\rho_N(0)$ projected from the left and right onto 
the ground-state manifold.} $|E_0\rangle$.

Taking partial traces of the above equation for the $N$-particle density operator, however, appears cumbersome to us. Instead, we find it technically 
more convenient to directly differentiate the RDMs with respect to (negative imaginary) time and perform manipulations similarly to the derivation of contracted Schr\"odinger
equations \cite{valdemoroII,mazziotti_contracted_1998,many_electr_and_red_damats,maziotti_rdm_book07}.
In Appendix \ref{app:imag_time}, we explicate this derivation for the 1-RDM. 
As in the case of contracted Schr\"odinger equations, one finds that the EOM for the $o$-RDM couples to both the
order $o+1$ and $o+2$, which can be traced back to the $N$-RDM EOM featuring an anti-commutator instead of the commutator occurring for 
real-time dynamics. With the help of an appropriate truncation approximation (see Section \ref{sec:truncation}), one can then
relax an initial guess for the $o$-RDM to the (approximate) ground-state
$o$-RDM. It would be very interesting to compare the performance of these
EOM, which includes also an
adaptive single-particle basis, to the conventional contracted Schr\"odinger equation 
approach \cite{valdemoroII,mazziotti_contracted_1998,many_electr_and_red_damats,maziotti_rdm_book07} and its anti-hermitian variant \cite{mazziotti_anti-hermitian_2006} 
(which could also used for calculating the initial $o$-RDM, of course).
Since we, however, focus on the properties of the (truncated) BBGKY equations
\eqref{eq:mctdhb_part_eom}, \eqref{eq:BBGKY_arb_gauge} for many-body dynamics here,
only situations with analytically known initial states are considered
in the applications of Section \ref{sec:appl}.

\subsection{Special representations of the BBGKY equations of motion}
\label{sec:BBGKY_EOM_rep}
Before discussing truncation approximations in Section \ref{sec:truncation},
we briefly inspect selected equivalent representations of the BBGKY
EOM \eqref{eq:mctdhb_part_eom}, \eqref{eq:BBGKY_arb_gauge} here, which
turns out to be useful for both computational purposes and conceptional insights.

\subsubsection{Single-particle Hamiltonian gauge}
While any chosen gauge $g_{ij}$ leads to the same solution for the $o$-RDM
$\hat\rho_o$ as argued before, we empirically found that the single-particle Hamiltonian gauge
$g_{ij}=h_{ij}$ is numerically more favorable for integrating the EOM (see
also \cite{MCTDH_BJMW2000} for a similar observation for MCTDH). We suspect
the following mechanism being responsible for this effect.
The commutator $[\hat{\tilde H},\hat\rho_o]$, expressed in the eigenbasis of
$\hat{\tilde H}$, results in terms being proportional to the difference
of two eigenenergies in the EOM \eqref{eq:BBGKY_arb_gauge}, which might lead to
stiff equations and small integrator time-steps. 
In the above gauge, however,
the single-particle terms are removed from the Hamiltonian \eqref{eq:H_tild} such
that the impact of these energy differences is reduced. Possibly, one might
boost the integration further by incorporating also a fraction of the interaction
energy into the constraint operator in an appropriate mean-field sense.

\subsubsection{Natural-orbital gauge}
Conceptional insights into the role of correlations can be gained by spectrally 
decomposing the $o$-RDM 
\begin{equation}
\label{eq:spectral_o_rdm}
 \hat\rho_o=\sum_{r=1}^{C^o_m}\lambda_r^{(o)}\,|\phi^o_r\rangle\!
 \langle\phi^o_r|
\end{equation}
and reformulating the BBGKY EOM as EOM for the so-called natural
populations (NPs)\footnote{The terms 'natural population' and
'natural orbital' have originally been introduced for the 1-RDM only \cite{loewdin_norb55}
but are employed for all orders in this work.} 
$\lambda_r^{(o)}$ and natural orbitals (NOs) $|\phi^o_r\rangle$ \cite{loewdin_norb55}.
For ultracold bosonic systems, the dynamics of the 1-RDM NPs
is of particular interest for diagnosing quantum depletion
and fragmentation into several Bose-Einstein condensates 
\cite{Onsager_Penrose_BEC_liquid_He_PR_1956,mueller06,rdm_Sakmann_PRA2008}.
In the context of the MCTDH theory, it is well-known that one can enforce 
the SPFs to coincide with the 1-RDM NO given that this coincidence is
also ensured initially \cite{MCTDH_based_on_natorb_Jansen_JCP1993,
comment_MCTDH_based_on_natorb_Manthe_JCP1994,MCTDH_BJMW2000} by an appropriate
choice of the constraint operator, which reads\footnote{We note that the real-valued diagonal elements $g_{ii}$ may
be chosen arbitrarily.} for indistinguishable particles
and two-body interactions
\begin{equation}\label{eq:no_constr}
 g_{ij} = h_{ij} - (1-\delta_{ij})
 \frac{\langle\phi_i^1|\,\hat I_1\big(\hat\rho_{2}\big) \,|\phi^1_j\rangle}{\lambda^{(1)}_i-\lambda^{(1)}_j}.
\end{equation}
Thereby, one finds that the 1-RDM NP dynamics is driven by the collision integral
\begin{equation}\label{eq:np1_eom}
 i\partial_t\lambda_r^{(1)}=\langle\phi^1_r|\,\hat I_1\big(\hat\rho_{2}\big) \,|\phi^1_r\rangle,
\end{equation}
and the corresponding NOs obey
\begin{align}
\label{eq:no1_eom}
 &i\partial_t |\phi^1_r\rangle = \hat h |\phi^1_r\rangle -
\sum_{\substack{l=1\\l\neq r}}^m
\frac{\langle\phi_l^1|\,\hat I_1\big(\hat\rho_{2}\big) \,|\phi^1_r\rangle}{\lambda^{(1)}_l-\lambda^{(1)}_r}\,|\phi^1_l\rangle
\\\nonumber
 &+\frac{(N-1)}{\lambda_r^{(1)}}[\mathds{1}-\hat{\mathbb{P}}]\sum_{q,p,s=1}^m f_{qp}f_{rs}\,
 \rho^2_{\uvec{q}+\uvec{p},\uvec{r}+\uvec{s}}\,[\hat v]_{sp}\,
 |\phi^1_q\rangle,
\end{align}
where the mean-field operator matrix $[\hat v]_{sp}$ has to be evaluated with respect to
the NO basis. Before we proceed, some comments are in order here. (i) The 
EOM \eqref{eq:np1_eom}, \eqref{eq:no1_eom} turn into the exact EOM for
the 1-RDM NPs and NOs \cite{pernal_time-dependent_2007,appel_phd2007,td_norbs_and_npops_Gross_EPL2010,
giesbertz_phd2010,brics_time-dependent_2013,rapp_equations_2014}
in the limit $m\rightarrow\infty$ where the last term
in \eqref{eq:no1_eom} vanishes. For a truncation of the single-particle basis to some
finite $m$, the above EOM describe 
the variationally optimal dynamics of the NOs (see also \cite{brics_single-photon_2017}).
(ii) The reciprocal eigenvalue $1/\lambda_r^{(1)}$ in \eqref{eq:no1_eom} has to be regularized
as usual (see footnote \ref{foot:regularization}). (iii) In the case of NP degeneracies,
both the constraint operator \eqref{eq:no_constr} and the NO EOM \eqref{eq:no1_eom}
can become undefined due to the ambiguity of the NOs within the degenerate subspace(s).
One can cope with this issue by setting the $g_{ij}$ to zero if
$\lambda^{(1)}_i=\lambda^{(1)}_j$ \cite{brics_single-photon_2017} or regularize
the difference $\lambda^{(1)}_i-\lambda^{(1)}_j$ in the equations \cite{MCTDH_BJMW2000}.
Alternatively, a Taylor expansion of $\hat\rho_1(t+\Delta t)$ up to second order
in $\Delta t$ as performed in \cite{manthe_multi-configurational_2015} should 
lift the ambiguity in many cases. In any case, initially non-degenerate NPs typically 
repel each other during the dynamics according to the Wigner-von-Neumann non-crossing 
rule \cite{crossing1929} with the time $t$ as the only ``external'' parameter, unless
symmetries lead to non-incidental crossings \cite{priv_comm1}.
Due to these technical subtleties, we do not employ the natural-orbital gauge
for simulations but only for analytical insights into the essential features of the
2-RDM which actually drive the NP dynamics according to
Eq.\ \eqref{eq:np1_eom} (see Section \ref{sec:trunc_role_corr}).

\subsubsection{Spectral representation on all orders}
While the constraint operator can only be used for deriving the spectral representation
at the order $o=1$, one may proceed for orders $o>1$ by inserting the representation
\eqref{eq:spectral_o_rdm} into the EOM \eqref{eq:BBGKY_arb_gauge} and projecting 
the result onto NOs (see \cite{appel_phd2007,td_norbs_and_npops_Gross_EPL2010,
giesbertz_phd2010,brics_time-dependent_2013,rapp_equations_2014} for the application of this
strategy to the order $o=1$). The result of this calculation is similar to
Eq.\ \eqref{eq:np1_eom} and the first line of Eq.\ \eqref{eq:no1_eom}, and reads
\begin{equation}\label{eq:npo_eom}
 i\partial_t\lambda_r^{(o)}=\langle\phi^o_r|\,\hat I_o\big(\hat\rho_{o+1}\big) \,|\phi^o_r\rangle,
\end{equation}
\begin{align}
\label{eq:no_o_eom}
 &i\partial_t \phi^o_{r;\vec n} = \langle\vec n|\hat{\tilde H}|\phi^o_r\rangle  -
\sum_{\substack{l=1\\l\neq r}}^m
\frac{\langle\phi_l^o|\,\hat I_o\big(\hat\rho_{o+1}\big) \,|\phi^o_r\rangle}{\lambda^{(o)}_l-\lambda^{(o)}_r}\,\phi^o_{l;\vec n}
\end{align}
where $\phi^o_{r;\vec n}\equiv\langle \vec n|\phi^o_r\rangle$. So again, only
the collision integral drives the non-unitary dynamics of the $o$-RDM NPs, as expected.
We note that the EOM \eqref{eq:npo_eom} will be the starting-point for 
our construction of a novel correction algorithm for the truncated BBGKY EOM, which
non-perturbatively
enforces necessary representability conditions such as the positive semi-definiteness (see Section
\ref{sec:corr_eom}).

\section{Truncation approximation and the role of correlations}\label{sec:truncation}
Having discussed the BBGKY hierarchy of EOM stemming from the MCTDHB theory without
further approximations, we investigate closure approximations for
truncating the hierarchy at order $\bar{o}$ here. This is to impose further 
approximations to the MCTDHB theory. While one effectively has to find an
approximation for the unknown collision integral $\hat I_{\bar o}$ only, we
pursue here the standard path of approximating the unknown $(\bar o+1)$ RDM 
by $\hat\rho_{\bar o+1}^{\rm appr}$ such that we obtain for the
approximate collision integral 
$\hat I_{\bar o}^{\rm appr}=\hat I_{\bar o}(\hat\rho_{\bar o+1}^{\rm appr})$.
The general strategy in the following is to appropriately decompose the $o$-RDM 
into a part which can be constructed from lower order RDMs and a rest, which {\it defines} 
$o$-particle correlations. Then, the truncation approximation consists
in neglecting the thereby defined $(\bar o+1)$ correlations. Such an approach
is expected to be appropriate for weak and intermediate interaction strengths,
e.g.\ for studying the emergence of correlations on top of a Bose-Einstein condensate
or fragmented condensates.

In the following, we first discuss requirements on such a closure approximation, which have
to be fulfilled for respecting important conservation laws (Section \ref{sec:trunc_cons_laws}).
Then, different cluster-expansion schemes and thereby
different definitions of few-particle correlations are critically discussed in
Section \ref{sec:trunc_cluster}.
After these rather technical considerations, we conceptually analyze the
role of two-particle correlations for the dynamics of 1-RDM natural
populations, i.e.\ for dynamical quantum depletion or fragmentation of
a bosonic ensemble (Section \ref{sec:trunc_role_corr}).

\subsection{Truncation approximation and conservation laws}
\label{sec:trunc_cons_laws}

While the bosonic symmetry is explicitly incorporated in our formal
framework and thus trivially conserved, other
symmetries and conservation laws are only obeyed by the 
truncated BBGKY EOM \eqref{eq:mctdhb_part_eom},
\eqref{eq:BBGKY_arb_gauge} if the closure approximation $\hat\rho_{\bar o+1}^{\rm appr}$
fulfills certain conditions. For analyzing these requirements, we 
partly
follow the lines of \cite{qua_kin_theo_bonitz} and \cite{akbari_challenges_2012}
while taking the time-dependence of the SPFs into account, whenever necessary.

First of all, the traces of the RDMs are conserved for any truncation approximation
$\hat\rho_{\bar o+1}^{\rm appr}$ due to the commutator structure of the EOM \eqref{eq:BBGKY_arb_gauge}.
Second, any hermitian closure approximation $\hat\rho_{\bar o+1}^{\rm appr}$ 
results in the conservation of hermiticity of the $o$-RDMs, again by virtue
of the commutator structure of their EOM. Third, the conservation of
compatibility can be studied by inspecting
\begin{align}\nonumber
i\partial_t\,\langle \vec{n}|[\tr_1(\hat\rho_{o+1})&-\hat\rho_o]|\vec{m}\rangle
=
\langle\vec{n}|[\hat{\tilde H},\tr_1(\hat\rho_{o+1})-
\hat\rho_o]
|\vec{m}\rangle\\
&+\kappa\,\langle\vec{n}|
\hat I_o\big(\tr_1(\hat\rho_{o+2})-\hat\rho_{o+1}\big)
|\vec{m}\rangle
\end{align}
with $\kappa=(N-o-1)/(N-o)$.
As in the case of the BBGKY hierarchy represented in some time-independent basis,
the compatibility of the closure approximation, 
$tr_1(\hat\rho_{\bar{o}+1}^{\rm appr})=\hat\rho_{\bar{o}}$ for all times,
constitutes a sufficient condition for the conservation of compatibility 
of all lower order RDMs, given that these RDMs are compatible at the initial time
of the propagation.

Fourth, we discuss energy conservation in the sense of 
$\frac{d}{dt}\langle\hat H\rangle_t=\langle(\frac{\partial}{\partial t}\hat H)\rangle_t$,
where the partial derivative on the right hand side relates to a potential explicit time
dependence of the Hamiltonian $\hat H$. Focusing on truncation orders $\bar o\geq 2$, one
obtains the same results for the EOM \eqref{eq:mctdhb_part_eom}, \eqref{eq:BBGKY_arb_gauge}
as found for the BBGKY EOM being represented in a time-independent basis 
\cite{akbari_challenges_2012}. Namely, if the total energy expectation value of the system
is calculated as 
\begin{equation}
 \langle\hat H\rangle_t=N\,\tr\big(\hat h_1\,\hat\rho_1\big)
+\frac{N(N-1)}{2}\,\tr\big(\hat v_{12}\,\hat\rho_2\big)
\end{equation}
then energy conservation is ensured by the bosonic symmetry of the RDMs,
independently of the chosen truncation approximation. If, however,
one alternatively computes  the energy expection value as
$\langle\hat H\rangle_t
=N\,\tr(\hat k_2\,\hat\rho_2)$ with
the auxiliary $2$-particle Hamiltonian 
$\hat k_2=[\hat h_1+\hat h_2+(N-1)\hat v_{12}]/2$ \cite{bopp59},
then energy conservation requires the truncation approximation to 
respect the compatibility requirement.

Fifth, single-particle symmetries are conserved as long as 
the truncation approximation respects this symmetry, which means the following.
Let $\hat\pi_\kappa$ denote a symmetry operation (e.g.\ parity transformation or
translation) 
acting on the $\kappa$th particle and  $\hat\Pi_n=\bigotimes_{\kappa=1}^n \hat\pi_\kappa$
the corresponding symmetry operation acting on $n$ particles. Furthermore, we
consider a Hamiltonian featuring this symmetry, i.e.\ $[\hat\Pi_N,\hat H]=0$,
and assume an initial state of definite symmetry.
By transferring the line of arguments of \cite{cao_multi-layer_2013}
to the current situation, one can show that the truncated
EOM \eqref{eq:mctdhb_part_eom}, \eqref{eq:BBGKY_arb_gauge}
conserve this symmetry, i.e.\ $[\hat\Pi_o,\hat\rho_o(t)]=0$ for $o=1,...,\bar o$,
if the following two conditions are met. (i)
All initial SPFs, i.e.\ also initially unoccupied ones, are of definite
symmetry, meaning $\hat \pi_1|\phi_j(t=0)\rangle=e^{i\theta_j}|\phi_j(t=0)\rangle$ for some
$\theta_j\in\mathbb{R}$. (ii) The reconstruction approximation $\hat\rho_{\bar o+1}^{\rm appr}$
features this symmetry,
$[\hat\Pi_{\bar o+1},\hat\rho_{\bar o+1}^{\rm appr}(t)]=0$ at time $t$,
given that the RDMs of lower order, from which $\hat\rho_{\bar o+1}^{\rm appr}(t)$
is constructed, commute with the corresponding $\hat\Pi_{o}$
transformation.

Sixth, one can show that the $g_{ij}$ gauge invariance of the EOM
\eqref{eq:mctdhb_part_eom}, \eqref{eq:BBGKY_arb_gauge}
remains untouched under truncation if the truncation approximation
$\hat\rho_{\bar o+1}^{\rm appr}(t)$
transforms as a bosonic $(\bar o+1)$-RDM under unitary transformation
of the single-particle basis. 
When discussing
the construction of compatible cluster expansions in Sections \ref{sec:trunc_sym_cluster} and 
\ref{sec:trunc_recursive_cluster}, this transformation behavior
turns out to be a subtlety which has to be carefully analyzed.

Finally, we refer the reader to \cite{bonitz_time_reversal17} for a
comprehensive discussion of the impact of closure approximations on the 
time-reversal invariance of the BBGKY hierarchy.

\subsection{Cluster expansions for finite bosonic systems}
\label{sec:trunc_cluster}

In the following, we first review the so-called cluster expansion for indistinguishable but
spinless particles (Section \ref{sec:trunc_cluster_spinless}) and analyze its
symmetrized variant for bosons (Section \ref{sec:trunc_sym_cluster}). When 
critically inspecting the resulting definition of few-particle correlations, 
also in comparison to the corresponding cluster expansion for fermions, we pinpoint 
an issue concerning size-extensitivity being related to a particularity of the bosonic
symmetrization operator. 
For this reason, we briefly touch upon an alternative 
cluster expansion, being outlined in more detailed
and critically discussed in Appendix \ref{app:trunc_alternative_cluster}.
Eventually, we arrive at a compatible, recursively formulated cluster expansion for
bosons, which allows for going
to large truncation orders (Section \ref{sec:trunc_recursive_cluster}).
It is this cluster expansion which we employ in the applications of
Section \ref{sec:appl}.

\subsubsection{Cluster expansion for indistinguishable spinless particles}
\label{sec:trunc_cluster_spinless}

Following e.g.\ \cite{qua_kin_theo_bonitz}, the cluster expansion for 
a system of indistinguishable but spinless particles reads
\begin{align}
\label{eq:cluster_indist_spinless}
 \hat\rho_2^{(1,2)}&=:\hat\rho_1^{(1)}\hat\rho_1^{(2)}+\hat c_2^{(1,2)}\\\nonumber
 \hat\rho_3^{(1,2,3)}&=:\hat\rho_1^{(1)}\hat\rho_1^{(2)}\hat\rho_1^{(3)}+[\hat c_2^{(1,2)}\hat\rho_1^{(3)}+
 \hat c_2^{(1,3)}\hat\rho_1^{(2)}\\\nonumber
 &\phantom{=:}+\hat c_2^{(2,3)}\hat\rho_1^{(1)}]
 +\hat c_3^{(1,2,3)}\\\nonumber
  \hat\rho_4^{(1,2,3,4)}&=:\hat\rho_1^{(1)}\hat\rho_1^{(2)}\hat\rho_1^{(3)}\hat\rho_1^{(4)}+[\hat c_2^{(1,2)}\hat\rho_1^{(3)}
  \hat\rho_1^{(4)}+...]\\\nonumber
 &+[\hat c_3^{(1,2,3)}\hat\rho_1^{(4)}+...]+
 [\hat c_2^{(1,2)}\hat c_2^{(3,4)}+...]+\hat c_4^{(1,2,3,4)}
\end{align}
etc. Here, the super-index in e.g.\ $\hat\rho_1^{(\kappa)}$ indicates onto which
particle the respective operator shall act. The occurring terms in this cluster expansion
have an intuitive interpretation: e.g.\ $\hat c_2^{(1,2)}\hat\rho_1^{(3)}$ 
describes the situation in which the first two particles are correlated while the third one
constitutes an independent ``spectator''. Let us now briefly summarize
the properties of this expansion and the resulting closure approximation: (i) Given 
a compatible family of RDMs
$\hat\rho_o^{(1,...,o)}$, it is straightforward to see that all cluster operators
are contraction-free, i.e.\ $\tr_1(\hat c_o^{(1,...,o)})=0$. This implies that 
the truncation approximation of setting $\hat c_{\bar o+1}^{(1,...,\bar o+1)}$
to zero
is compatible so that the truncated BBGKY EOM would 
conserve the compatibility of the RDMs. (ii) It is moreover easy to see that the above expansion
is even termwise compatible, meaning that for each class of terms  on the 
right hand side of order $(o+1)$ (indicated by square brackets), there exists a corresponding class of terms at order $o$ that constitutes its partial
trace, e.g.\ $\tr_1([\hat c_2^{(1,2)}\hat\rho_1^{(3)}+...])=\hat c_2^{(1,2)}$ if
the partial trace is taken over the ``third'' particle. (iii) Given that the RDMs are invariant under the
symmetry operation $\hat\Pi^{(1,...,o)}_o=\bigotimes_{\kappa=1}^o \hat\pi^{(\kappa)}$
where $\hat\pi$ denotes a single-particle symmetry operator, i.e.\ $[\hat\Pi_o^{(1,...,o)},\hat\rho_o^{(1,...,o)}]=0$, one 
can show by induction that also the clusters (and thus also the
reconstruction functional at order $(\bar o+1)$) features this symmetry $[\hat\Pi_o^{(1,...,o)},\hat c_o^{(1,...,o)}]=0$.
(iv) While the clusters and therefore the corresponding reconstruction functional
do not single out any particle, meaning $\hat P_\pi\hat c_o^{(1,...,o)}\hat P_\pi=
\hat c_o^{(1,...,o)}$ for any permutation $\pi\in S(o)$ and 
$\hat P_\pi$ denoting the corresponding particle-permutation operator,
they lack bosonic symmetry, i.e.\
$\hat P_\pi\hat c_o^{(1,...,o)}\neq\hat c_o^{(1,...,o)}$ in general.
Thus, projecting onto the bosonic sector is in order here (see
\cite{qua_kin_theo_bonitz} and references therein for a more detailed line of argument),
which is discussed in the following Section.

\subsubsection{Symmetrization of the cluster expansion for indistinguishable bosons}
\label{sec:trunc_sym_cluster}
Since the clusters as defined in Eq.\ \eqref{eq:cluster_indist_spinless} commute with
the respective particle-transposition operators, it is sufficient to apply the respective symmetrization
operator $\hat S_o=\sum_{\pi\in S(o)}\hat P_\pi/o!$ only from the right
\begin{align}
\label{eq:cluster_symm}
 \hat\rho_2&=:\hat\rho_1^{(1)}\hat\rho_1^{(2)}\hat S_2+\hat c_2\\\nonumber
 \hat\rho_3&=:\big(\hat\rho_1^{(1)}\hat\rho_1^{(2)}\hat\rho_1^{(3)}+[\hat c_2^{(1,2)}\hat\rho_1^{(3)}+...]
 \big)\hat S_3+\hat c_3\\\nonumber
  \hat\rho_4&=:\big(\hat\rho_1^{(1)}\hat\rho_1^{(2)}\hat\rho_1^{(3)}\hat\rho_1^{(4)}+[\hat c_2^{(1,2)}\hat\rho_1^{(3)}
  \hat\rho_1^{(4)}+...]\\\nonumber
 &\phantom{=:}+[\hat c_3^{(1,2,3)}\hat\rho_1^{(4)}+...]+
 [\hat c_2^{(1,2)}\hat c_2^{(3,4)}+...]\big)\hat S_4+\hat c_4
\end{align}
etc. Apparently, the number of terms in this expansion increases rapidly with increasing order.
While we provide a recursive scheme for efficiently evaluating clusters of high 
order in Section \ref{sec:trunc_recursive_cluster}, we address here 
the following more fundamental problems and the corresponding properties of the
above cluster expansion: (i) ideal BECs of fixed particle number
as correlation-free reference states, (ii) compatibility, (iii) invariance under
symmetries and (iv) size-extensitivity. Finally, we briefly
comment on differences to the corresponding cluster expansion for fermions.

First, it is natural to require that a correlation measure shall not 
testify correlations if the $N$-particle system is fully condensed, i.e.\
if the system is in a Gross-Pitaevskii mean-field state
$|\Psi\rangle=\otimes_{j=1}^N|\phi\rangle$ with the condensate
wavefunction $|\phi\rangle$. Here, we show that this is indeed the case 
for the clusters $\hat c_o$ defined by \eqref{eq:cluster_symm}. Obviously, 
the $1$-RDM of such a BEC reads $\hat\rho_1=|\phi\rangle\!\langle\phi|$.
Evaluating the first class of terms on the right-hand-side of Eq.\ \eqref{eq:cluster_symm},
we find at order $o$ that $(\hat\rho_1^{(1)}...\,\hat\rho_1^{(o)})\hat S_o=|\phi...\phi\rangle\!\langle\phi...\phi|$,
i.e.\ the projector onto the $o$-fold Hartree product of the condensate wavefunction,
which equals exactly $\hat \rho_o$. Thereby, all clusters vanish for this state.

We have explicated this illustrative calculation here only in order to demonstrate
why we have decided to use the 
idempotent symmetrization operator
$\hat S_o=\hat S_o^2$ and trace-one RDMs in the expansion \eqref{eq:cluster_symm}.
This is namely in contrast to most other works which typically use $D^o_{(i_1,...,i_o),(j_1,...,j_o)}$
(featuring trace $N!/(N-o)!$) as RDMs and $o!\,\hat S_o$ as the
symmetrization operator for the cluster expansion, which is then also called
 cumulant expansion\footnote{The cumulants can be calculated as
derivatives of the generating function 
$\xi(\{\alpha_r\},\{\alpha_r^*\})=\ln(\langle\exp(\sum_r\alpha_r\ad{r})
\exp(-\sum_r\alpha_r^*\a{r})\rangle)$ with respect to $\alpha_i$ and $\alpha_j^*$
and setting all $\alpha$'s to zero (see e.g.\ \cite{maziotti_rdm_book07}).}
\cite{qua_kin_theo_bonitz,many_electr_and_red_damats,maziotti_rdm_book07}.
While the cumulant expansion is perfectly suitable for systems with vanishing
chemical potential, e.g.\ photons \cite{kira_cluster-expansion_2008},
it testifies non-vanishing correlations on all orders
for an ideal BEC with a fixed number $N$ of atoms, even if $N$ becomes
large \cite{kira_excitation_2014}.
Thereby, this approach is not suitable for systematically 
taking correlations into account on top of a BEC. In numerical experiments, we have indeed observed that 
the truncated BBGKY EOM become
almost immediately 
exponentially instable and give wrong results 
if the cumulant expansion is used for the truncation 
(data not shown). To cure this flaw,
we employ the trace-one RDM and the idempotent symmetrization operator 
for the cluster expansion in this work. In passing, we note that recently
also an alternative solution to this problem based on a non-unitary
transformation into the so-called excitation picture of a BEC has
been developed \cite{kira_excitation_2014,kira_hyperbolic_2015,kira_coherent_2015}.

Second, in order to conserve the compatibility of the initial RDMs, a cluster
expansion should ideally respect compatibility, i.e.\ its clusters should be
contraction-free. In contrast to the case of identical but spinless particles (Section
\ref{sec:trunc_cluster_spinless}), however, neither is the expansion \eqref{eq:cluster_symm}
termwise compatible nor are the thereby defined clusters contraction-free in general.
This can be easily seen by inspecting the second order, for which a 
straightforward calculation gives $\tr_1(\hat c_2)=(\hat\rho_1-\hat\rho_1^2)/2$ (see also \cite{cassing_self-consistent_1992}).
Thus, the partial trace
of $\hat c_2$ vanishes only if $\hat\rho_1$ is idempotent, which is equivalent 
to the total system being in a Gross-Pitaevskii mean-field state where
all clusters vanish anyway (see also \cite{gessner_reduced_2016}). In Section \ref{sec:trunc_recursive_cluster}, 
we restore the compatibility of the cluster expansion \eqref{eq:cluster_symm}
by means of a unitarily invariant decomposition of the cluster.

Third, as in the case of indistinguishable spinless particles,
the clusters $\hat c_o$ defined by Eq.\ \eqref{eq:cluster_symm}
commute with the symmetry operator $\hat \Pi_o$ given that the
state of the total system features such a symmetry. This is an immediate consequence
of $[\hat \Pi_o,\hat S_o]=0$.

Fourth, a cluster expansion should ideally be size-extensive in the sense that
it does not testify correlations between two subsystems which feature
no mode-entanglement between each other. 
Even in the absence of mode-entanglement,
the bosonic particle-exchange symmetry does in general
induce correlations between particles, which should be appropriately described by our
methodological approach, of course.
Such correlations should, however, be excluded from the correlation
{\it definition}, on which a cluster expansion is based, so that higher order
clusters can be neglected without impeding physical mechanisms that are induced
by such bosonic-symmetry induced correlations.
Here, we relax this requirement and only
demand that a system consisting of two independent ideal BECs, i.e.\ the simplest case 
of a two-fold fragmented condensate, shall not feature
$\hat c_o$
correlations. While
a single BEC is correlation-free as discussed above, the cluster expansion
\eqref{eq:cluster_symm} unfortunately diagnoses correlations between these two independent BECs,
which can be seen as follows. Suppose that $N_{A/B}$ atoms reside in the 
condensate wavefunction $|\phi_{A/B}\rangle$
such that the total system state reads $|\Psi\rangle=|N_A,N_B\rangle$. A straightforward
calculation shows that two-particle correlations $\hat c_2$ are present in this case, even in 
the infinite particle limit $N=N_A+N_B\rightarrow\infty$ with
$\lambda=N_A/N$ kept constant where one obtains the following expression
\begin{equation}
 \hat c_2 = \lambda(1-\lambda)|\phi_{AB}^+\rangle\!\langle\phi_{AB}^+|+\mathcal{O}(1/N)
\end{equation}
with $|\phi_{AB}^+\rangle=(|\phi_A\phi_B\rangle+|\phi_B\phi_A\rangle)/\sqrt{2}$.
So the correlation measure $\hat c_2$ testifies correlations between the two independent
condensates, which stem solely from the bosonic particle-exchange symmetry.

In order to approximately cure this flaw of lacking size-extensitivity, 
we have explored the construction of an alternative bosonic cluster expansion, which we, however, discard in the
end due to a mathematical subtlety. Let us nevertheless briefly report on the concepts as well as pitfall here.
Inspired by \cite{castin_les_houches2001}, the central idea is to modify the different classes of terms 
of the expansion \eqref{eq:cluster_symm} such that termwise compatibility is ensured and
so-called multi-orbital mean-field states \cite{best_mean-field2003,td_multiorbital_mf_2006}
possess vanishingly small correlations. The latter means that 
there are an occupation-number vector $\vec{k}$ and single-particle basis states
such that the total wavefunction can be represented by a single permanent $|\Psi\rangle=|\vec{k}\rangle$.
The second order of this alternative cluster expansion has been discussed in \cite{thesis_skroenke,schurer_capture_2015}
as well as applied for an in-depth analysis of quantum many-body dynamics far-off equilibrium \cite{schurer_capture_2015}.
In Appendix \ref{app:trunc_alternative_cluster}, we exemplarily outline the construction of this expansion.
Unfortunately, however, it turns out that the
thereby defined cluster operators may depend on the choice of the single-particle basis
in the case of NP degeneracies,
which hinders us to utilize this approach for truncating the BBGKY hierarchy.
For this reason, we stick to the symmetrized cluster expansion \eqref{eq:cluster_symm} 
and make it compatible (see Section \ref{sec:trunc_recursive_cluster}). 

Finally, we briefly compare the above properties to the fermionic case (with fixed particle
number $N$). 
Here, the cumulant expansion is the appropriate approach since it ensures
that Hartree-Fock states do not feature correlations \cite{semicond_qo_Kira2011} (whereas using
the idempotent anti-symmetrization operator and trace-one RDMs
leads to correlations in this case). Analogously to the bosonic case,
the cumulants turn out to be only contraction-free if the system is 
in a Hartree-Fock state. Yet surprisingly,
the cumulants prove to be size-extensive \cite{Mukherjee99,herbert_thesis03,maziotti_rdm_book07}.

\subsubsection{Recursive formulation of the compatible symmetrized cluster expansion}
\label{sec:trunc_recursive_cluster}

We now come back to the symmetrized cluster expansion \eqref{eq:cluster_symm},
make it compatible by means of a unitarily invariant decomposition \cite{coleman1974,coleman_reduced_1980,au-chin_characteristic_1983,sun_unitarily_1984}
and finally
give a recursive formulation allowing for an efficient evaluation at high orders.

Apparently, the cluster $\hat c_{\bar o+1}$ defined by \eqref{eq:cluster_symm} contains 
information about the RDMs of lower order such that neglecting it violates compatibility.
These important pieces of information can be identified by 
the so-called unitarily invariant decomposition (UID) of hermitian bosonic operators \cite{au-chin_characteristic_1983,sun_unitarily_1984},
which allows for uniquely decomposing any $o$-body operator $\hat B_o\in\mathcal{B}_o$
into $\hat B_o=\hat B_o^{\rm red}\oplus\hat B_o^{\rm irr}$ where $\hat B_o^{\rm red}$ 
contains all information about the partial traces of $\hat B_o$, i.e.\
$\tr_1(\hat B_o^{\rm red})=\tr_1(\hat B_o)$, and $\hat B_o^{\rm irr}$ covers
what may be termed irreducible $o$-particle properties. This decomposition is unique
in the sense of being invariant under unitary transformations of the single-particle
basis. We further note that $\hat B_o^{\rm red}$ is a linear functional
in all partial traces $\tr_k(\hat B_o)$ of $\hat B_o$, which we explicate in
Appendix \ref{app:uid}.

Analogously to \cite{lackner_propagating_2015,lackner_high-harmonic_2017}
dealing with fermions, we now define the $(\bar o+1)$-particle correlations which
are neglected in the truncation approximation to be the irreducible, i.e.\ contraction-free
component of the cluster $\hat c_{\bar o+1}$ of the expansion \eqref{eq:cluster_symm} (see \cite{cassing_self-consistent_1992}
for an alternative approach for ensuring compatibility).
If we abbreviate the approximation for $\hat \rho_{\bar o+1}$ as induced by \eqref{eq:cluster_symm}
by $\hat \eta_{\bar o+1}:=\hat \rho_{\bar o+1}-\hat c_{\bar o+1}$,
we obtain the following compatible closure approximation
\begin{equation}
\label{eq:clos_appr_compatible}
 \hat \rho_{\bar o+1}^{\rm appr}:=
 \hat \eta_{\bar o+1}+\hat c_{\bar o+1}^{\rm red}=
 \hat \rho_{\bar o+1}^{\rm red}\big[\hat \rho_{1},...,\hat \rho_{\bar o}
 \big]\oplus\hat \eta_{\bar o+1}^{\rm irr}.
\end{equation}
Practically, this means that we have to calculate (i)
$\hat \eta_{\bar o+1}$, which equals the right-hand-side of \eqref{eq:cluster_symm}
when neglecting the unknown $\hat c_{\bar o+1}$, (ii) its contraction-free component
$\hat \eta_{\bar o+1}^{\rm irr}$ via the UID and (iii) the
reducible component $\hat \rho_{\bar o+1}^{\rm red}$ of
the unknown $\hat \rho_{\bar o+1}$, which, however, depends only on its known
partial traces, i.e.\ the RDMs which are propagated via the truncated
BBGKY EOM. 

In this way, the truncation approximation consists in replacing the
exact $\hat \rho_{\bar o+1}^{\rm irr}$ by $\hat \eta_{\bar o+1}^{\rm irr}$. In passing, we note
that the UID ensures only compatibility but not termwise compatibility as fulfilled by the
alternative cluster expansion outlined in
Appendix \ref{app:trunc_alternative_cluster}.
In contrast to this alternative cluster expansion however, the closure approximation
\eqref{eq:clos_appr_compatible} is invariant under unitary transformations of the SPFs
as a consequence of the UID. Thus, the gauge invariance
of the truncated EOM \eqref{eq:mctdhb_part_eom}, \eqref{eq:BBGKY_arb_gauge} with respect
to the constraint operator $g_{ij}$  is ensured by \eqref{eq:clos_appr_compatible}.

In order to construct the closure approximation \eqref{eq:clos_appr_compatible}
also at high truncation orders $\bar o$, we finally state an efficient
recursive algorithm for evaluating the clusters $\hat c_o$ of the expansion
\eqref{eq:cluster_symm}. The key idea here is to find computation rules 
for the different classes of expansion terms which are indicated in \eqref{eq:cluster_symm}
by square brackets. If we define the one-body cluster by $\hat c_1\equiv\hat \rho_1$, 
we can abbreviate the class of terms at order $o$ which involves 
$K$ different clusters, where the cluster $\hat c_{\sigma_r}$ occurs $n_r$-times ($r=1,...,K$),
by the symbol
\begin{widetext}
\begin{align}
\label{eq:cluster_symbol}
 \hat F_{\,\sigma_1,...,\sigma_K}^{n_1,...,n_K}\equiv&
 \Big[\,\hat c_{\sigma_1}^{(1,...,\sigma_1)}
 \hat c_{\sigma_1}^{(\sigma_1+1,...,2\sigma_1)}\;...\;
 \hat c_{\sigma_1}^{([n_1-1]\sigma_1+1,...,n_1\sigma_1)}
 \hat c_{\sigma_2}^{(n_1\sigma_1+1,...,n_1\sigma_1+\sigma_2)}
 \;...\;
 \hat c_{\sigma_K}^{(o-\sigma_K+1,...,o)}\;+\\\nonumber
 &+\;\text{all distinguishable permutations of the particle labels}\,
 \Big]\,\hat S_o,
\end{align}
\end{widetext}
where $o=\sum_{r=1}^Kn_r\sigma_r$. Here, we assume the ordering
$0<\sigma_1<\sigma_2<....<\sigma_K$ as well as $n_r>0$ for all $r=1,...,K$.
Now we may express the cluster $\hat c_o$ of the expansion \eqref{eq:cluster_symm}
as
\begin{equation}\label{eq:cluster_def_with_symb}
 \hat c_o = \hat \rho_o-\sum_{K=1}^{o-1} 
 \sum_{\substack{n_1,...,n_K>0\\
 0<\sigma_1<\sigma_2<....<\sigma_K}}^{\sum_{r=1}^Kn_r\sigma_r=o}
 \hat F_{\,\sigma_1,...,\sigma_K}^{n_1,...,n_K}.
\end{equation}
As a matter of fact, this sum runs over as many symbols as there are integer partitions of
the number $o$ minus one, i.e.\ $P(o)-1$ with $P(\cdot)$ denoting the partition function,
such that we can use an algorithm which generates integer partitions for
labeling the symbols of a given order.
In appendix \ref{app:recursion_rules}, we prove the following two 
computation rules, which are sufficient for evaluating the symbol 
$\hat F_{\,\sigma_1,...,\sigma_K}^{n_1,...,n_K}$ in terms of the symbols of lower orders:
\begin{align}
 \label{eq:cluster_rule1}
 \hat F_{\,\sigma}^n &=\frac{1}{n}\hat J_{(n-1)\sigma}^{\sigma}\big(\hat F_{\,\sigma}^{n-1},\hat c_\sigma \big)\\
  \label{eq:cluster_rule2}
 \hat F_{\,\sigma_1,...,\sigma_K}^{n_1,...,n_K}&=
\hat J_{\,o-n_K\sigma_K}^{n_K\sigma_K}\big(\hat F_{\,\sigma_1,...,\sigma_{K-1}}^{n_1,...,n_{K-1}},\hat F_{\,\sigma_K}^{n_K} \big),
\end{align}
where again $o=\sum_{r=1}^{K}n_r\sigma_r$ and $\hat J_{o_1}^{o_2}(\cdot,\cdot)$ denotes the 
joining super-operator introduced in Appendix \ref{app:superoperators}. In our software implementation,
we store for each symbol $\hat F_{\,\sigma_1,...,\sigma_K}^{n_1,...,n_K}$ 
at the order $o$ the (integer-partition based) labels of the symbols of lower order
that are needed for applying the respective computation rule. The required joining operations \eqref{eq:join_op}
are implemented in a highly efficient manner by using the combinadic-number
based labeling of bosonic number states \cite{general_mapping_Streltsov10}
in combination with mapping tables \cite{cao_multi-layer_2013} for easily addressing the label
of the $(o_1+o_2)$-particle number state $|\vec{a_1}+\vec{a_2}\rangle$ given 
the $o_1$-particle number state $|\vec{a_1}\rangle$ and the $o_2$-particle number state $|\vec{a_2}\rangle$. 

Having recursively calculated all clusters $\hat c_o$ up to the truncation order $\bar o$,
the (incompatible) auxiliary closure approximation $\hat \eta_{\bar o+1}$ can be constructed,
from which we finally obtain the compatible closure approximation
\eqref{eq:clos_appr_compatible} via the UID (see Eq.\ \eqref{eq:red_comp_UID}).
This is how we truncate the BBGKY hierarchy in the numerical simulations of 
Section \ref{sec:appl}.

\subsection{On the role of two-particle correlations for dynamical quantum depletion}
\label{sec:trunc_role_corr}

After the above technical considerations on how to properly define and evaluate
few-particle correlations for constructing a cluster expansion, 
we investigate here the impact of two-particle correlations on the
1-RDM natural populations $\lambda_r^{(1)}$, which is highly relevant for understanding the mechanisms 
underlying dynamical quantum depletion and fragmentation of Bose-Einstein 
condensates \cite{Onsager_Penrose_BEC_liquid_He_PR_1956}. We address this problem 
from two perspectives. 

First, we analyze the role of the irreducible component $\hat\rho_2^{\rm irr}$
on the 1-RDM dynamics. By inserting the decomposition
$\hat\rho_2 = \hat\rho_2^{\rm red}\oplus\hat\rho_2^{\rm irr}$ together with the explicit expression \eqref{eq:red_comp_UID} 
for $\hat\rho_2^{\rm red}$ 
into \eqref{eq:BBGKY_arb_gauge}, we find the following
\begin{equation}
 i\partial_t\,\langle\varphi_q|\hat\rho_1|\varphi_p\rangle=
 \langle\varphi_q|\big([\hat h_{\rm eff},\hat\rho_1]+\hat I_1(\hat\rho^{\rm irr}_{2})\big)|\varphi_p\rangle,
\end{equation}
where the effective single-particle Hamiltonian reads $\hat h_{\rm eff}=
\hat h-\hat g +\frac{N-1}{m+2}\tr_1[\hat v_{12}(\mathds{1}+\hat P_{12})]$
and $\hat P_{12}$ permutes the particle labels 1 and 2.
So the coupling of a single atom being in the state $\hat\rho_1$ to
the remaining $(N-1)$ atoms via the collision integral $\hat I_1(\hat\rho_{2})$ has a two-fold impact. 
While the reducible component $\hat\rho_2^{\rm red}$ only leads to a renormalization of the 
single-particle Hamiltonian to $\hat h_{\rm eff}$, an effect sometimes called Lamb shift 
in the context of open quantum-systems \cite{Breuer_theory_open_quantum_systems},
non-unitary dynamics of the 1-RDM can only be induced by the irreducible component $\hat\rho_2^{\rm irr}$.
Thus, only these correlations can drive the dynamics of the NPs $\lambda^{(1)}_r(t)$. We note
that this result does not depend on whether the RDMs are represented 
in the dynamically adapted MCTDHB SPF basis or with respect to some time-independent 
basis. It is only important that the single-particle basis is finite, which is a 
technical requirement for the UID \cite{coleman1974,coleman_reduced_1980,au-chin_characteristic_1983,sun_unitarily_1984}.

Second, we explicate the EOM \eqref{eq:np1_eom} for the NPs  $\lambda^{(1)}_r(t)$
\begin{equation}\label{eq:NP_eom_comp}
   \partial_t\lambda_r^{(1)} = 2(N-1)\sum_{i,j,k=1}^m f_{ik}f_{jr}
\,\mathfrak{Im}\big(\, v_{rjik}\,\rho^2_{\uvec{i}+\uvec{k},\uvec{r}+\uvec{j}}\,\big),
\end{equation}
where we remind the reader about the definition $f_{qp}=\sqrt{(\delta_{qp}+1)/2}$.
Employing the hermiticity of $\hat v_{12}$ as well as $[\hat P_{12},\hat v_{12}]=0$,
one easily verifies that the diagonal elements $\rho^2_{\vec{a},\vec{a}}$
do not contribute to the right hand side of Eq.\ \eqref{eq:NP_eom_comp}. As a result, the NP evolution
can only be driven by 
the coherences $\rho^2_{\vec{a},\vec{b}}$ with $\vec a\neq \vec b$ of the
2-RDM represented in permanents with respect the instantaneous NOs $|\phi^1_s\rangle$. 
More precisely, only such coherences in the irreducible component
$\hat\rho_2^{\rm irr}$ can induce non-trivial dynamics of the 1-RDM NPs.

Besides being of conceptual interest, this insight has also consequences
for truncation approximations. Truncating the BBGKY hierarchy at the 
first order using the recipe outlined in Section \ref{sec:trunc_recursive_cluster}
means using the following closure approximation
\begin{align}\nonumber
 \hat\rho_2^{\rm appr}&=\sum_{q,p=1}^m(1+\delta_{qp})\,\lambda_q^{(1)}\,\Big(\frac{\lambda^{(1)}_p}{2}
 +\frac{1-\lambda^{(1)}_q}{m+2}\Big)\,\\\label{eq:rho2_approx}
 &\phantom{=}|\uvec{q}+\uvec{p}\rangle\!\langle \uvec{q}+\uvec{p}|-\frac{1-\tr(\hat\rho_1^2)}{(m+2)(m+1)}\mathds{1}^+_2,
\end{align}
where the number-states are given with respect to the instantaneous NOs $|\phi^1_s\rangle$.
In this basis, $\hat\rho_2^{\rm appr}$ turns out to be diagonal implying that the
NPs $\lambda^{(1)}_r(t)$ are constant in time. Thus, when using the truncation scheme
of Section \ref{sec:trunc_recursive_cluster}, the truncation order $\bar o$ must be 
larger than one in order to account for
dynamical quantum depletion (see also \cite{appel_phd2007,td_norbs_and_npops_Gross_EPL2010,brics_time-dependent_2013} for a similar discussion
for the fermionic case).

Similarly, if the total system is in a multi-orbital mean-field state, i.e.\ a single permanent (cf. Eq.\
\eqref{eq:MMF_o_RDM} of Appendix \ref{app:trunc_alternative_cluster}) or if one truncates the BBGKY hierarchy at order $\bar o=1$
by means of the alternative cluster expansion outlined in Appendix \ref{app:trunc_alternative_cluster},
the 2-RDM entering the right hand side of Eq.\ \eqref{eq:NP_eom_comp} is diagonal 
in the NO number-state basis, leading again to stationary NPs \cite{thesis_skroenke}. Consequently, for a
system being initially prepared in a single permanent (as in the case of two independent
BECs discussed in Section \ref{sec:trunc_sym_cluster}), a Taylor expansion of 
$\lambda^{(1)}_r(t)$ about $t=0$ lacks the linear term. This is a consequence
of $|\Psi_t\rangle$ being continuous in time, which requires a continuous admixture of further 
number-states\footnote{As a side remark, this can be seen
as a deeper reason for the fact that the so-called time-dependent
multi-orbital mean-field theory \cite{td_multiorbital_mf_2006} has to rely
on stationary occupations of the dynamically optimized orbitals.} (with respect to the instantaneous NOs) for having a well-defined 
finite time-derivative of $\lambda^{(1)}_r$.

\section{RDM representability and purification strategies}\label{sec:rep_corr}

While the BBGKY EOM truncated by virtue of the cluster expansion discussed in 
Section \ref{sec:trunc_recursive_cluster} conserve the trace and compatibility
of the RDMs, other properties of RDMs are 
not ensured. As we shall investigate in detail in Section \ref{sec:appl}, the initial
RDM lose e.g.\ their positive semi-definiteness in the course of the time evolution due to the applied 
truncation approximation, which has also been observed in \cite{akbari_challenges_2012,lackner_propagating_2015}
when truncating the BBGKY hierarchy for fermionic problems at order $\bar o=2$.
In this Section, we first review important necessary representability conditions which
$\hat\rho_o$ has to fulfill in order to represent an $o$-RDM of a bosonic $N$-particle
system. Thereafter, we discuss purification strategies for preventing
representability defects in the solution of the truncated BBGKY EOM.

\subsection{Necessary representability conditions}\label{sec:rep_cond}
Besides being compatible and of unit trace, which shall be assumed in the following, 
there are further important 
necessary representability conditions on the $o$-RDM.
For reviewing them, we assume the total $N$-particle system
to be in some pure state $|\Psi\rangle$ and follow the lines of 
\cite{garrod_1964,mazziotti_uncertainty_2001,gidofalvi_boson_2004}. Then
 one
has $\langle\Psi|\hat A_o^\dagger\hat A_o|\Psi\rangle\geq0$ 
for an arbitrary (not necessarily hermitian)
polynomial $\hat A_o$  of order $o$ in the annihilation and creation operators
$\a{r}^{(\dagger)}$, e.g.\ 
\begin{align}\nonumber
 \hat A_2&=\sum_{i,j=1}^m\big(c^{(1)}_{ij}\a{i}\a{j} 
 +c^{(2)}_{ij}\ad{i}\ad{j} 
 +c^{(3)}_{ij}\a{i}\ad{j}
 +c^{(4)}_{ij}\ad{i}\a{j}\big)\\
 &\phantom{=}+\sum_{i=1}^m\big(c^{(5)}_{i}\a{i}
 +c^{(6)}_{i}\ad{i}\big)+c^{(7)}
\end{align}
with the $c^{(\kappa)}$'s being arbitrary complex numbers.
Setting certain $c^{(\kappa)}$'s to zero while allowing the remaining ones
to take arbitrary values, the inequality $\langle\Psi|\hat A_o^\dagger\hat A_o|\Psi\rangle\geq0$  implies the positive semi-definiteness
of various matrices such as the $o$-RDM
$D^o_{(i_1,...,i_o),(j_1,...,j_o)}=\langle\Psi|\ad{j_1}\dots\ad{j_o}\a{i_o}
 \dots\a{i_1}|\Psi\rangle$ or the so-called $o$-hole RDM
 $Q^o_{(i_1,...,i_o),(j_1,...,j_o)}=\langle\Psi|\a{j_1}
 \dots\a{j_o}\ad{i_o}\dots\ad{i_1}|\Psi\rangle$. By normal ordering, all these 
 matrices can be expressed in terms of the $o$-RDM and RDMs of lower order such
 that the required positive semi-definiteness of these matrices effectively induce
 necessary representability conditions for the RDMs. Fulfilling all these 
 conditions implies that the Heisenberg uncertainty relation
 $\langle(\hat A-\langle\hat A\rangle)^2\rangle
 \langle(\hat B-\langle\hat B\rangle)^2\rangle\geq|\langle
 [\hat A,\hat B]\rangle|/2$, with the expectation values
 being evaluated with respect to $\hat\rho_o$, is 
 fulfilled for any observables $\hat A$, $\hat B$ involving
 at most $\lfloor o/2\rfloor$-body operators
 \cite{mazziotti_uncertainty_2001}. In contrast to this, violations of
 these conditions can imply an unphysical violation of the uncertainty 
 relation between two such observables.
 
Since most observables of interest in the field of ultracold atoms
involve at most two-body operators, we focus on representability
conditions for the $1$- and $2$-RDM here. Whereas being positive semi-definite
is necessary and sufficient for the representability of $\hat\rho_1$,
the known necessary and sufficient representability conditions for the $2$-RDM
are not useful in practice \cite{garrod_1964} (see \cite{gidofalvi_boson_2004}, which deals with a model system,
for an exception). Therefore, we consider here only the
important necessary $D$-, $Q$- and $G$-condition for representability \cite{garrod_1964,maziotti_rdm_book07},
which can be derived as outlined above. 
In the following, we assume $\hat\rho_1$ to be representable.
Then, the $D$-condition, i.e.\ the positive
semi-definiteness of $D^2_{(i_1,j_1),(i_2,j_2)}$, directly implies
the positive semi-definiteness of the two-hole RDM
 $Q^2_{(i_1,j_1),(i_2,j_2)}$ (the $Q$-condition) \cite{gidofalvi_boson_2004}. In contrast to this, 
 the positivity
of the one-particle-one-hole RDM $G^2_{(i_1,j_1),(i_2,j_2)}
=\langle\Psi|
\ad{i_1}\a{j_1}\ad{j_2}\a{i_2}|\Psi\rangle$ (the $G$-condition) is not inherited
from the $D$-condition. 
This is because $G^2_{(i_1,j_1),(i_2,j_2)}$ 
is not related to the $D_2$-matrix but its partial transposed via
$G^2_{(i_1,j_1),(i_2,j_2)}
=D^2_{(i_2,j_1),(i_1,j_2)}+\delta_{j_1j_2}D^1_{i_2,i_1}$ \cite{gidofalvi_boson_2004}.
Thereby, the $G$-condition constitutes an independent representability 
requirement on the $2$-RDM, which can be crucial as highlighted by
the numerical results of e.g.\ \cite{gidofalvi_boson_2004}. Besides the $D$-condition, we, however, do not employ 
the above $G$-condition but its more restrictive original variant \cite{garrod_1964},
which demands the positive semi-definiteness of the following matrix
\begin{equation}\label{eq:K_matrix}
 K^2_{(i_1,j_1),(i_2,j_2)}=\frac{1}{\mathcal{N}_K}\Big(
 G^2_{(i_1,j_1),(i_2,j_2)}-D^1_{j_1,i_1}D^1_{i_2,j_2}
 \Big).
\end{equation}
In contrast to \cite{garrod_1964}, however,
we have included the normalization factor $\mathcal{N}_K=N(N+m-1)-\tr(\hat D_1^2)$
in the definition in order to enforce\footnote{We note that
this normalization factor is derived under the assumption of representability of the $2$-RDM. Representability
defects might lead to (slight) deviations from trace one.} unit trace, $\sum_{i,j}K^2_{(i,j),(i,j)}=1$, such that
the eigenvalues of $\hat\rho_2$ and $K^2_{(i_1,j_1),(i_2,j_2)}$ attain comparable values.
This $K$-condition\footnote{Originally, this matrix was denoted as $G$ in \cite{garrod_1964},
but in order to distinguish it from the one-particle-one-hole matrix we decided
to use the letter $K$ in this work.} can be obtained by
the above recipe by setting all $c^{(\kappa)}$ to zero except for 
 $c^{(3)}_{ij}$ and $c^{(7)}$. Finally, we remark that violating the $K$-condition can have
 severe impact on the predictions for density-density correlations being readily
 accessible in ultracold quantum gas experiments (see e.g.\ \cite{hung_extracting_2011}).
 Namely in this case, the positive semi-definiteness of the density-fluctuation covariance matrix
 $C(x,y)=\langle \delta\hat\rho(x)\delta\hat\rho(y)\rangle$, with
 $\delta\hat\rho(x)=\hat\psi^\dagger(x) \hat\psi(x)-\langle\hat\psi^\dagger(x) \hat\psi(x)\rangle$
 and $\hat\psi(x)$ denoting the field operator, is not guaranteed (see Eq.\ 
 (7.14) of \cite{garrod_1964}).

\subsection{Correction strategies}\label{sec:corrections}

Since the truncated BBGKY EOM do not respect the above necessary representability constraints, we discuss here correction strategies.
These strategies can be viewed as an attempt to approximately compensate 
that we neglect $\hat I_{\bar o}(\hat c^{\rm irr}_{\bar o+1})$
in the closure approximation (see Eq.\ \eqref{eq:clos_appr_compatible}). First, we discuss how to correct the solution
after propagating the truncated BBGKY EOM for a small time-step $\Delta t$. Thereafter, we summarize a strategy for correcting 
the truncated BBGKY EOM themselves.

\subsubsection{Purification of the solution}\label{sec:purif_sol}

Originally designed for correcting a $2$-RDM with slight representability defects
in the context of contracted Schr\"odinger equations, the iterative purification scheme by Mazziotti
\cite{mazziotti_purification_2002}
has been employed for correcting the truncated BBGKY EOM for electronic problems
in \cite{lackner_propagating_2015}, which is known as dynamical purification.
Since our approach relies partly on these concepts, we briefly review the main 
ideas of that dynamical purification scheme. 

Assuming that $\hat\rho_2(t)$ is representable, propagating the truncated BBGKY EOM for a fixed, small time-step $\Delta t$
results in $\hat\rho_2(t+\Delta t)$ which features a slight representability defect, i.e.\
slightly violates a necessary representability condition. 
In the following, we shall assume that its partial trace,
$\hat\rho_1(t+\Delta t)$, is representable, i.e.\ positive semi-definite. Otherwise
an appropriate purification scheme for $\hat\rho_1(t+\Delta t)$ has to be
applied and the reducible part $[\hat\rho_2(t+\Delta t)]^{\rm red}$ has to be updated
accordingly \cite{mazziotti_purification_2002}.

Now the idea
is to iteratively update the $2$-RDM by adding a contraction-free correction term $\hat{\mathcal{C}}_2$,
i.e.\
$\hat\rho_2(t+\Delta t)\rightarrow \hat\rho_2(t+\Delta t)+\hat{\mathcal{C}}_2$,
such that its partial trace remains invariant. The Mazziotti purification scheme
relies on an ansatz for $\hat{\mathcal{C}}_2$. Namely for correcting a lack 
of positive semi-definiteness of $\hat\rho_2(t+\Delta t)$, one 
assumes $\hat{\mathcal{C}}_2=\sum_{i\in I}a_i[|\phi^2_i\rangle\!\langle\phi^2_i|]^{\rm irr}$,
where $I$ denotes the set of indices of all NOs whose NPs $\lambda^{(2)}_i$ lie below
a small negative threshold of e.g.\ $\epsilon=-10^{-10}$. The coefficients $a_i$ are determined
such that $\hat{\mathcal{C}}_2$ raises all negative NPs of $\hat\rho_2(t+\Delta t)$
to zero in first order perturbation theory, i.e.\ $\lambda^{(2)}_i+\langle\phi^2_i|
\hat{\mathcal{C}}_2|\phi^2_i\rangle=0$, which constitutes a system of linear equations
for the $a_i$'s. The ansatz for $\hat{\mathcal{C}}_2$ can similarly be extended\footnote{
In the case of fermions, where the $Q$-condition is 
independent of the $D$-condition, this ansatz can be extended to also correct slightly
negative eigenvalues of the $2$-hole RDM \cite{mazziotti_purification_2002}.
}
to also improve negative eigenvalues of $G^2_{(i_1,j_1),(i_2,j_2)}$ \cite{lackner_dipl_thesis}.
The updated RDM $\hat\rho_2(t+\Delta t)+\hat{\mathcal{C}}_2$
may still violate a representability condition such that this update
scheme has to be iterated. 

While the dynamical purification based on the Mazziotti scheme has proven to be successful for the dynamics
of electrons in atoms
\cite{lackner_propagating_2015,lackner_high-harmonic_2017},
it is not minimal invasive by construction and may violate energy conservation.
For this reason, we liberate the above dynamical-purification scheme from its underlying
ansatz for the correction $\hat{\mathcal{C}}_2$, which is determined by
solving an optimization problem in this work. 
First, we note that updating 
the $2$-RDM to $\hat\rho_2(t+\Delta t)+\hat{\mathcal{C}}_2$ also implies an
update of the $K$-operator $\hat K_2=\sum_{i_1,i_2,j_1,j_2}K^2_{(i_1,j_1),(i_2,j_2)}
|\varphi_{i_1}\varphi_{j_1}\rangle\!\langle\varphi_{i_2}\varphi_{j_2}|$ to
$\hat K_2+\hat \Delta_2$ due to the relationship between
$\hat K_2$ and $\hat D_2$. We explicate the operator $\hat\Delta_2$ in Appendix
\ref{app:corr_rdm} where also all the details for the following purification 
scheme are provided. Now, we determine the update $\hat{\mathcal{C}}_2$ by minimizing
the $p$-norm
\begin{equation}\label{eq:p_norm}
 |\hat{\mathcal{C}}_2|_p\equiv\sum_{\vec n,\vec m|2}|{\mathcal{C}}^2_{\vec n,\vec m}|^p
\end{equation}
under the linear constraints of being (i) hermitian, (ii) contraction-free, $\tr_1(\hat{\mathcal{C}}_2)=0$,
(iii) energy conserving, $\tr(\hat v_{12}\hat{\mathcal{C}}_2)=0$, (iv) symmetry
respecting if existing, $[\hat\Pi_2,\hat{\mathcal{C}}_2]=0$, (v) 
raising negative NPs below a threshold $\epsilon$  to zero in first order perturbation theory,
$\lambda^{(2)}_i+\langle\phi^2_i|
\hat{\mathcal{C}}_2|\phi^2_i\rangle=0$ for all $i$ with $\lambda^{(2)}_i<\epsilon$
and (vi) raising negative eigenvalues $\xi_i$ of $\hat K_2$ below a threshold $\epsilon$ 
to zero in first order perturbation theory,
$\xi_i+\langle\Xi_i|
\hat\Delta_2|\Xi_i\rangle=0$ for all $i$ with $\xi_i<\epsilon$, where 
$|\Xi_i\rangle$ denotes the
eigenvector of $\hat K_2$ corresponding to $\xi_i$.
Having solved this optimization problem, we judge whether the updated $2$-RDM fulfills
the $D$- and $K$-condition and iterate the updating procedure if necessary.

Specifically, we have performed numerical experiments on the $1$-norm as well as the $2$-norm (also called Frobenius norm). 
Mathematically, the case $p=1$ leads to the
so-called basis pursuit problem \cite{chen_atomic_2001,Lorenz_2015_SBP},
which we solve by the linear-program solver SOPLEX of the SCIP optimization suite
\cite{SCIP}. The case $p=2$ results in a quadratic program, which we solve with
Lagrange multipliers for the constraints. In our numerical simulations,
however, we found that the $1$-norm scheme has often a harder time to
converge to a $2$-RDM which respects the $D$- and the $K$-condition and
results in a much more noisy time-evolution
of e.g.\ the NPs $\lambda^{(2)}_i$, as compared to the Frobenius norm. 
This finding indicates that the correction operator $\hat{\mathcal{C}}_2$ is not
a sparse matrix (as favored by the $1$-norm).
For this
reason, we only apply the $p=2$ approach in Section \ref{sec:appl}.

This minimal-invasive dynamical purification scheme can conceptually be extended 
for also purifying higher order RDMs,
which, however, becomes computationally harder because of the resulting higher-dimensional
minimization problems. For simplicity, we use the
optimization approach for the $2$-RDM only and make the Mazziotti ansatz for 
achieving $\hat\rho_o\geq0$ at orders $o>2$. For putting purifications on different
orders together, we purify the lowest order where a purification defect
has been observed first, update accordingly
the reducible part of the next order RDM and continue with their purification (if necessary),
etc.

Finally, let us remark that asking for a (small) correction $\hat{\mathcal{C}}_2$
which makes a given indefinite $\hat\rho_2(t+\Delta t)$ positive semi-definite
constitutes a non-linear problem. Both the Mazziotti and our minimal invasive scheme
replace this problem by a linear one (plus iteration) due to the requirement
on the shift of eigenvalues in first order perturbation theory. Thereby, 
these two approaches are perturbative in some sense, which can
hinder these iterative schemes to converge to a fixed point fulfilling the 
posed representability conditions as observed in Section \ref{sec:appl}. 
Therefore, we provide next a non-perturbative strategy aiming at correcting and stabilizing 
the truncated BBGKY EOM themselves.

\subsubsection{Correction of the EOM}\label{sec:corr_eom}

The central idea of this correction strategy is to allow slight representability 
defects in the RDM but modify its EOM in a minimal invasive way such
that these defects are exponentially damped and the EOM thereby stabilized. 
As above, we first describe the correction strategy for the EOM of the $2$-RDM
with all technical details covered by Appendix \ref{app:corr_eom}
and then comment on an extension to the EOM of higher order RDM.

Let us abbreviate the EOM \eqref{eq:BBGKY_arb_gauge} for the $o$-RDM as follows
$\partial_t\rho^o_{\vec n,\vec m}=\langle \vec n|\hat R_o|\vec m\rangle$ (note
that the negative imaginary unit is absorbed in $\hat R_o$). Our aim now is 
to correct $\hat R_2$ by $\hat R_2+\hat{\mathcal{C}}_2$ such that negative eigenvalues
of $\hat\rho_2$ or $\hat K_2$ are exponentially damped to zero. First, we note
that a correction of $\hat R_2$ also implies a modification of the EOM for
$\hat K_2$, which shall be denoted as $\partial_tK^2_{(i_1,j_1),(i_2,j_2)}
=\langle\varphi_{i_1}\varphi_{j_1}|\hat T_2|\varphi_{i_2}\varphi_{j_2}\rangle$, namely to
$\hat T_2\rightarrow\hat T_2+\hat \Delta_2$. This is due to the relationship
between $\hat K_2$ and $\hat D_1$, $\hat D_2$. In Appendix \ref{app:corr_eom},
we explicate the respective expressions.

As for the RDM purification scheme discussed above, we determine the 
$\hat{\mathcal{C}}_2$ by minimizing the $2$-norm \eqref{eq:p_norm} under
certain linear constraints. (i) We demand $\hat{\mathcal{C}}_2$ to be hermitian
because of $\hat R_2^\dagger=\hat R_2$. (ii) The correction term
shall be contraction-free, $\tr_1(\hat{\mathcal{C}}_2)=0$, because the conservation
of compatibility as ensured by our truncation approximation should not be
affected by the EOM correction. Another way to motivate demand (ii) is to
view $\hat{\mathcal{C}}_2$ as an effective approximation of 
 the neglected term $-i\,\hat I_{2}(\hat c^{\rm irr}_{3})$ (when truncating
 at $\bar o=2$). In Appendix \ref{app:UID_coll_int},
we prove that the collision integral $I_o(\hat A_o)$ with a contraction-free
argument $\hat A_o$ is itself contraction-free. (iii) The energy conservation
as ensured by the truncation approximation shall not be affected, which amounts
to $\tr(\hat v_{12}\hat{\mathcal{C}}_2)=0$ because of $\hat{\mathcal{C}}_2$ being
contraction-free. (iv) Symmetries if existent shall be respected by the correction,
i.e.\ $[\hat \Pi_2,\hat{\mathcal{C}}_2]=0$.
(v) In order to damp negative NPs $\lambda^{(2)}_r$, we make use of the NP
EOM $\partial_t\lambda^{(2)}_r=-i\langle\phi^2_r|\hat I_2(\hat\chi_3)|\phi^2_r\rangle$
(see Section \ref{sec:BBGKY_EOM_rep}),
where $\hat\chi_3$ stands for $\hat\rho_3^{\rm appr}$ if $\bar o=2$
and for $\hat\rho_3$ if $\bar o>2$. Upon EOM correction,
this equation is modified to
$\partial_t\lambda^{(2)}_r=\langle\phi^2_r|[\hat{\mathcal{C}}_2-i\hat I_2(\hat\chi_3)]|\phi^2_r\rangle$,
which we require to equal $-\eta\lambda^{(2)}_r$ for all negative NPs below a small
threshold, $\lambda^{(2)}_r<\epsilon$. Thereby, all these negative NPs
are damped with a damping rate $\eta$ as  $\lambda^{(2)}_r(t+\tau)=\lambda^{(2)}_r(t)\exp(-\eta\tau)$
for all $t$ and $\tau\geq0$ where $\lambda^{(2)}_r(t+\tau)$ is smaller than $\epsilon$. 
At the same time, 
also the NPs above the threshold are forced to move such that the 
trace of the $2$-RDM stays unity. Here, the damping constant $\eta$ should be chosen
to be much larger than any system frequency of  physical relevance.
(vi) Analogously, one can show that the modified EOM for the $\hat K_2$ eigenvalues
reads $\partial_t\xi_r=\langle\Xi_r|[\hat T_2+\hat \Delta_2]|\Xi_r\rangle$ which 
is again set to $-\eta\xi_r$ given that $\xi_r<\epsilon$.

As in the case of the RDM purification, one can in principle extend this 
minimal-invasive correction scheme of the EOM also to higher orders, incorporating
various necessary respresentability conditions (see e.g.\ \cite{mazziotti_uncertainty_2001}
for the cases $o=3,4$). In this work, however, we only make
the Mazziotti ansatz
 $\hat{\mathcal{C}}_o=\sum_{i |\lambda^{(o)}_i<\epsilon}a_i[|\phi^o_i\rangle\!\langle\phi^o_i|]^{\rm irr}$
for orders $o>2$
and determine the $a_i$ coefficients such that (v) is fulfilled. Again,
the corrections of the EOM on different orders can be combined in a bottom-up approach
by successively updating the reducible part of the right-hand-side of the next-order EOM.

\section{Applications}\label{sec:appl}
In the following, we apply the above methodological framework to two examples in 
order to analyze the accuracy and stability of this BBGKY approach in dependence
on the truncation order. The scenarios involve tunneling dynamics in a double-well in Section
\ref{sec:BH_dimer} as well as 
interaction quenches in a harmonic trap in Section \ref{sec:breath}, while details about the numerical
integration of the truncated BBGKY EOM are given in Appendix \ref{app:eom_intr}.
\subsection{Tunneling dynamics in a Bose-Hubbard dimer}\label{sec:BH_dimer}

In this scenario, we assume that $N$ bosonic atoms are loaded into an effectively
one-dimensional double-well potential. Preparing the system in an initial state featuring a
particle-number imbalance with a left and the right well allows for studying
the tunneling dynamics of such a many-body system, which has been subject of 
numerous studies covering both mean-field \cite{raghavan_coherent_1999,smerzi_quantum_1997} and many-body calculations 
taking correlations into account \cite{milburn_quantum_1997,exact_quantum_dynamics_Cederbaum_PRL2009,sakmann_quantum_2010,univ_frag_dw,gertjerenken_beyond-mean-field_2013}. Effects unraveled in such a realization
of a bosonic Josephson junction cover macroscopic tunneling and self-trapping \cite{raghavan_coherent_1999,smerzi_quantum_1997,direct_obs_boson_JJ}
as well a decay of tunneling oscillations due to the dephasing of populated
many-body eigenstates of the post-quench Hamiltonian \cite{milburn_quantum_1997,exact_quantum_dynamics_Cederbaum_PRL2009,sakmann_quantum_2010,univ_frag_dw,gertjerenken_beyond-mean-field_2013}.

For sufficiently deep wells, the microscopic many-body Hamiltonian of this
system can be well
approximated by a two-site Bose-Hubbard Hamiltonian within the 
lowest-band tight-binding approximation
\begin{align}
\label{eq:BH_hamilt}
 \hat H = &-J(\ad{L}\a{R}+\ad{R}\a{L})\\\nonumber
 &+\frac{U}{2}\big[\hat n_L(\hat n_L-1)
 +\hat n_R(\hat n_R-1)
 \big]
\end{align}
where $\a{L/R}$ annihilates a boson in the lowest-band Wannier state localized in
the left/right well and $\hat n_{i}\equiv\ad{i}\a{i}$ denotes the corresponding 
occupation-number operator of the site $i\in\{L,R\}$. The first
term in \eqref{eq:BH_hamilt} describes tunneling between the two wells
weighted with the hopping amplitude $J>0$. The second term refers to
on-site interaction of strength $U$ and stems from the 
short-range van-der-Waals interaction between the atoms. For convenience,
we take the hopping amplitude as our energy scale and state times
in units of $1/J$.

The Bose-Hubbard dimer features an almost trivial computational complexity since
the full many-body wavefunction depends only on $C^N_2=N+1$ complex-valued 
coefficients such that the corresponding time-dependent Schr\"odinger equation can be 
numerically exactly solved for very large atom numbers. So there is no
need for an alternative computational approach here. On the other hand,
this system can serve as a good playground for analyzing the properties of the
truncated BBGKY approach because (i) the corresponding numerically exact solution
is available and (ii) we can easily represent RDMs of large order without using
a dynamically optimized truncated single-particle basis. This allows for 
systematically investigating the accuracy of our results solely in dependence on
the truncation order $\bar o$.

In the following, 
we consider the initial state $|\Psi_0\rangle=|N,0\rangle$ with all atoms located in the
left well and focus on the tunneling regime by setting the dimensionless interaction parameter
$\Lambda=U(N-1)/(2J)$ to $0.1$, i.e.\ well below the critical value $\Lambda_{\rm crit}=2$
for self-trapping \cite{raghavan_coherent_1999,smerzi_quantum_1997}. In the weak interaction
regime $\Lambda\ll1$, beyond mean-field effects such as
the aforementioned collapse of tunneling oscillation \cite{milburn_quantum_1997} and the universal formation of 
a two-fold fragmented condensate out of a single condensate \cite{univ_frag_dw}
are expected to
play a significant role after the time-scale $t_{\rm mf}\approx\sqrt{2N+1}/(J\Lambda)$,
the so-called quantum break time
\cite{gertjerenken_beyond-mean-field_2013}.

Most of the following calculations deal with $N=10$ atoms such that $t_{\rm mf} \approx 46/J$.
For comparison, we also increase $N$ to $100$ atoms while keeping
$\Lambda$ constant, which results in a longer quantum break time of $t_{\rm mf}  \approx 142/J$.
We analyze the accuracy of the truncated BBGKY hierarchy approach in three steps. First,
we inspect the particle-number imbalance, a highly-integrated quantity 
characterizing the tunneling dynamics, second turn to the eigenvalues of the lowest-order RDMs, which constitute a highly
sensitive measure for correlations, and third compare the whole
lowest order RDMs to the corresponding exact results. For a deeper interpretation
of these findings, we thereafter analyze the exact results for the
whole $N$-particle wavefunction as well as for the corresponding $o$-particle
correlations. Finally, we investigate the performance of the
correction strategies outlined in Section \ref{sec:corrections}.

\subsubsection{Particle-number imbalance}

\begin{figure}[t]
\centering
 \includegraphics[width=0.495\textwidth]{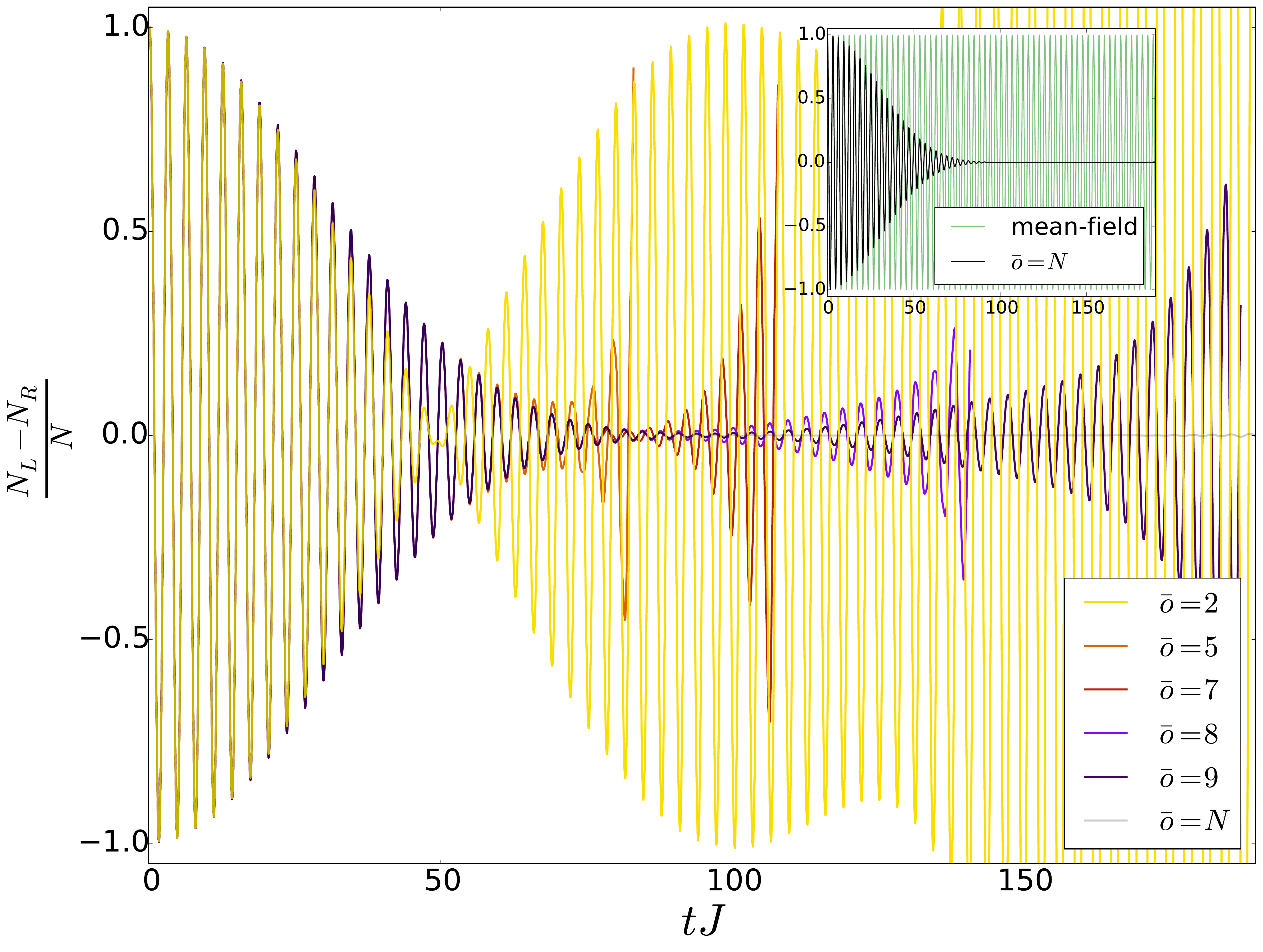}
 \caption{(color online) Time evolution of the particle-number imbalance $(N_L-N_R)/N$
 with $N_i\equiv\langle\hat n_i\rangle$, $i\in\{L,R\}$ for various
 truncation orders $\bar o$. Inset: numerically exact solution of the many-body Schr\"odinger
 equation in comparison to the corresponding mean-field calculation. Parameters: $N=10$ atoms located initially in the left
 well, dimensionless interaction parameter $\Lambda=0.1$.}
 \label{fig:BH_imbalance}
 \end{figure}

In order to study the tunneling dynamics, the imbalance of the particle 
numbers between the left and right well, $[\langle\hat n_L\rangle-\langle\hat n_R\rangle]/N$,
is depicted in Figure \ref{fig:BH_imbalance} for $N=10$ atoms.
Focusing first on the inset, which shows the numerically exact results,
we see the expected collapse of tunneling oscillations due to a dephasing
of the populated post-quench Hamiltonian eigenstates. Indeed, this collapse
happens on the time-scale $t_{\rm mf} \approx 46/J$, while a 
corresponding Gross-Pitaevskii
mean-field simulation (see inset of Figure \ref{fig:BH_imbalance}) 
reveals undamped tunneling oscillations. After $t\sim 200/J$, a
revival of the tunneling oscillations emerges in the numerically exact
calculation (not shown). 

Turning now to the truncated BBGKY approach, we see that all truncation orders
$\bar o\geq2$ give good results for the first $\sim 8$ tunneling oscillations.
Thereafter, the $\bar o=2$ curves departs from both the exact
and the higher truncation-order results, and features a premature maximal
suppression of tunneling oscillations at $t\sim50/J$. In
the subsequent premature revival of tunneling oscillations unphysical 
values $|\langle\hat n_L\rangle-\langle\hat n_R\rangle|/N>1$ are reached
at about $t=100/J$, indicating a lack of $1$-RDM representability.

These findings suggest that higher-order correlations than $\hat c_2$ play a 
significant role. Increasing the truncation order $\bar o$ stepwise up to
the maximally reasonable order $\bar o=N-1=9$,
we clearly see that the accuracy of our results improves systematically.
The larger $\bar o$ is, the more accurate is the collapse of the tunneling
oscillations described. However, all non-trivial truncations $\bar o<N$
predict a premature revival of the tunneling oscillations, which goes
hand and in hand with a maximal suppression of the tunneling-oscillation
amplitude to small but noticeable values (while the exact results
do not feature noticeable oscillations at the corresponding times).
We note that for $2<\bar o<10$ the simulations suffer from drastic instabilities
of the EOM,
being discussed in the subsequent Section, such that
we had to stop them after a certain time. This is why the corresponding
curves in Figure \ref{fig:BH_imbalance} are not provided for the
whole range of depicted times.

\subsubsection{Natural populations} 
 \begin{figure}[t]
   \includegraphics[width=0.495\textwidth]{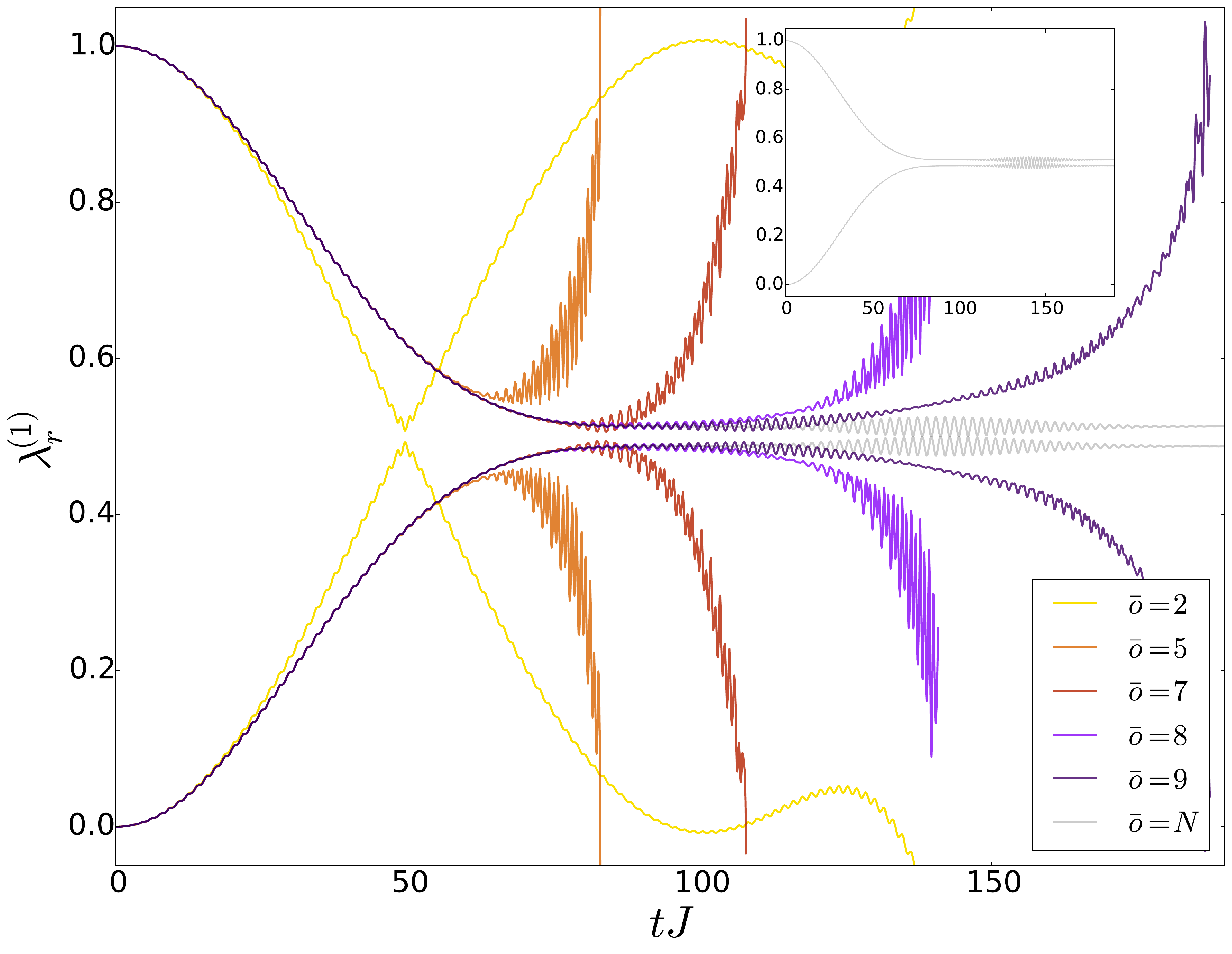}
   \caption{(color online) Natural populations of the $1$-RDM for various
   truncation orders $\bar o$. Inset: numerically exact solution of the many-body Schr\"odinger
 equation. Parameters: same as in Figure \ref{fig:BH_imbalance}.}
   \label{fig:BH_npop1}
\end{figure}
 \begin{figure}[t]
   \includegraphics[width=0.495\textwidth]{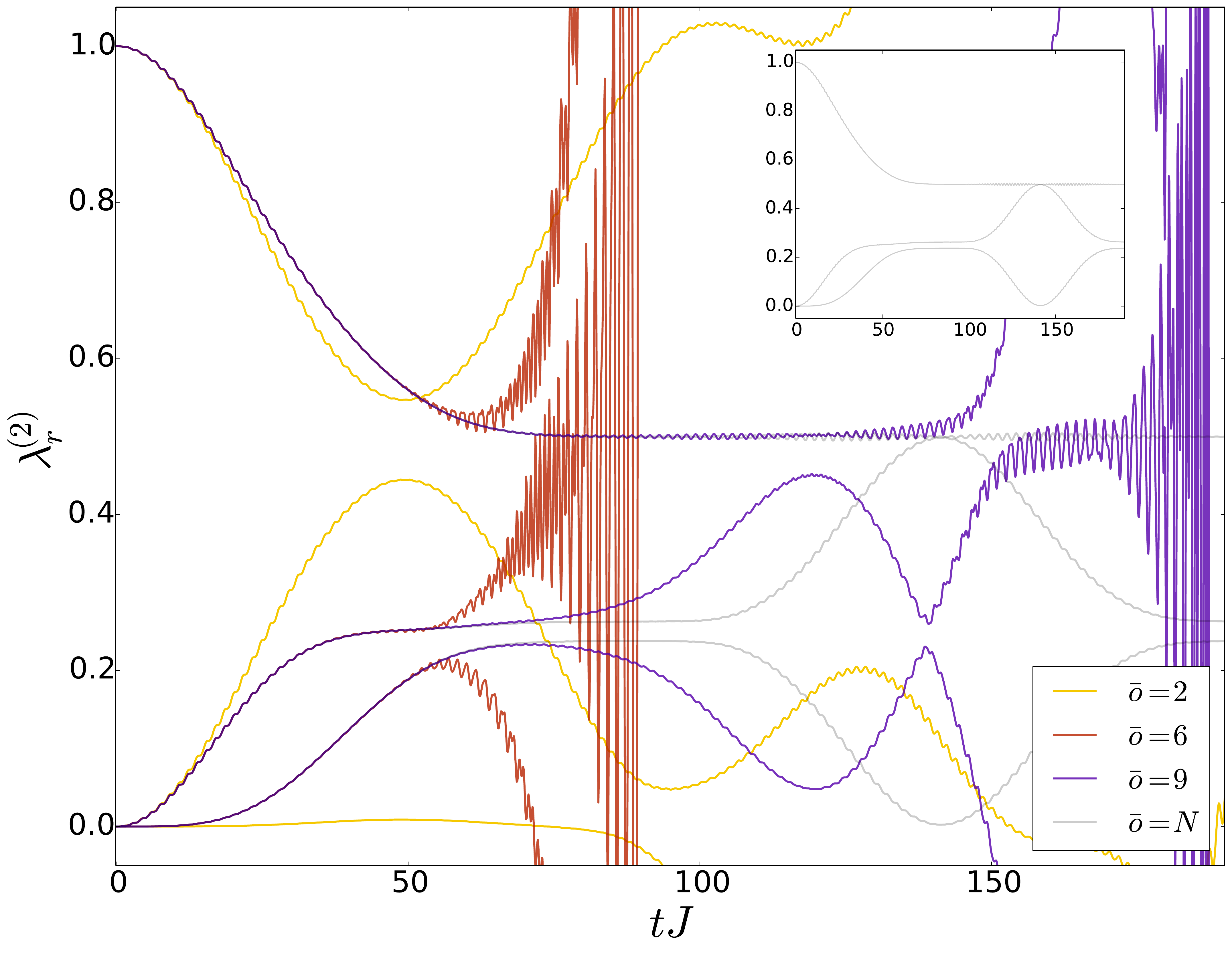}
   \caption{(color online) Natural populations of the $2$-RDM for various
   truncation orders $\bar o$. Inset: numerically exact solution of the many-body Schr\"odinger
 equation. Parameters: same as in Figure \ref{fig:BH_imbalance}.}
   \label{fig:BH_npop2}
\end{figure}

Next we analyze the NPs of the 1-RDM
in Figure 
\ref{fig:BH_npop1}, which can diagnose beyond mean-field behavior. 
The numerically exact results (see corresponding inset)
reveal dynamical quantum depletion leading to 
a two-fold fragmented condensate for $t\gtrsim 80/J$ with almost equal population
of the corresponding NOs, $\lambda_1^{(1)}\approx0.5\approx\lambda_2^{(1)}$ (see also e.g.\
\cite{univ_frag_dw}). Strikingly fast oscillations in these NPs emerge and decay
around $t\sim140/J$, which we can connect to the periodical emergence and decay
of a NOON state of the total system (see below).

The corresponding results of the truncated BBGKY approach feature a similar 
dependence on the truncation order $\bar o$ as the particle-number imbalance does.
While the $\bar o=2$ prediction starts to deviate noticeably from the exact results already
for $t\gtrsim25/J$, we obtain trustworthy results for a longer time, the larger
$\bar o$ is chosen. In particular, the truncated BBGKY approach can accurately
determine the achieved mean degree of fragmentation 
(see $\bar o=8,9$ results at $t\sim100/J$). Even for the largest truncation order $\bar o=9$, however,
the truncated BBGKY simulations 
predict a premature and very fast 
revival of condensation (i.e. $\lambda_1^{(1)}\approx 1$), while 
this process starts only after $t\sim 200/J$ in the exact calculation and happens
more slowly (not shown).
Most importantly, this unphysical fast re-condensation overshoots the
range of valid NPs such that the $1$-RDM ceases to be positive semi-definite, indicating
an exponential-like instability of the EOM. 

While we have so far only studied the prediction of the truncated BBGKY approach
for one-particle properties, 
we now inspect the NPs of the $2$-RDM in Figure
\ref{fig:BH_npop2}, also called
natural geminal populations \cite{rdm_Sakmann_PRA2008}. 
The exact dynamics (see the inset) features two important aspects, which we have
also observed for the NPs of higher-order RDMs (not shown). (i)
The dominant NP $\lambda_1^{(2)}$ first loses weight in favor for 
the other NPs. (ii) At about $t\sim140/J$, all NPs are suppressed except for
$\lambda_1^{(2)}\approx 0.5\approx\lambda_2^{(2)}$. Having observed
the latter feature for the NPs of all orders $o\in\{1,...,9\}$, we may conclude that 
in this stage of the dynamics a subsystem of $o$ particles occupies approximately
only two $o$-particle states with almost equal probabilities. As we will see below,
this finding is caused by the periodical emergence and decay
of a NOON state of the total system which is discussed below.

Turning now to the predictions of the truncated BBGKY approach, we 
see again a systematic improvement of accuracy with increasing truncation
order $\bar o$. The maximal time for which the highest truncation order 
$\bar o=9$ gives reliable results, however, has reduced from $t\sim110/J$ 
for the $1$-RDM NPs (see Figure \ref{fig:BH_npop1}) to 
$t\sim70/J$ for the $2$-RDM NPs (see Figure \ref{fig:BH_npop2}). Thereafter,
the largest NP $\lambda_1^{(2)}$ is well described until $t\sim130/J$,
while the other two NPs already show strong deviations: it seems that the emergence of 
the feature (ii) discussed above happens premature, name at about $t\sim120/J$.
Furthermore, we also witness the exponential-like instabilities leading to 
$2$-RDM NPs outside the interval $[0,1]$.

\begin{figure*}[t]
 \includegraphics[width=0.495\textwidth]{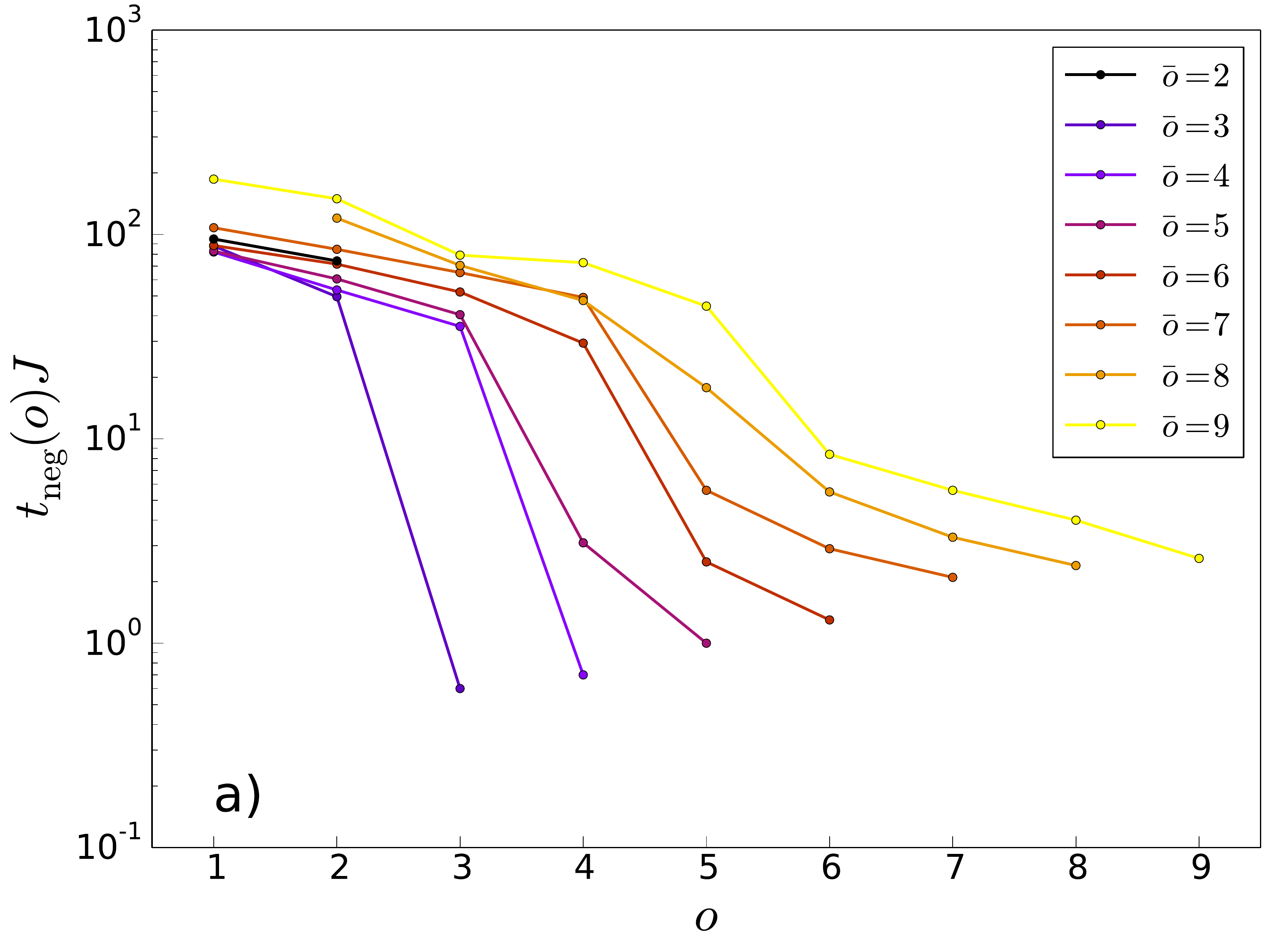}
 \includegraphics[width=0.495\textwidth]{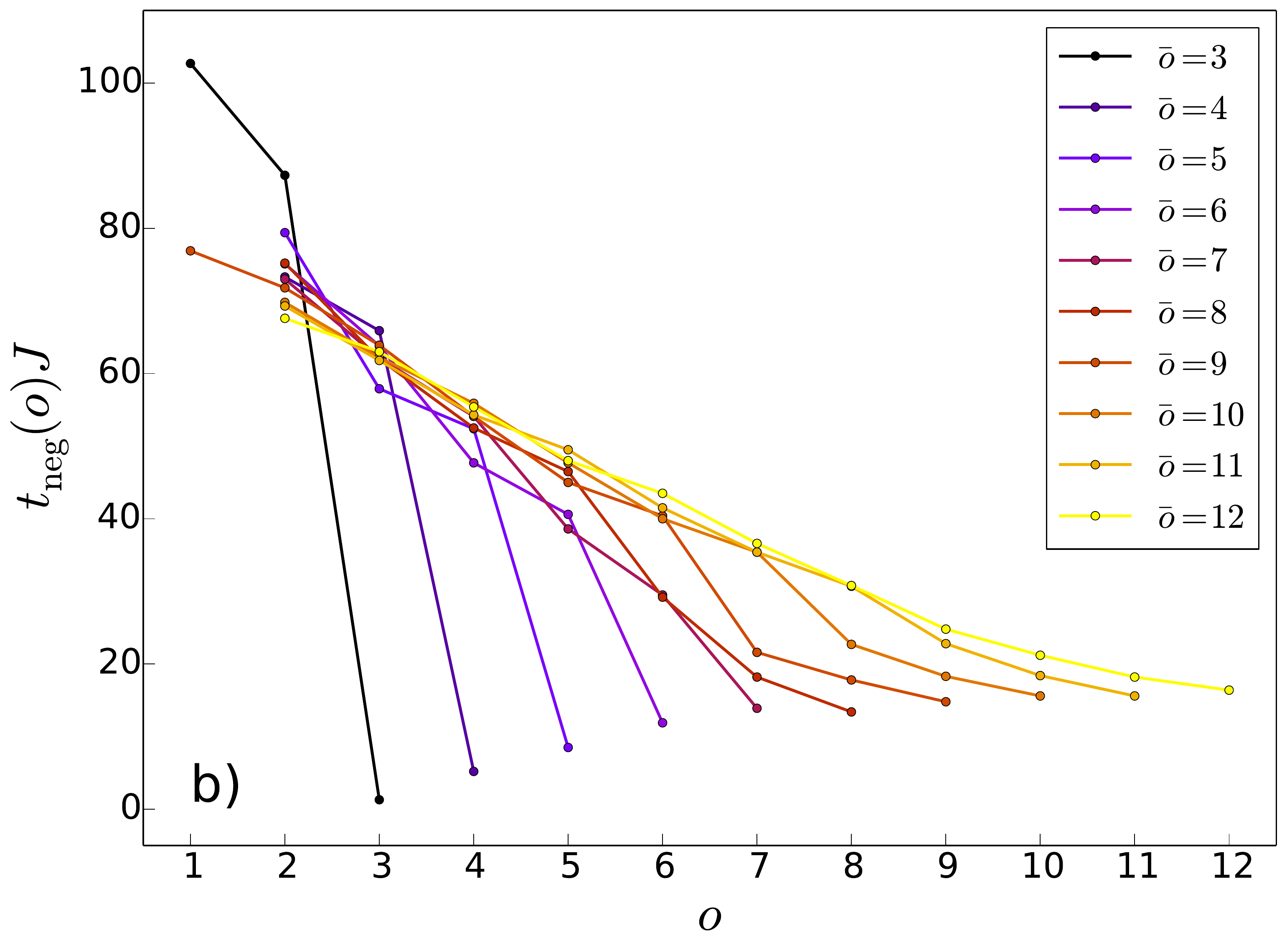}
 \caption{(color online) First time $t_{\rm neg}(o)$ when the lowest $o$-RDM NP is smaller than $\epsilon=-10^{-10}$
 in dependence on $o$ for various truncation orders $\bar o$.
 a) Same parameters as in Figure \ref{fig:BH_imbalance}. b) Same as a) but for the atom number $N$ increased to $100$
 while keeping the interaction parameter $\Lambda=0.1$ constant. The results for $\bar o=2$ are not plotted in b)
 and read $t_{\rm neg}(1)\approx216/J$ as well as  $t_{\rm neg}(2)\approx168/J$. }
 \label{fig:BH_neg_t}
\end{figure*}

In order to analyze how this unphysical behavior emerges, we depict the first
time $t_{\rm neg}(o)$ when the lowest $o$-RDM NP $\lambda^{(o)}_{o+1}$ 
is smaller than the threshold $\epsilon=-10^{-10}$ for various
$o$ and different truncation orders $\bar o$ in Figure \ref{fig:BH_neg_t} a).
For fixed truncation order $\bar o$, $t_{\rm neg}(o)$ decreases 
with increasing order $o$. This means that the representability defect
of $\hat\rho_o$ lacking positive semi-definiteness starts at the
truncation order $o=\bar o$ and propagates then successively to
lower orders due to coupling via the collision integral. For most
orders $o$, we moreover find that $t_{\rm neg}(o)$ increases with
increasing truncation order $\bar o$, which fits to the above findings regarding
enhanced accuracy for larger $\bar o$ (exceptions occur at order $o=1,2$  in particular
for $\bar o=2$). 

Increasing the number of atoms to $N=100$ while keeping the dimensionless interaction parameter
$\Lambda=0.1$ constant, we again find a monotonous decrease of $t_{\rm neg}(o)$ with increasing $o$
for fixed truncation order $\bar o$ (see Figure \ref{fig:BH_neg_t} b)). This confirms
the above finding that the lack of positivity successively propagates from higher to lower orders.
In contrast to the $N=10$ case, we only find an enhancement of $t_{\rm neg}(o)$ with increasing
$\bar o$ for orders $o\geq6$. In particular, we see that the largest truncation order considered,
$\bar o=12$, features the smallest $t_{\rm neg}(o=2)$. It is 
quite possible that the an ``enhancement'' of non-linearity with increasing truncation
order $\bar o$ (note that the applied closure approximation, cf.\ Section \ref{sec:trunc_recursive_cluster}, is a polynomial of degree $(\bar o+1)$ 
in $\hat\rho_1$ and of degree $\lfloor(\bar o+1) /o\rfloor$ in the cluster $\hat c_o$) is the
reason why the BBGKY EOM are more prone to these instabilities for larger $\bar o$.

Having compared so far only certain aspects of $o$-particle properties, 
we finally aim at comparing the prediction of the truncated BBGKY approach for 
the whole $o$-RDM to the exact results.

\subsubsection{Reduced density operators}

\begin{figure*}[t]
 \includegraphics[width=0.495\textwidth]{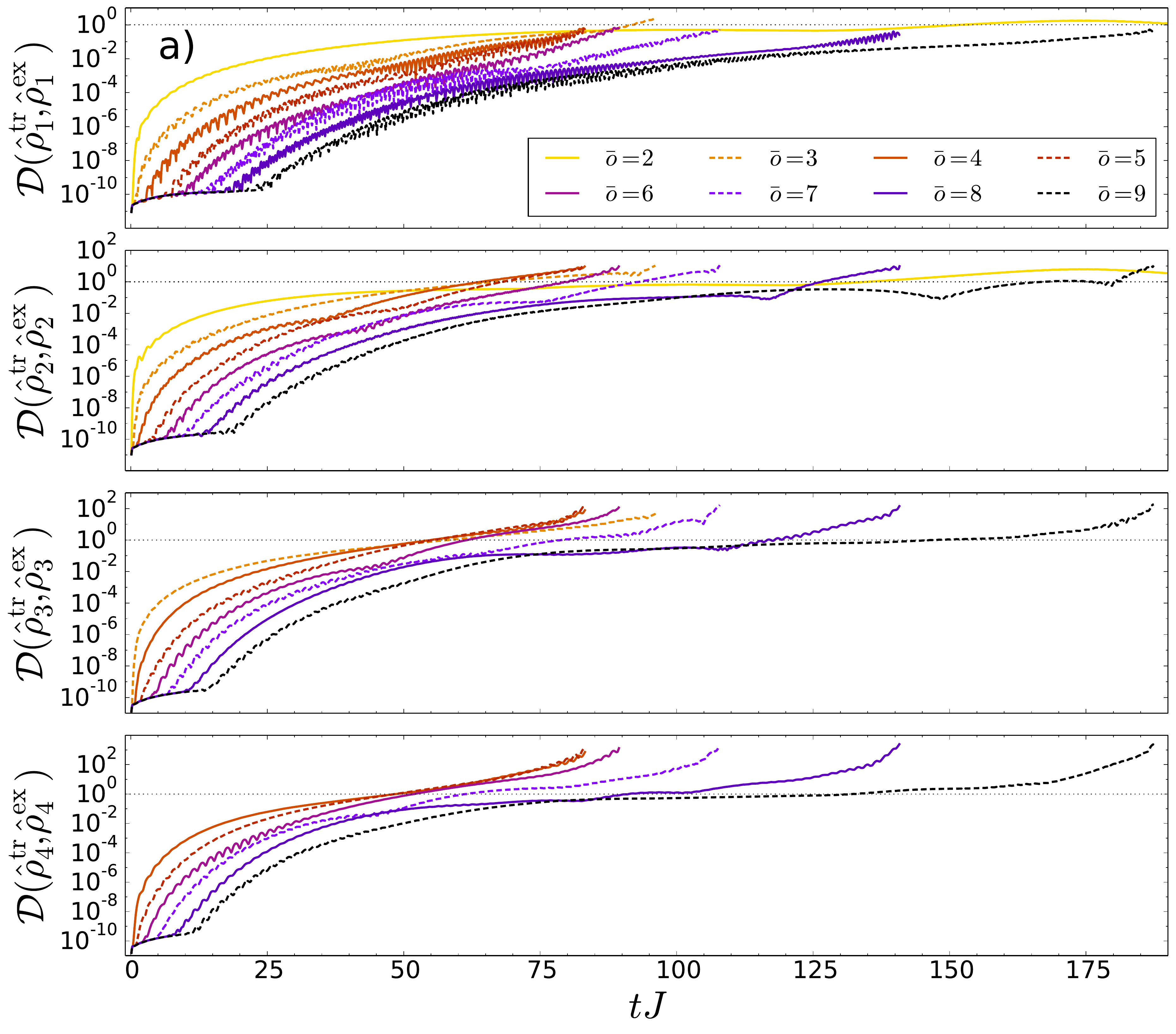}
  \includegraphics[width=0.495\textwidth]{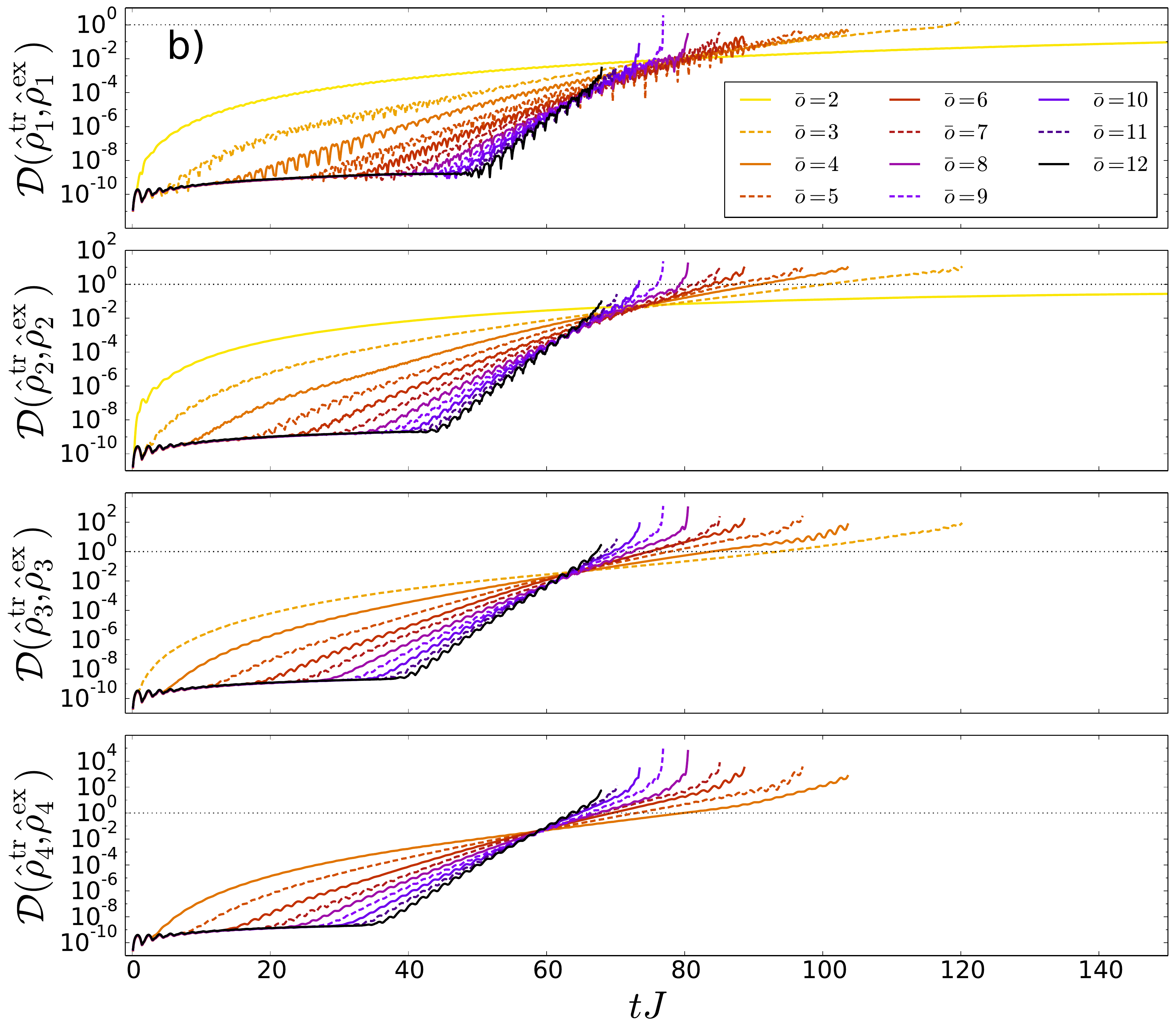}
 \caption{(color online) Time evolution of the trace distance $\mathcal{D}(\hat\rho_o^{\rm tr},\hat\rho_o^{\rm ex})$
 between the exact result and the truncated BBGKY prediction for the $o$-RDM ($o=1,...,4$) and various truncation
 orders $\bar o$. The dotted horizontal lines at unity ordinate value indicate the upper bound for the trace distance
 between two density operators (see main text).
 a) Same parameters as in Figure \ref{fig:BH_imbalance}. b) Same as a) but for the atom number $N$ increased to $100$
 while keeping the interaction parameter $\Lambda=0.1$ constant.}
 \label{fig:BH_rdm_comp}
\end{figure*}

For this purpose, we 
take the trace distance
$\mathcal{D}(\hat\rho_o^{\rm tr},\hat\rho_o^{\rm ex})\equiv||\hat\rho_o^{\rm tr}-\hat\rho_o^{\rm ex}||_1/2$
\cite{nielsen_chuang_book} as a measure for deviations between the
truncated BBGKY prediction for the $o$-RDM denoted by $\hat\rho_o^{\rm tr}$ and the
numerically exact result $\hat\rho_o^{\rm ex}$. Here, $||\cdot||_1$ refers to the trace-class norm (also 
called Schatten-1 norm) being defined as $||\hat A||_1\equiv\tr(\sqrt{\hat A^\dagger\hat A})$
for any trace-class operator $\hat A$. For hermitian operators $\hat A$, $||\hat A||_1$ 
equals the sum of absolute values of $\hat A$'s eigenvalues. 
One can easily prove the inequality $|\tr(\hat A_o\hat\rho_o^{\rm tr})-\tr(\hat A_o\hat\rho_o^{\rm ex})|
\leq 2||\hat A_o||_1\,\mathcal{D}(\hat\rho_o^{\rm tr},\hat\rho_o^{\rm ex})$ where $\hat A_o$ 
denotes an arbitrary $o$-body observable. This means that $\mathcal{D}(\hat\rho_o^{\rm tr},\hat\rho_o^{\rm ex})$
provides an upper bound for the deviations in the expectation value predictions for $\hat A_o$. 
Moreover,
given that its arguments are density operators (i.e. hermitian, positive semi-definite
and trace one), the trace-distance is bounded by
$\mathcal{D}(\hat\rho_o^{\rm tr},\hat\rho_o^{\rm ex})\in[0,1]$ and
can be interpreted as the
probability that these two quantum states can be distinguished by the outcome of a single measurement 
\cite{nielsen_chuang_book}. 

In Figure \ref{fig:BH_rdm_comp}, we depict $\mathcal{D}(\hat\rho_o^{\rm tr},\hat\rho_o^{\rm ex})$
for the orders $o=1,...,4$ and various truncation orders $\bar o$, where subfigures a) and b) refer
to the $N=10$ and $N=100$ case with the same interaction parameter 
$\Lambda=0.1$, respectively. For fixed truncation order $\bar o$, we clearly see that the accuracy of the 
truncated BBGKY prediction for the $o$-RDM decreases with increasing order $o$. Up to a
certain time, which depends on the order $o$, we moreover find $\mathcal{D}(\hat\rho_o^{\rm tr},\hat\rho_o^{\rm ex})$
to decrease with increasing truncation order $\bar o$. 

The instabilities of the truncated BBGKY EOM manifest themselves in the trace distant exceeding
its upper bound $\mathcal{D}(\hat\rho_o^{\rm tr},\hat\rho_o^{\rm ex})\leq 1$ for density operators,
implying that $\hat\rho_o^{\rm tr}$ lacks to have trace one or to be positive semi-definite. Since 
the conservation of the initial RDM trace is ensured by the truncated BBGKY approach, violations
of $\tr(\hat\rho_o^{\rm tr})=1$ can at most occur numerically if the system gets deep into the
exponential-like instability (where we observe the truncated BBGKY EOM to become stiff
such that the integrator has a hard time). Thus, exceeding the upper bound on the trace distance
is connected to a lack of positive semi-definiteness and can be observed to
happen earlier for increasing order $o$ and fixed truncation order $\bar o$.

For the case of $N=100$ atoms (see Figure \ref{fig:BH_rdm_comp} b)), we observe the additional
particularity that in the vicinity of $t\sim63/J$ the accuracy of the truncated BBGKY prediction
for the $o$-RDM does not depend on the truncation order $\bar o$, which happens slightly earlier
for larger $o$. Before this point, a systematic increase of accuracy is observed for increasing 
truncation order $\bar o$. Thereafter, lower truncation orders give (slightly) better results than
higher ones. Furthermore, while in the $N=10$ case one-body properties (such as e.g.\ the particle-number
imbalance) can be described with reasonable accuracy up to $t\sim2\, t_{\rm mf}$ (when the collapse
of tunneling oscillations has already taken place), the instabilities hinders us to obtain
accurate results for $t$ larger than about $0.56\, t_{\rm mf}$ in the case of $N=100$ (at this time,
the tunneling oscillation amplitude is still significant).

\subsubsection{Many-body state and $o$-particle correlations: exact results}

\begin{figure*}[t]
 \includegraphics[width=0.7\textwidth]{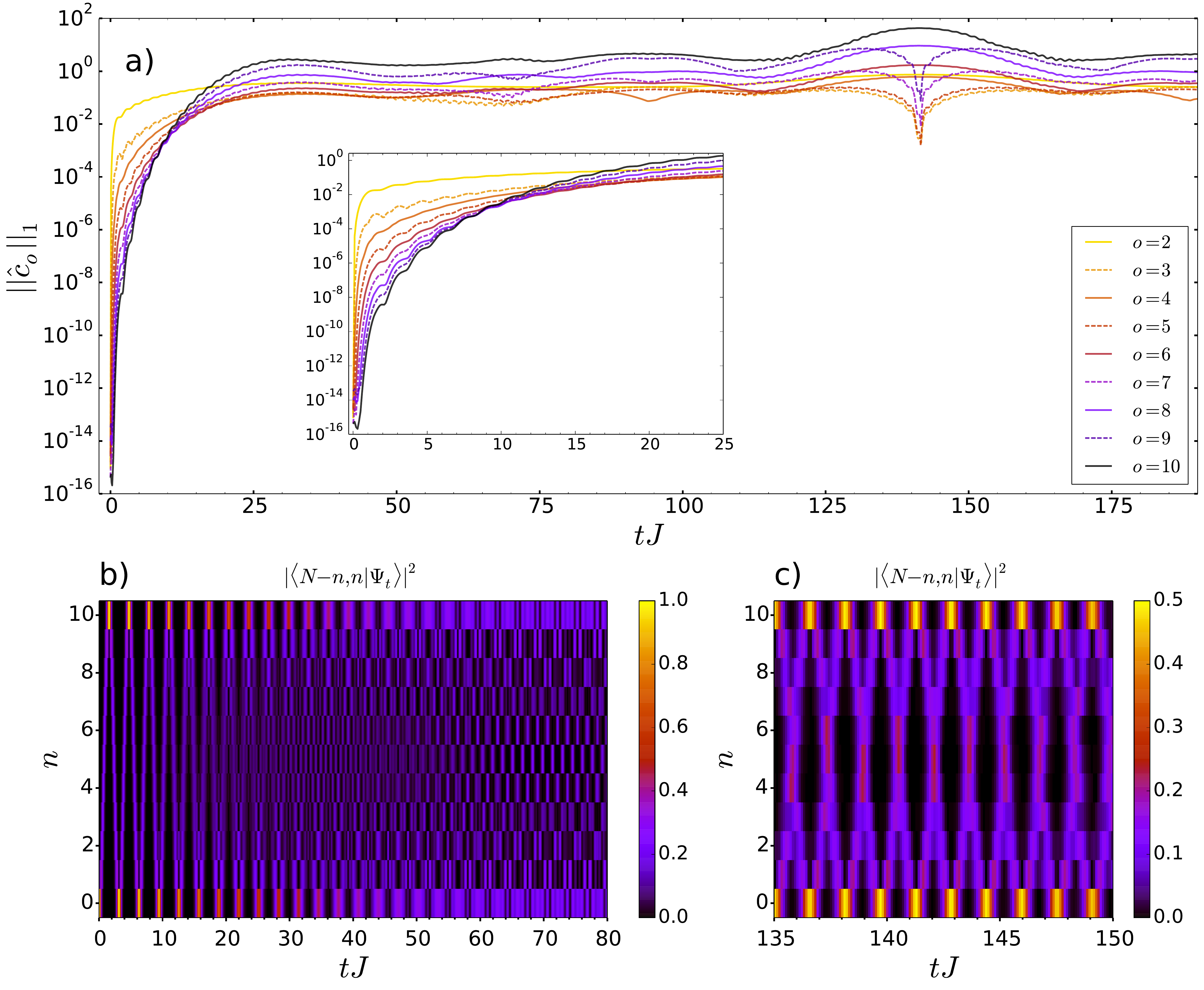}
 \caption{(color online) a) Time evolution of the cluster's trace-class norm $||\hat c_o||_1$
 for all orders $o$, obtained from the numerically exact solution of the
 time-dependent Schr\"odinger equation. Inset: zoom into early time dynamics. b) and c) Probability
 to find $n$ atoms in the right and $(N-n)$ atoms in the left well, $|\langle N-n,n|\Psi_t\rangle|^2$, versus time
 for two characteristic stages of the dynamics. Parameters: same as in Figure \ref{fig:BH_imbalance}.}
  \label{fig:BH_cluster_dynamics}
\end{figure*}

In order to obtain physical insights into the above findings, we finally come back
to the numerically exact results for $N=10$ and measure the strength of $o$-particle correlations
in terms of $||\hat c_o||_1$ in
Figure \ref{fig:BH_cluster_dynamics} a). From the inset, we infer that the correlations
initially build up in a hierarchical manner. First, only two-particle correlations  start to play a role, 
then three-particle correlations and so on. This hierarchy in $||\hat c_o||_1$ holds, however, only until
$t\sim8/J$, when the ordering of the $||\hat c_o||_1$'s with respect to the
order $o$ starts to become reversed. After a certain point, $N$-particle correlations become the most dominant
ones. 
This holds in particular in the vicinity of $t\sim140/J$, where we have observed fast oscillations
in the NPs $\lambda^{(1)}_{1/2}$ and found for all orders $o=1,...,9$ that
the RDMs feature approximately only two finite NPs $\lambda^{(o)}_{1}\approx0.5\approx\lambda^{(o)}_{2}$.
At this stage of the dynamics, all clusters $\hat c_o$ of odd order $o$ are strongly suppressed.

For connecting the above findings regarding $o$-particle correlations to the full many-body state, 
we depict in the Subfigures \ref{fig:BH_cluster_dynamics} b)
and c) the probability $|\langle N-n,n|\Psi_t\rangle|^2$ of finding $n$ atoms in the right and $(N-n)$ atoms in the left well.
For the early dynamics, we witness how the system
becomes delocalized in the Fock space such
that the tunneling oscillations become suppressed (Subfigure \ref{fig:BH_cluster_dynamics} b)). 
At later times, around
$t\sim140/J$, we, however, find the system to periodically oscillate between 
a NOON state $(|N,0\rangle+e^{i\theta}|0,N\rangle)/\sqrt{2}$ (with some phase $\theta\in\mathbb{R}$)
and some broad distribution being approximately symmetric with respect to its maximum 
at about $n=5$ (Subfigure \ref{fig:BH_cluster_dynamics} c)). Due to this approximate symmetry of the 
distribution around $n=5$, the particle-number imbalance approximately equals
$[\langle\hat n_L\rangle-\langle\hat n_R\rangle]/N\approx0.5$, i.e.\ tunneling
oscillations are still suppressed. This approximate symmetry moreover leads to
a doubling of the oscillation frequency compared to the initial tunneling-oscillation frequency,
which is most probably linked to the fast oscillations in  $\lambda^{(1)}_{1/2}$.
Finally, one can analytically show  that 
the $n$-RDM of the above mentioned NOON state reads $\hat\rho_n=(|n,0\rangle\!\langle n,0|+|0,n\rangle\!\langle 0,n|)/2$,
meaning that the state of an $n$-particle subsystem is an incoherent statistical mixture with
all particles residing in the left (right) well with probability $0.5$.
Thereby, we can directly connect the fact that the RDMs of all orders feature approximately only two finite
NPs of approximately equal value to the underlying many-body state. Coming back to the findings for  $||\hat c_o||_1$
of Figure \ref{fig:BH_cluster_dynamics} a), we may conclude that a NOON state 
leads to strong high-order correlations $\hat c_o$ such that 
truncating the BBGKY hierarchy by means of the applied cluster expansion 
cannot be expected to give accurate results.

In summary, we have seen following.  (i) While the truncated BBGKY approach gives highly accurate
results for short times with controllable accuracy via the truncation order $\bar o$,
 the BBGKY approach
shows deviations at longer times. (ii) Exponential-like instabilities, induced
by the non-linear truncation approximation, propagate form high to low orders
and lead to unphysical results at a certain point. 
(iii) $o$-particle correlations arise very fast
in this tunneling scenario and soon cease to be in decreasing order with respect
to $o$. (iv) The system evolves into a NOON state being dominated by $N$-particle
correlations.

There appear to be at least two plausible causes why the BBGKY approach fails at a certain point:
First, the number of terms in the cluster expansion
\eqref{eq:cluster_def_with_symb} drastically increases with the order $o$,
which implies that clusters should decay fast for a
controllable approximation. For example, at the largest truncation order
considered above, $\bar o=12$, the truncation approximation $\hat\rho_{13}^{\rm appr}$ 
already involves $100$ classes of terms. Our findings (iii) and (iv), however, 
might indicate that this system is not suitable for 
a truncation based on the $o$-particle correlations defined in
Section \ref{sec:trunc_recursive_cluster}. Other truncation approximations
might by more suitable.

Second, the exponential-like instabilities, being connected to a 
lack of representability, might be the main cause for the failure of the
BBGKY EOM at longer times. This hypothesis is supported by the 
fast break-down of the BBGKY approach in the $N=100$ case for the truncation
order $\bar o=12$. For this reason, we analyze next the performance of the 
correction strategies outlined in Section \ref{sec:corrections}.

\subsubsection{Performance of the correction algorithms}

\begin{figure*}[t]
 \includegraphics[width=0.9\textwidth]{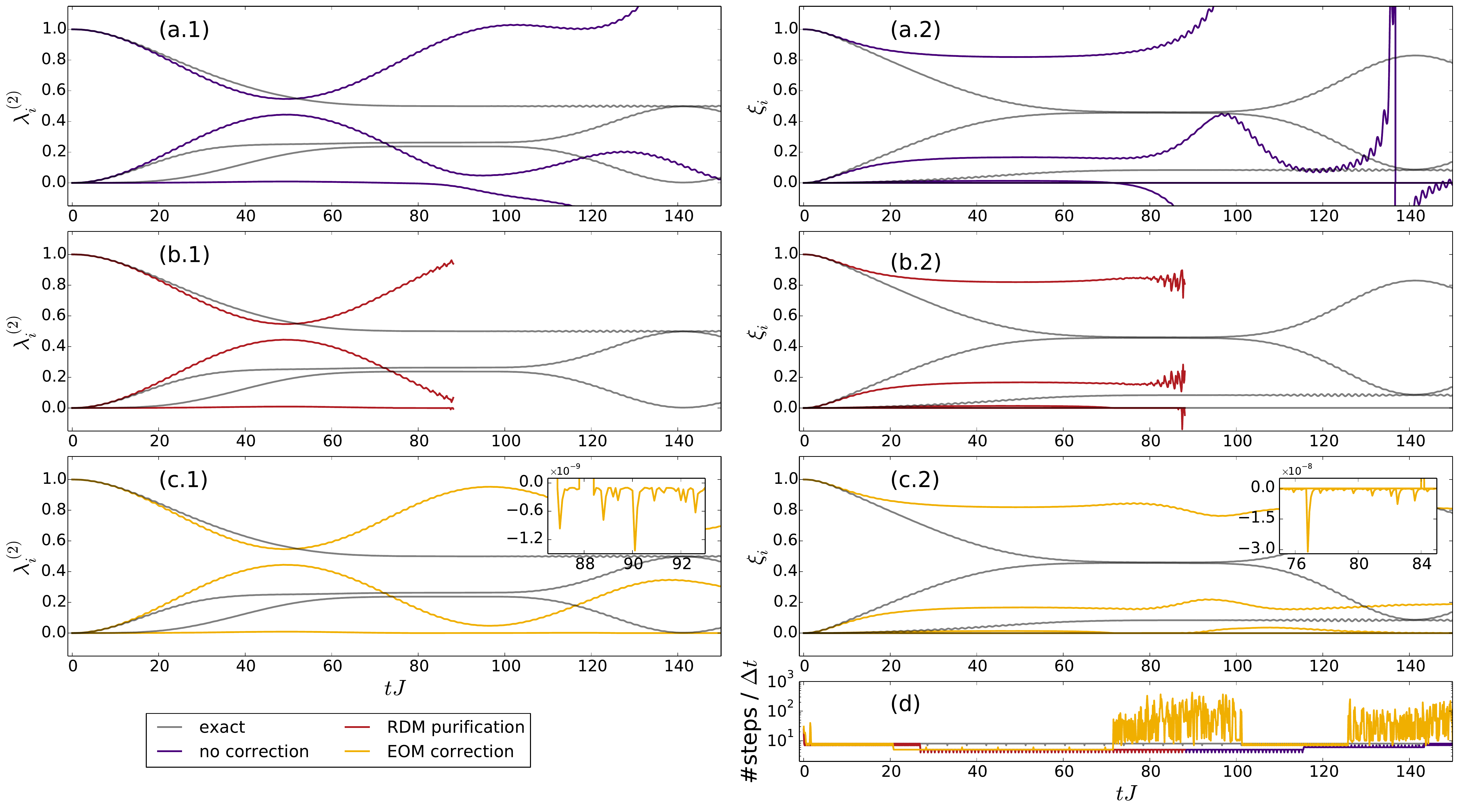}
 \caption{(color online) Comparison of the correction
 strategies outlined in Section \ref{sec:corrections} (for the BBGKY hierarchy
 truncated at $\bar o=2$). Left column:
 Time evolution of the NPs $\lambda^{(2)}_i$. Right column [except for (d)]:
 Time evolution of the $\hat K_2$ eigenvalues $\xi_i$. First row: Truncated BBGKY
 results without correction versus exact ones. Second row: Truncated BBGKY
 results with iterative minimal-invasive purification of the $2$-RDM after each
 $\Delta t=0.1/J$ (maximal number of iterations: $500$). Third row:
  Truncated BBGKY results with minimal-invasive correction of the $2$-RDM EOM
  (damping rate of negative eigenvalues: $\eta=10J$). Insets of (c.1) and (c.2):
  close-ups showing the imposed exponential damping of negative eigenvalues.
  For both correction strategies, eigenvalues are regarded as negative if they 
  are smaller than the threshold $\epsilon=-10^{-10}$. Subfigure (d): number 
  of integrator steps \cite{zvode} per write-out time-step $\Delta t$.
  Parameters: same as in Figure \ref{fig:BH_imbalance}.}
  \label{fig:BH_correction_alg}
\end{figure*}

In the following, we first focus on the correction algorithms applied 
to the BBGKY hierarchy truncated at $\bar o=2$. Thereafter, we comment on
the performance of these algorithms if extended to larger truncation orders
by means of a corresponding ansatz for the correction operator (see Section \ref{sec:corrections}).

Figure \ref{fig:BH_correction_alg} depicts the time evolution of the NPs $\lambda^{(2)}_i$
and the $\hat K_2$ eigenvalues $\xi_i$ for the truncated BBGKY results without correction,
with the iterative minimal invasive purification of the $2$-RDM and with the minimal
invasive correction of the $2$-RDM EOM in comparison to the exact results.
Apparently, all cases deviate significantly from the exact results after 
$t\gtrsim17/J$ so that we shall concentrate here solely on the stabilization
performance of the correction algorithms. 

Inspecting first the uncorrected
results [Figures \ref{fig:BH_correction_alg} (a.1), (a.2)], we observe that 
the $K$-condition (i.e.\ $\hat K_2\geq0$) diagnoses earlier a lack of representability
compared to the $D$-condition (i.e.\ $\hat\rho_2\geq0$). For both
operators, the falling of an eigenvalue below zero is accompanied by an avoided
crossing which involves the next-larger eigenvalue (this is hardly visible 
in the case of the $\hat K_2$ eigenvalue where the avoided crossing happens at 
about $t\sim71.5/J$). In fact, we have observed that level-repulsion ``pushes'' eigenvalues below zero 
in various other situations (see also Section
\ref{sec:breath}).

Now turning to the minimal invasive correction algorithms based on the $2$-norm minimization
of the correction operator $\hat{\mathcal{C}}_2$,
we set the threshold $\epsilon$ below 
which an eigenvalue is regarded as negative to $-10^{-10}$. Let us first inspect the dimensionality
of the optimization problem underlying both our purification algorithm
of the $2$-RDM and the correction algorithm of its EOM (see Appendix \ref{app:corr_rdm} for the details).
The bosonic hermitian correction operator $\hat{\mathcal{C}}_2$ can be parametrized by
$m^2(m+1)^2/4=9$ real-valued parameters. Requiring $\hat{\mathcal{C}}_2$ to be contraction-free
and energy-conserving imposes $m^2+1=5$ constraints such that the system of linear
equations corresponding to the constraints is underdetermined as long
as the numbers of negative $\hat\rho_2$ eigenvalues $d$ and negative $\hat K_2$ eigenvalues
$d'$ obey $d+d'<4$.

Figure  \ref{fig:BH_correction_alg} (b.1) and (b.2) depict the results if 
the iterative minimal-invasive purification algorithm is applied after each $\Delta t=0.1/J$. 
Clearly, we see that this correction
algorithm induces strong noise in the $\hat K_2$ eigenvalues when the smallest
eigenvalue $\xi_i$ has reached significant negative values in the
uncorrected BBGKY calculation [see subfigure (a.2)]. Actually, after $t=86.5/J$, the 
iterative purification algorithm fails to converge after the maximal number of $500$ steps. 
Thus, this iterative scheme 
fails to prevent that smallest eigenvalue is pushed
to negative values due to level repulsion.

In a certain sense, we may view the iterative purification algorithm of the $2$-RDM 
as being based on a fixed stepsize as well as perturbative. In each iteration step namely, 
we update $\hat\rho_2(t)
\rightarrow\hat\rho_2(t)+\hat{\mathcal{C}}_2$ with $\hat{\mathcal{C}}_2$ 
shifting negative eigenvalues to zero in first-order perturbation theory. In the correction
algorithm for the $2$-RDM EOM, we effectively allow for variable update stepsizes by
coupling the correction scheme to the integration of the EOM, i.e.\ to the
employed integrator ZVODE \cite{zvode} featuring adaptive stepsizes. Moreover, by imposing constraints 
on the time-derivative of negative eigenvalues, we realize a non-perturbative correction
scheme.

This can nicely be inferred from the insets of Figure \ref{fig:BH_correction_alg} (c.1) and (c.2)
showing a close-up of slightly negative eigenvalues. These are exponentially damped to zero, namely
as e.g.\ $\xi_i(t+\tau)=\xi_i(t)\exp[-\eta\tau]$ for $t$ and $\tau$ such that $\xi_i(t+\tau)<\epsilon$,
with the chosen damping constant $\eta=10J$. As a consequence, the truncated BBGKY EOM 
becomes stabilized and we have observed that the $D$- and $K$-representability condition
are fulfilled to a good approximation for at least $t\leq1000/J$ (times
later than $t=150/J$ not shown in Figure \ref{fig:BH_correction_alg}).
When enforcing negative eigenvalues to be damped to zero, one might 
fear that eigenvalues accumulate in the range $[\epsilon,0]$. This, however, is not the 
case as shown in the insets of Figure \ref{fig:BH_correction_alg} (c.1) and (c.2)
because no constraint on the time-derivative of an eigenvalue is enforced if its value exceeds the threshold $\epsilon$
such that the (corrected) EOM may lift this eigenvalue above zero. We finally
remark that the number of integrator steps per $\Delta t$
significantly increases
in the vicinity of avoided crossings of 
$\hat\rho_2$ or $\hat K_2$ eigenvalues close to zero 
[see Figure
\ref{fig:BH_correction_alg} (d)]. This finding confirms the non-perturbative, adaptive 
nature of the EOM correction algorithm and at the same time highlights 
the significance of controlling such avoided crossings for a successful stabilization of the
truncated BBGKY EOM.

Without showing additional graphical illustrations, let us now briefly comment on the behavior of the correction algorithms for truncation
orders $\bar o>2$, using the Mazziotti ansatz \cite{mazziotti_purification_2002} for the correction operators $\hat{\mathcal{C}}_o$
on orders $o>2$ (see Section \ref{sec:corrections}).
Focusing first on the RDM purification, we have observed that $\hat\rho_{\bar o}$
can be kept positive semi-definite up to a few tens $1/J$ longer (compared to the uncorrected case) before
this iterative correction algorithm fails to converge after $500$ steps. Due 
to the losing of positive semi-definiteness in decreasing sequence
with respect to the RDM order (see Figure \ref{fig:BH_neg_t}), 
we found for $\bar o\geq 4 $ that also  $\hat\rho_{\bar o-1}\geq0$ 
is valid for somewhat longer times compared to the
uncorrected case. Unfortunately, however, this correction scheme fails to converge so
early that it does not improve the timescale, on which 
the most important RDMs for making predictions for ultracold quantum gas
experiments, namely $\hat\rho_1$ and $\hat\rho_2$, obey the considered representability 
conditions. 

Extending the EOM correction scheme to higher truncation orders $\bar o>2$
by means of the Mazziotti ansatz for the higher-order correction operators unfortunately
proved to be quite unsuccessful. This failure manifests itself in an enormous increase
of integrator steps per $\Delta t$, i.e.\ the EOM becoming stiff, in combination
with the quadratic optimization problem for determining $\hat{\mathcal{C}}_2$ having
no solution, i.e.\ constraints contradicting one another. Unfortunately, we
cannot tell whether the latter is a fundamental problem or whether it is 
only induced by the EOM to become stiff 
due to an inappropriate ansatz of $\hat{\mathcal{C}}_o$ for $o>2$, potentially
leading to integration errors.

To sum up, while we can successively stabilize the BBGKY EOM  truncated at
order $\bar o=2$ by enforcing the $D$- and $K$-representability condition
via a minimal invasive correction of the $2$-RDM EOM, the issue of 
higher-order correlations becoming dominant after a certain time remains unsolved
in this example. Since this tunneling scenario might well be
unsuitable for a closure approximation based on neglecting certain few-particle correlations,
we now turn to an example, where a BEC becomes only slightly depleted in the course of the
quantum dynamics.
\subsection{Interaction-quench induced breathing dynamics of harmonically trapped bosons}\label{sec:breath}

In this application, we are concerned with collective excitations of
$N$ ultracold bosons confined to a quasi one-dimensional harmonic trap. In harmonic
oscillator units (HO units), the corresponding Hamiltonian reads
\begin{align}
 \hat H = \sum_{i=1}^N\frac{\hat p_i^2+\hat x_i^2}{2}+g\sum_{1\leq i<j\leq N}\delta(\hat x_i-\hat x_j)
\end{align}
where we model the short-range van-der-Waals interaction by the contact potential \cite{Pethick_Smith2008}
of strength $g$. Initially, we assume all atoms to reside in the ground state of 
the single-particle Hamiltonian, i.e.\ a Gaussian orbital, which
is the exact many-body ground state in the absence of interactions. Then, the interaction strength
is instantaneously quenched to $g=0.2$ such that the ideal BEC becomes slightly 
depleted and its density performs breathing oscillations, i.e.\ expands and contracts periodically.
This so-called breathing mode has been investigated theoretically as well as experimentally in different settings (see e.g.\ \cite{Stringari2002,Moritz2003,bauch_2009,Bonitz2014,Schmitz2013,Tschischik2013,Bouchoule2014} for single-component
systems
and e.g.\ \cite{pyzh_spectral_2017} for mixtures), and measuring its frequency proves to be useful
for characterizing the interaction regime \cite{Haller2009}.

\begin{figure}[t]
 \includegraphics[width=0.495\textwidth]{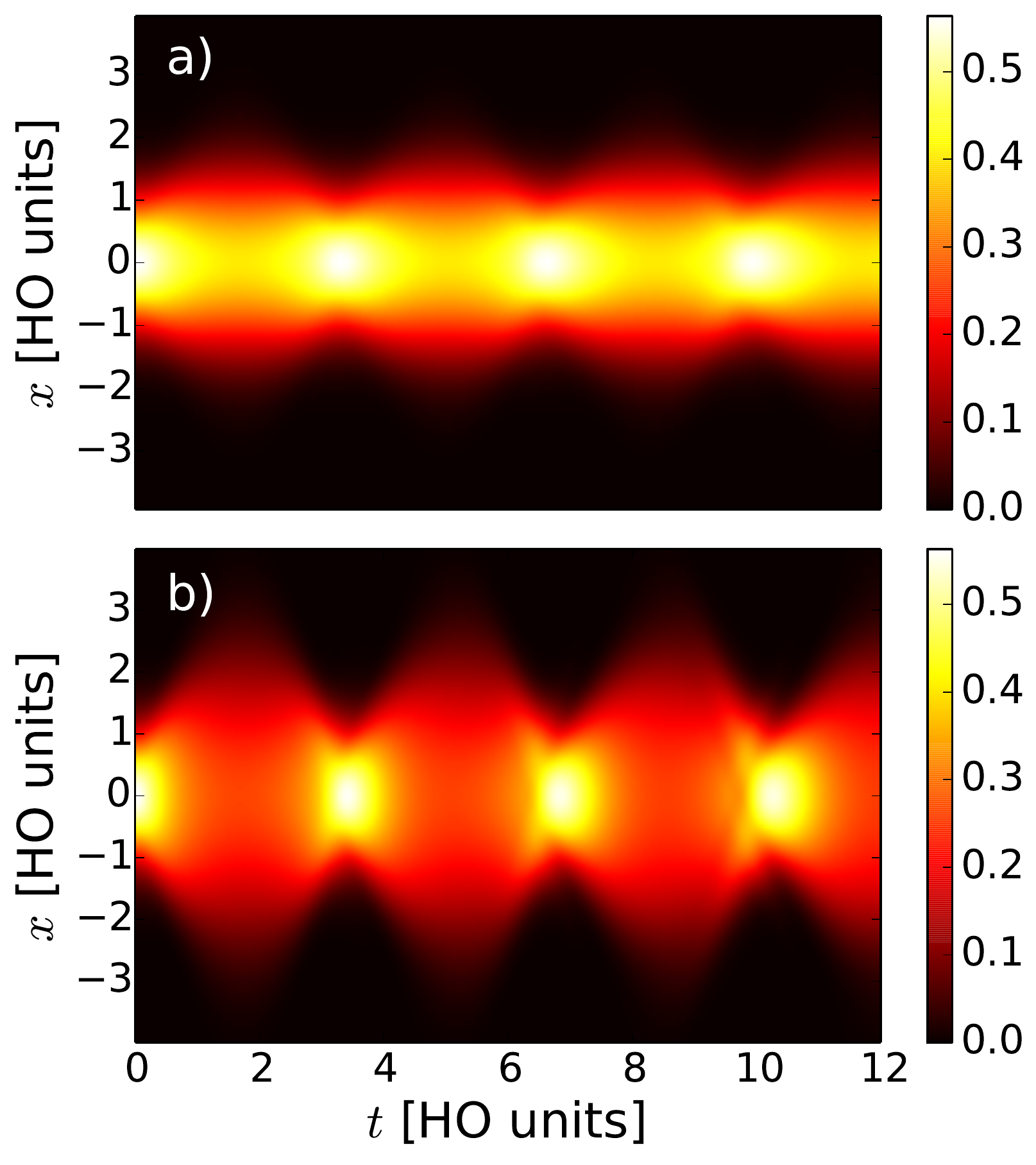}
 \caption{(color online) Time-evolution of the reduced one-body density $\rho_1(x;t)=\langle x|\hat \rho_1(t)| x\rangle$
 for $N=10$ [subfigure (a)] and $N=30$ [subfigure (b)] bosons quenched from the non-interacting ground-state
 to a contact-interaction strength of $g=0.2$. These results are obtained by MCTDHB simulations with $m=4$ dynamically
 optimized SPFs.}
  \label{fig:breath_dens_ex}
\end{figure}

Before we discuss the results of the truncated BBGKY approach, let us first
inspect the results of MCTDHB simulations with $m=4$ dynamically optimized SPFs,
which we obtain by our implementation
\cite{cao_multi-layer_2013,kroenke_multi-layer_2013,cao_unified_2017}. 
In fact, we find that the 
smallest natural population of the 1-RDM attains a maximal value of about $1.6\cdot 10^{-3}$,
which provides a good indicator in praxis that the contribution of this orbital is almost negligible in the 
calculation. One could improve the accuracy further by increasing the number of
SPFs, of course. Yet since we aim at benchmarking the truncated BBGKY approach,
which can be viewed as an additional approximation to the MCTDHB approach,
it is sufficient to take the MCTDHB simulations with $m=4$ SPFs as reference results.
For representing the SPFs,
a harmonic discrete variable representation \cite{dvr_and_their_utilization_Light_Carrington_2000,MCTDH_BJMW2000} with $n=256$ ($n=320$) grid points
is employed for case of $N=10$ ($N=30$) particles.

In Figure \ref{fig:breath_dens_ex},
we depict the time evolution of the reduced one-body density, i.e.\ the diagonal 
of the $1$-RDM in position representation  $\rho_1(x;t)=\langle x|\hat \rho_1(t)| x\rangle$,
for $N=10$ and $N=30$ bosons. In both cases, we clearly see that the atomic density 
periodically expands and contracts. Since the interaction quench leads 
to a more than three times larger interaction energy per particle of the ensemble of $N=30$ atoms
compared to $N=10$ (at $t=0$), the density of the former expands much further
into the outer parts of the trap. In contrast to this, the density of the $N=10$ atom
ensemble seems to stay Gaussian (with a time-dependent width) to a good approximation,
indicating that we operate in the linear-response regime here. In both cases,
the quench leads only to a slight quantum depletion of at most $3\%$ (see below).

In the following, we first show that the truncated BBGKY approach leads to stable
results in the $N=10$ case, whose accuracy can be systematically improved by
increasing $\bar o$. Thereafter, we turn to the $N=30$ case where
we again encounter instabilities of the EOM and thus apply correction algorithms.
We stress that for both
cases we  operate with $m=4$ dynamically optimized SPFs,
solving the truncated BBGKY EOM coupled to the MCTDHB EOM for the SPFs, 
which is in contrast to the Bose-Hubbard tunneling scenario of Section
\ref{sec:BH_dimer}.

\subsubsection{Breathing dynamics of $N=10$ bosons}

\begin{figure}[t]
   \includegraphics[width=0.495\textwidth]{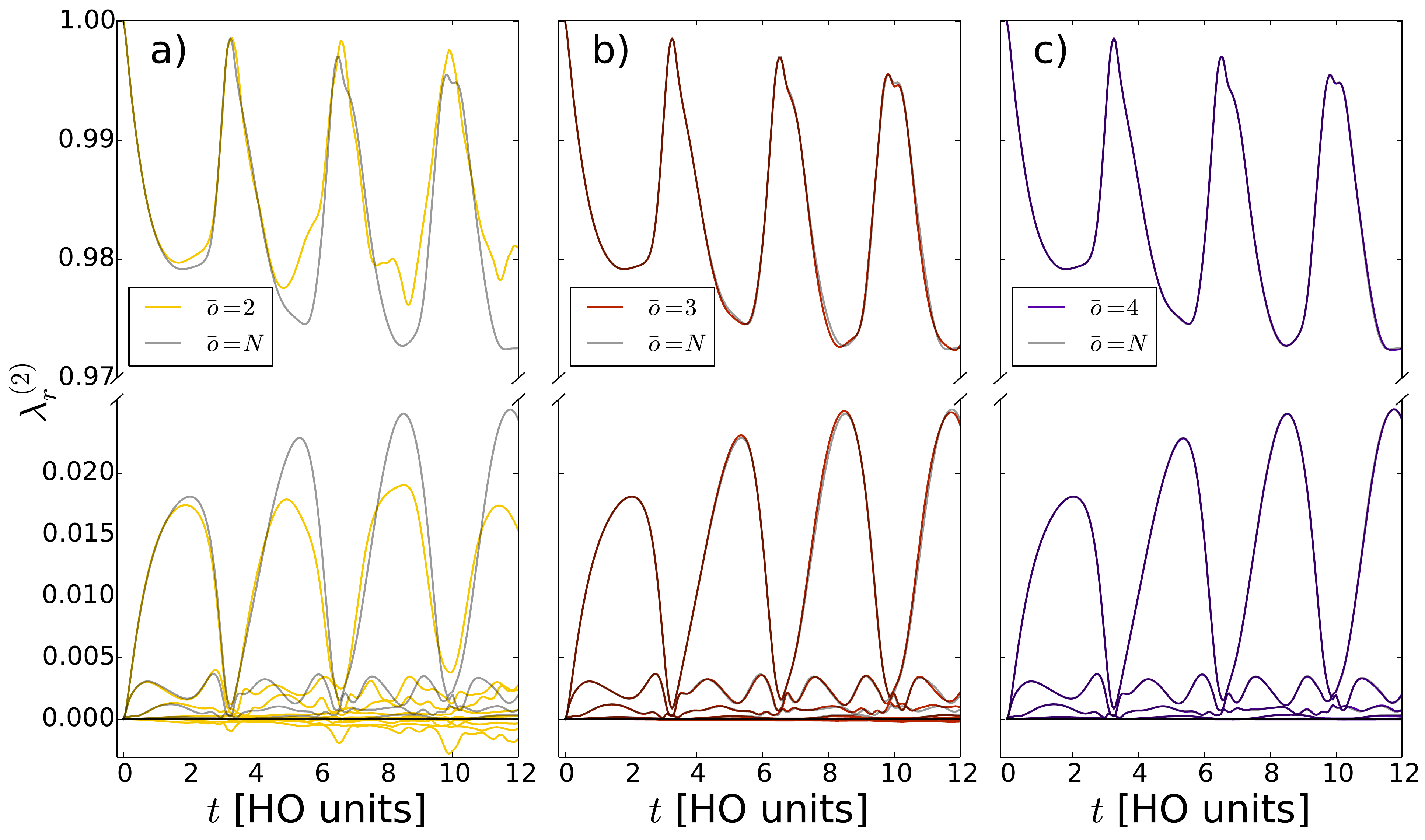}
   \caption{(color online) Natural populations of the $2$-RDM for 
   the truncation orders $\bar o=2$ [a)], $\bar o=3$ [b)] and $\bar o=4$ [c)]
   in comparison to the \mbox{MCTDHB} results. Parameters: $N=10$ atoms, post-quench interaction
   strength $g=0.2$, $m=4$ dynamically
   optimized SPFs.}
   \label{fig:breathing_NP2_N10}
\end{figure}

In Figure \ref{fig:breathing_NP2_N10}, we show the time-evolution of the 
$2$-RDM NPs for various truncation orders. Focusing first on the MCTDHB results,
we see that correlations (in the sense of deviations from a Gross-Pitaevskii mean-field
state where on all orders $o$ 
there is only one finite NP $\lambda_1^{(o)}=1$ and all other NPs vanish) repeatedly emerge and decay. The deviations from the NP distribution
of a Gross-Pitaevskii mean-field state is approximately most pronounced when the 
density is most spread-out and become almost negligible when the density has approximately recovered
its initial width [see Figure \ref{fig:breath_dens_ex} a)].

While the truncated BBGKY results for $\bar o=2$ feature significant deviations
from the MCTDHB results, the results drastically improve when going to $\bar o=3$
and become practically indistinguishable from the MCTDHB results already
at the truncation order $\bar o=4$. Actually, convergence of the $1$-RDM NPs 
$\lambda^{(1)}_i$ is reached even at $\bar o=3$ (not shown). Coming back to $\bar o=2$,
we point out that the $2$-RDM quickly becomes indefinite where small negative eigenvalues
are in particular pushed further to larger negative values when the density
contracts to its initial width and small but positive NPs approach zero. As in 
the case of the above tunneling scenario, we interpret this finding as ``induced'' by
level repulsion. Upon increasing the truncation order, we see that the $2$-RDM 
stays positive semi-definite on the considered time-interval, which is a nice 
example for how increasing the accuracy of the closure approximation
also stabilizes the truncated BBGKY EOM.

\begin{figure*}[t]
 \includegraphics[width=0.495\textwidth]{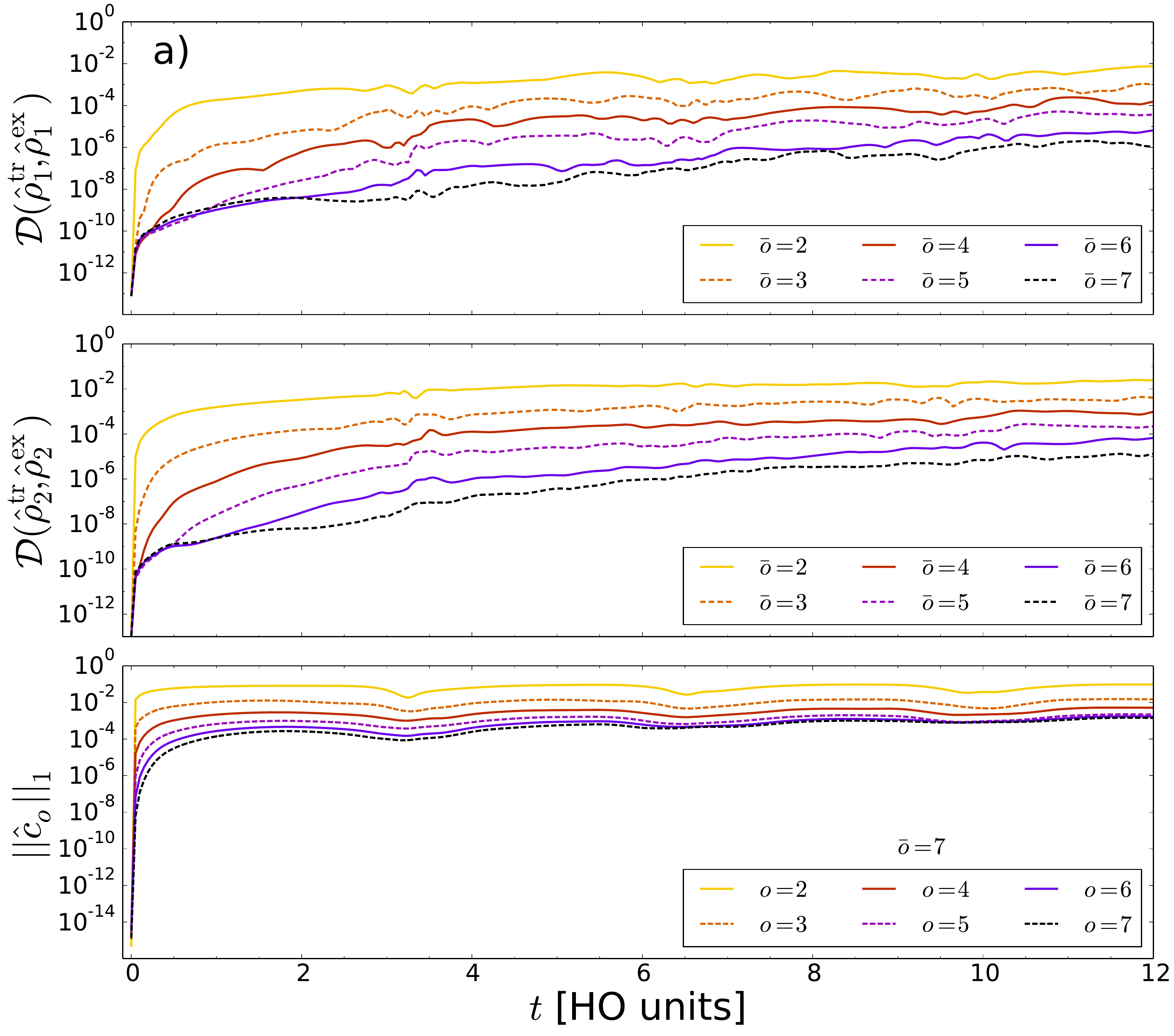}
  \includegraphics[width=0.495\textwidth]{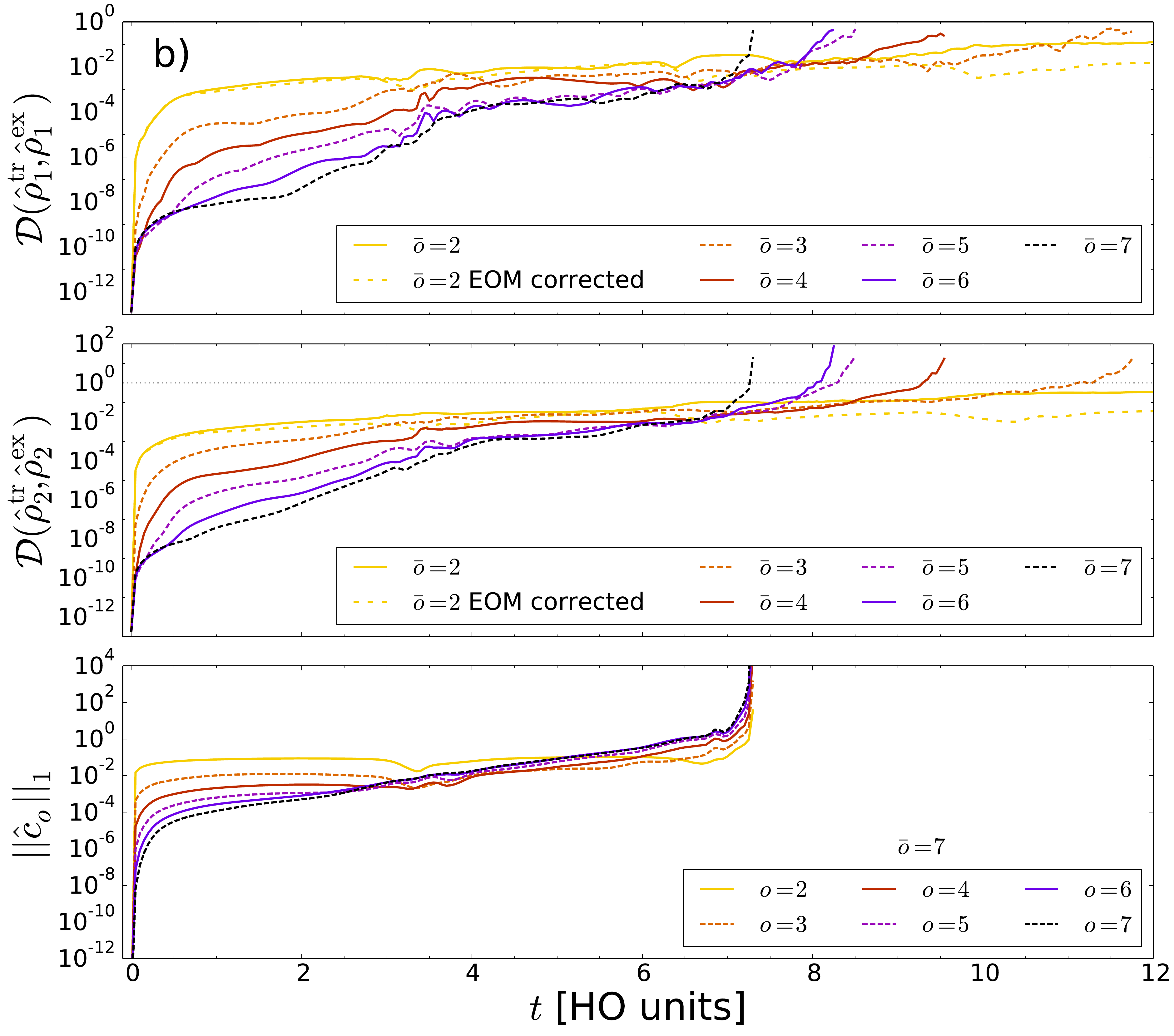}
 \caption{(color online) 
 First and second row:
 Time evolution of the trace distance $\mathcal{D}(\hat\rho_o^{\rm tr},\hat\rho_o^{\rm ex})$
 between the MCTDHB and the truncated BBGKY prediction for the $o$-RDM ($o=1,2$) and various truncation
 orders $\bar o$. The dotted horizontal line at unity ordinate value indicate the upper bound for the trace distance
 between two density operators. Third row: Time-evolution of the cluster's trace-class norm $||\hat c_o||_1$
 for $o=1,...,7$ obtained from the data of the $\bar o=7$ simulations. Left column: $N=10$ atoms. Right columns: $N=30$.
 Otherwise, same parameters as in Figure \ref{fig:breathing_NP2_N10}.}
 \label{fig:breathing_rdm_comp}
\end{figure*}

For a systematic comparison, we next compare the trace-class distance
$\mathcal{D}(\hat\rho_o^{\rm tr},\hat\rho_o^{\rm ex})$
between the truncated BBGKY result for the $o$-RDM, $\hat\rho_o^{\rm tr}$, and 
the corresponding MCTDHB result, $\hat\rho_o^{\rm ex}$, in Figure 
\ref{fig:breathing_rdm_comp} a). We remark that although the SPFs of the truncated BBGKY approach 
obey the same EOM \eqref{eq:mctdhb_part_eom} as the
dynamically optimized SPFs of the MCTDHB method, we cannot expect these two
sets of SPFs to coincide because the $1$- and $2$-RDM entering
the SPF EOM differ in general, which has to be taken into account
when calculating $\mathcal{D}(\hat\rho_o^{\rm tr},\hat\rho_o^{\rm ex})$.
In stark contrast to the tunneling scenario, we see that the accuracy of the
truncated BBGKY results for the $1$- and $2$-RDM 
systematically improves upon increasing $\bar o$ for all considered times.

Finally, we quantify the strength of few-particle correlations
in terms of $||\hat c_o||_1$, as extracted from the $\bar o=7$ calculation [see
Figure \ref{fig:breathing_rdm_comp} a)]. Here, we see that the correlations 
stay bounded on the considered time interval and are ordered in a
clear hierarchy, i.e.\ $||\hat c_{o+1}||_1(t)<||\hat c_{o}||_1(t)$.
Apparently, these are ideal working conditions for the truncated BBGKY approach.

\subsubsection{Breathing dynamics of $N=30$ bosons}
Next, let us increase the quench-induced excitation energy per particle by more than
a factor of three when going to $N=30$ bosons and keeping the post-quench 
interaction strength $g=0.2$ the same. Similarly to the tunneling scenario,
we first inspect the natural populations, then compare lowest order RDMs and 
finally evaluate the performance of the correction algorithms under discussion.

\paragraph{Natural populations}

\begin{figure*}[t]
 \includegraphics[width=0.9\textwidth]{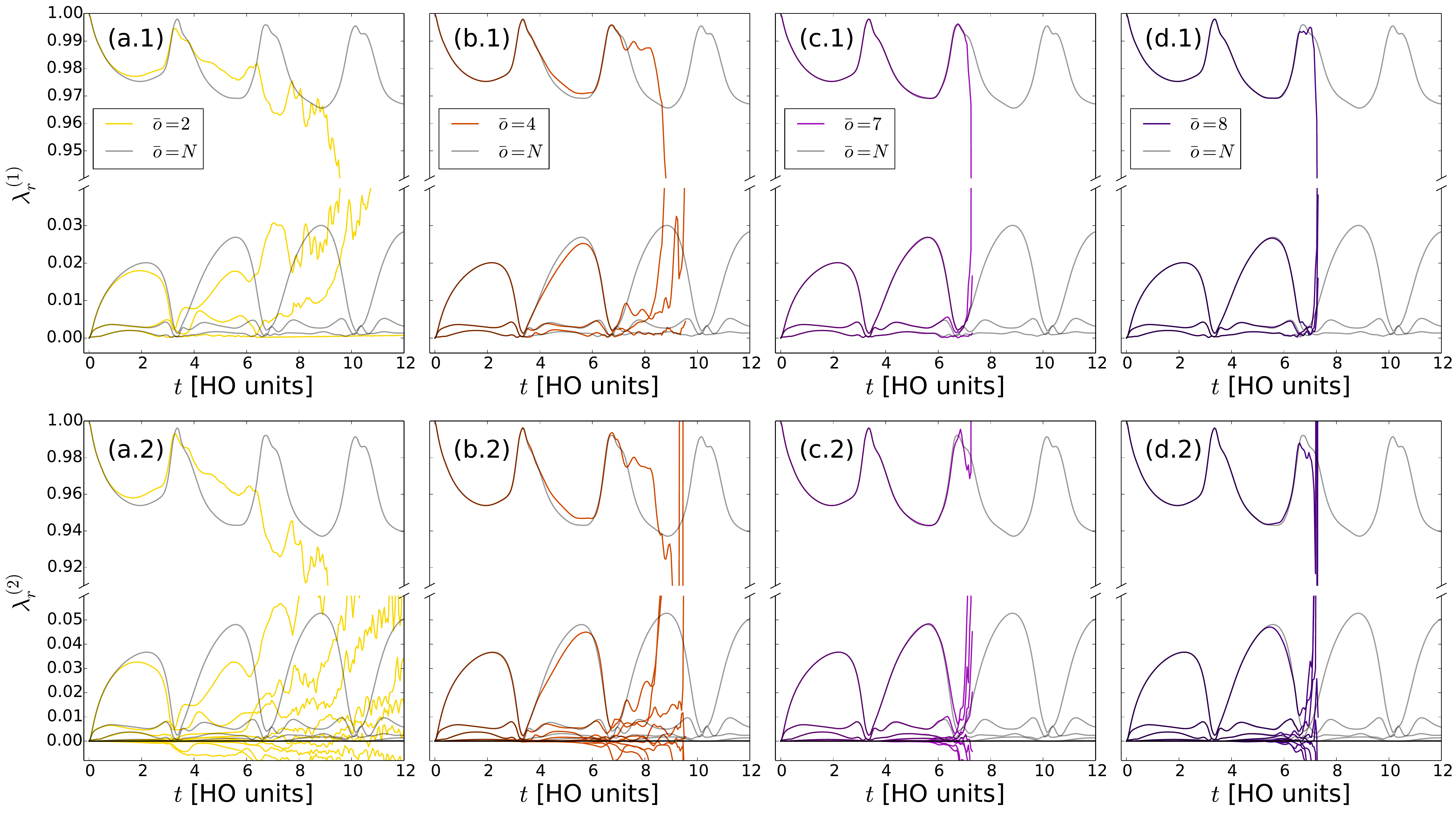}
 \caption{(color online) Top (bottom) row: Time-evolution of the $1$-RDM ($2$-RDM) NPs 
 for various truncation orders $\bar o$ in comparison to the MCTDHB results. 
 We note that the ordinates are broken into two parts for covering the whole 
 range of relevant values. In some cases, this leads to discontinuous curves, see
 e.g.\ the $\bar o=2$ curve in (a.2).
 Number
 of bosons: $N=30$.
  Otherwise, same parameters as in Figure \ref{fig:breathing_NP2_N10}.}
  \label{fig:breathing_NPs_N30}
\end{figure*}

In Figure \ref{fig:breathing_NPs_N30}, we show the NPs of the $1$- and $2$-RDM for various
truncation orders $\bar o$ in comparison
to the MCTDHB results. Similarly to the $N=10$ case, we see how the NP distributions as obtained
from MCTDHB
oscillates between the characteristics of the Gross-Pitaevskii mean-field state
and a (slightly) correlated one, which is approximately synchronized to the strongest
contraction and expansion of the density, respectively [see Figure 
\ref{fig:breath_dens_ex} b)]. In contrast to the former case, however,
we can converge the NPs to the MCTDHB results upon increasing the
truncation order $\bar o$ only for times $t\lesssim 5$ HO units. 
For all considered truncation
orders, we witness an exponential-like instability in the $2$-RDM NPs resulting in large negative
eigenvalues while the $1$-RDM stays positive semi-definite for the 
considered time-span. Fixing $\bar o$, we have observed also for this scenario that 
the lack of positive semi-definiteness of the $o$-RDMs  happens in 
decreasing sequence with respect to the order $o$ (not shown). 
Moreover,
these instabilities in the $2$-RDM NPs seem to be triggered by small positive NPs approaching zero
from above, namely when the density approximately shrinks to its initial width,
see e.g.\ Figure \ref{fig:breathing_NPs_N30} (c.2).
Finally, we have observed for the case $\bar o=2$ that increasing the number of SPFs from $m=4$ to $m=8$
slightly enhances the time-scale on which the instability of the $2$-RDM NPs 
takes place (not shown). This finding is reasonable since the
projector $(\mathds{1}-\hat{\mathbb{P}})$, occurring in the SPF EOM \eqref{eq:mctdhb_part_eom},
projects onto a smaller subspace when increasing $m$ such that the impact
of the non-linearity in the SPF EOM is effectively reduced.

\paragraph{Reduced density operators}
Comparing the BBGKY prediction for the complete $1$- and $2$-RDM with the corresponding MCTDHB results in terms of the trace-class distance in Figure \ref{fig:breathing_rdm_comp} b),
we see that deviations emerge much faster as compared to the $N=10$ case.
 At longer times, we also
observe that the accuracy of the BBGKY results does not monotonously increase anymore
with increasing $\bar o$. Moreover, the above mentioned instabilities 
also partly manifest themselves in $\mathcal{D}(\hat\rho_2^{\rm tr},\hat\rho_2^{\rm ex})$
attaining unphysical values above unity. Finally, we also
depict $||\hat c_o||_1$ as a measure for correlations in Figure 
\ref{fig:breathing_rdm_comp} b). While the correlations
are hierarchically ordered in decreasing sequence with respect to the order $o$ up
to $t\sim 2.7$ HO units, this ordering becomes  reversed later on.
This finding, however, is not conclusive, i.e.\ might be unphysical and 
related to the observed instability, since the values of $||\hat c_o||_1$ have been extracted 
from the BBGKY data with $\bar o=7$ (in contrast to the tunneling 
scenario where the numerically exact $\hat c_o$ have been used).

At this point, we shall remark that we expect a much better agreement for the
$N=30$ case when quenching to much lower interaction
strengths $g\ll0.2$ and thereby reducing the overall excitation energy.

\paragraph{Performance of the correction algorithms}
\begin{figure*}[t]
 \includegraphics[width=0.9\textwidth]{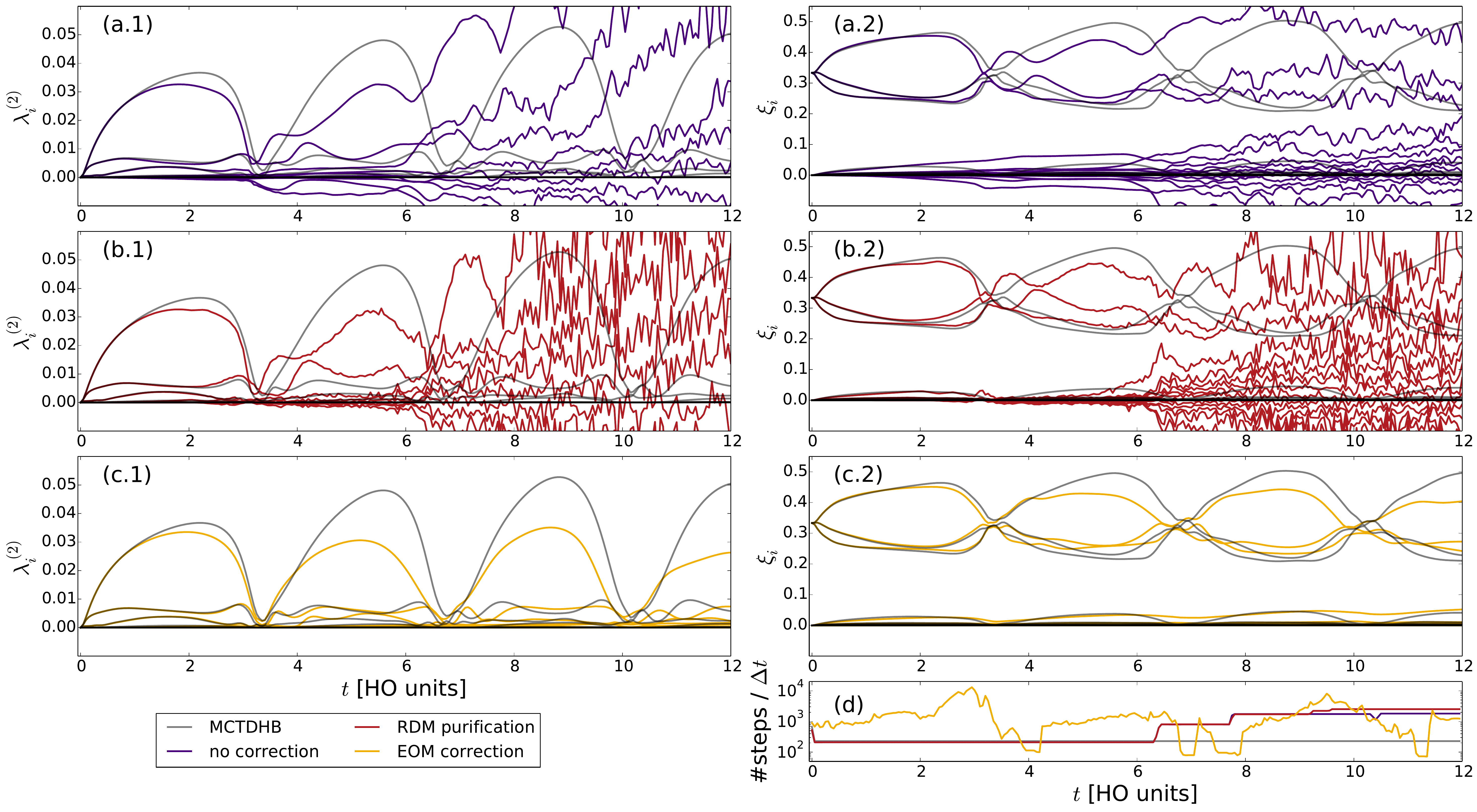}
 \caption{(color online)  
 Comparison of the correction
 strategies outlined in Section \ref{sec:corrections} (for the BBGKY hierarchy
 truncated at $\bar o=2$), i.e.\ same as Figure \ref{fig:BH_correction_alg}
 but for the interaction-quench scenario with $N=30$ bosons. Parameters:
 threshold $\epsilon=-10^{-10}$, damping constant $\eta=10$ HO units, write-out
 time-step $\Delta t=0.05$ HO units, maximal number of iterations: $500$. 
  Otherwise, same parameters as in Figure \ref{fig:breathing_NP2_N10}.}
  \label{fig:breathing_correction_alg}
\end{figure*}

\begin{figure}[]
 \includegraphics[width=0.495\textwidth]{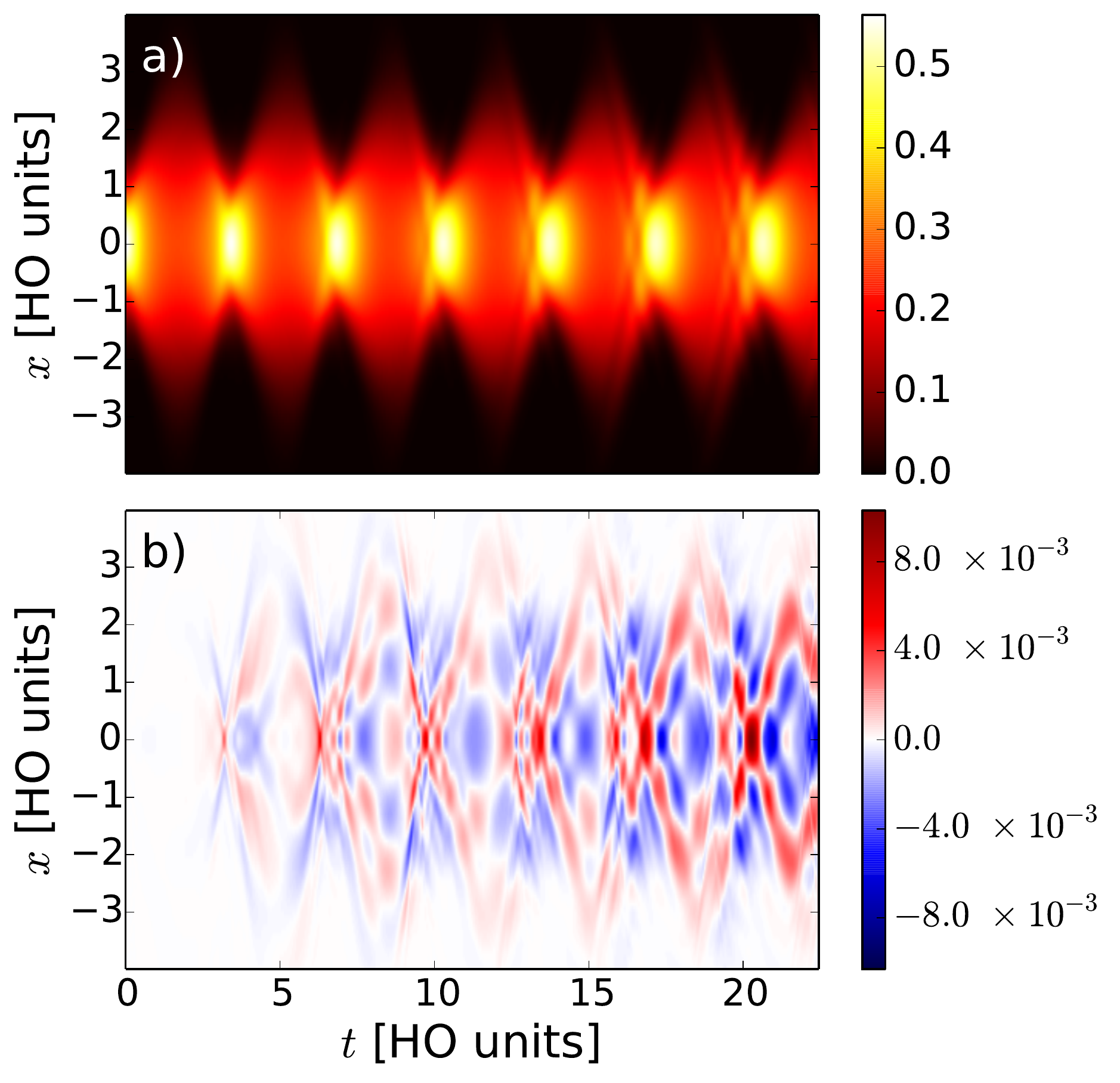}
 \caption{(color online) Subfigure a): Time-evolution of the reduced one-body density $\rho_1^{\rm ex}(x;t)$
 as obtained from \mbox{MCTDHB} on a longer time-scale. 
 Subfigure b): Absolute deviation $\rho_1^{\rm tr}(x;t)-\rho_1^{\rm ex}(x;t)$
 of the reduced one-body density $\rho_1^{\rm tr}(x;t)$ 
 as obtained from the BBGKY approach truncated at $\bar o=2$ and 
 stabilized by the minimal-invasive EOM correction algorithm with
  $\epsilon=-10^{-10}$ and $\eta=10$ HO units. Number of bosons: $N=30$. 
  Otherwise, same parameters as in Figure \ref{fig:breathing_NP2_N10}.}
  \label{fig:breathing_dens_comp_N30}
\end{figure}

Here, we again focus mainly on the performance of the correction algorithms
applied to the $\bar o=2$ BBGKY approach and comment later on larger truncation 
orders. In Figure \ref{fig:breathing_correction_alg}, we depict the spectrum\footnote{
We note that the $\hat K_2$ spectra at $t=0$ in the tunneling and the interaction-quench 
scenario differ although the system is initially a fully condensed BEC in both cases
(see Figures \ref{fig:BH_correction_alg} and \ref{fig:breathing_correction_alg}).
This is due to the fact that the $\hat K_2$ spectrum is sensitive to the total number
of SPFs $m$ even if not all of them are occupied.} of
$\hat K_2$ as well as a close-up to the $2$-RDM spectrum in the vicinity of zero
for the uncorrected BBGKY approach, the minimal invasive RDM purification algorithm
and the minimal invasive EOM correction algorithm. In the minimization problem
underlying both correction algorithms, we have to find the optimal
$\hat{\mathcal{C}}_2$ which depends on $m^2(m+1)^2/4=100$ real-valued parameters.
Being contraction-free and energy conserving leads to $m^2+1=17$ constraints. 
Moreover, $\hat{\mathcal{C}}_2$ has to obey the parity symmetry of our problem
imposing $m^4/8+m^3/4=48$ further constraints (see Appendix \ref{app:corr_rdm}). Thereby,
our system of linear constraints remains underdetermined as long 
as the number $d$ of negative $\hat\rho_2$ eigenvalues and number $d'$ of negative
$\hat K_2$ eigenvalues obey $d+d'<35$.

In Figure \ref{fig:breathing_correction_alg}, we see that the minimal-invasive
RDM purification algorithm clearly suppresses significant negative eigenvalues until
$t\sim2.5$ HO units. Thereafter, noticeably negative eigenvalue emerge but stay
bounded from below until $t\sim6$ HO units when a drastic
instability kicks in. Thus, this iterative algorithm soon fails to converge
after the maximal number of $500$ iteration steps. In order to understand
the deeper reason of this failure, we have analyzed the spectrum
of the updated operators $\hat\rho_2(t)+\alpha\,\hat{\mathcal{C}}_2$
and $\hat K_2(t)+\alpha\,\hat{\Delta}_2$
for $\alpha\in[0,1]$ and the first few iteration steps at such an instant in time (not shown). Thereby, we
have found that while the tangent on a negative eigenvalue (with respect to $\alpha$)
indeed crosses zero as imposed by our constraints, level repulsion 
with other (in most cases negative) eigenvalues often hinders this negative eigenvalue
to significantly move towards zero. We cannot rigorously prove that this is indeed
the only mechanism for the breakdown of this iterative purification algorithm, of course.

Yet at least, this finding gives a useful hint why our non-perturbative, adaptive approach,
 the minimal invasive correction scheme of the $2$-RDM EOM, gives very
 stable results [see Figure \ref{fig:breathing_correction_alg} (c.1) and (c.2)].
 Actually, we observe that the $D$- and $K$-conditions are fulfilled to
 a good approximation much longer, namely for at least $t\leq 36$ HO unit (not shown).
 From Figure \ref{fig:breathing_correction_alg} (d), we furthermore infer that the integrator
 variably adapts its step-size, but in contrast to the Bose-Hubbard tunneling scenario
 no systematic enhancement of integrator steps is observed when $\hat\rho_2$ or
 $\hat K_2$ eigenvalues avoid each other in the vicinity of zero. Apparently, the
 though stabilized result features noticeable deviations from the MCTDHB
 results for the respective eigenvalues.
 Yet, we find that the overall accuracy of the $\bar o=2$ results for the $1$-
 and $2$-RDM as measured by the trace-class distance is systematically
 improved for most times by correcting the $2$-RDM EOM, 
 as one can infer from Figure \ref{fig:breathing_rdm_comp} b).

 In order to judge the accuracy of the EOM-corrected $\bar o=2$ simulation
more descriptively, we depict the deviations of its prediction for the 
reduced one-body density from the MCTDHB results in Figure \ref{fig:breathing_dens_comp_N30}.
Note that this plot covers a longer time-span compared to the previous ones.
As expected, we find that the deviations increase in time. Compared to the absolute values of 
the density, these deviations are, however, small and, most importantly, somewhat
smaller than the deviations of corresponding Gross-Pitaevskii mean-field simulation from
the $m=4$ MCTDHB results (not shown). Finally, let us connect
the errors in the one-body density to the errors measured by the 
trace-class distance $\mathcal{D}(\hat\rho_1^{\rm tr},\hat\rho_1^{\rm ex})$
as depicted in Figure \ref{fig:breathing_rdm_comp} b). For this purpose, we note that
the density at position $x$ can be expressed as the expectation value
of the one-body observable $\hat A_1=|x\rangle\!\langle x|$. 
Thereby, we can estimate $|\rho_1^{\rm tr}(x;t)-\rho_1^{\rm ex}(x;t)|\leq 2
||\hat A_1||_1
\mathcal{D}(\hat\rho_1^{\rm tr}(t),\hat\rho_1^{\rm ex}(t))=2\mathcal{D}(\hat\rho_1^{\rm tr}(t),\hat\rho_1^{\rm ex}(t))$,
which is consistent with the results depicted in Figure \ref{fig:breathing_dens_comp_N30}.

Going to higher truncation orders by making the Mazziotti ansatz for the 
corresponding higher-order correction operators $\hat{\mathcal{C}}_o$ unfortunately does
not improve the BBGKY results, as already observed in the tunneling scenario.
While the iterative RDM purification 
scheme fails to prevent the instabilities, we observe the same obstacle 
for the EOM correction algorithm as previously encountered, namely the optimization problem
at order $o=2$ lacking a solution (results not shown). Yet due to the very promising
results of the EOM correction algorithm when truncating the BBGKY hierarchy at order
$\bar o=2$, we believe that extending the EOM correction algorithm to higher orders
without employing the Mazziotti ansatz for the correction operator
is a highly promising direction to go.

\section{Conclusions}\label{sec:concl}
In this exploratory work, we have developed a novel 
methodological framework
for simulating the quantum dynamics of finite ultracold bosonic systems.
Instead of solving
the time-dependent Schr\"odinger equation for the complete many-body system,
our goal is to truncate the BBGKY hierarchy of equations of motion in order to
obtain a closed theory for the dynamics of the low-order reduced 
density operators (RDMs). Here, we focus in particular on an efficient formulation
of the underlying theory, which allows us to systematically study the impact
of the truncation order on the accuracy and stability of the numerical results.

For this reason, we do not derive the BBGKY equations of motion from the exact 
von-Neumann equation by partial tracing but take the well-established variational Multi-Configuration
Time-Dependent Hartree method for Bosons (MCTDHB) \cite{MCTDHB_PRA08} for {\it ab-initio}
wavefunction propagation as our starting-point. Thereby, we use time-dependent 
variationally optimized single-particle functions (SPFs) as a our truncated single-particle
basis, while being still able to recover the exact results if we formally let the number of SPFs tend to infinity.
By expanding the RDMs with respect to bosonic number states using the dynamically optimized SPFs as the underlying
basis states, we obtain a highly efficient representation of these high-dimensional objects, which
also leads to an efficient and compact formulation of the corresponding BBGKY equations of
motion (EOM) in the second quantization picture. These EOM are coupled to the SPF EOM of the MCTDHB theory.

By a careful analysis, we show that this coupled system of EOM features
all properties, which are known for the BBGKY hierarchy as derived from the von-Neumann
equation of the total system. Although being deduced from the zero-temperature
MCTDHB theory, we find that the derived EOM for the RDM are also variationally optimal
in a certain sense if the total many-body system is in a mixed initial state, which 
opens a promising route for including low-temperature effects in the simulation
of ultracold atoms. Thus, truncating this BBGKY hierarchy of EOM can be
viewed as on the one hand introducing an additional approximation to the MCTDHB 
approach for simulating larger particle numbers with more SPFs 
and on the other hand as an extension of the zero-temperature MCTDHB
theory to finite temperatures.

We truncate the hierarchy of BBGKY EOM by using a reconstruction
functional for the unknown RDM $\hat\rho_{\bar o+1}$ where $\bar o$ denotes the truncation order.
While the commonly employed cluster (cumulant) expansion for truncating the BBGKY 
hierarchy  for fermionic systems is very well suited for taking correlations on top 
of a Hartree-Fock state into account \cite{qua_kin_theo_bonitz}, its corresponding bosonic variant 
has proven to be unfavorable for bosonic systems with a fixed number 
of particles since the correspondingly defined clusters diagnose that even an ideal Bose-Einstein
condensate features few-particle correlations on all orders \cite{kira_excitation_2014}.
In this work, we cure this flaw by simply replacing the RDMs and symmetrization operators in
this standard approach by the corresponding RDMs of unit trace and idempotent 
symmetrization operators, respectively. Since neglecting the complete cluster $\hat c_{\bar o+1}$
in the truncation approximation violates the compatibility to the lower order RDMs,
we use the so-called unitarily invariant decomposition of bosonic operators 
\cite{coleman1974,coleman_reduced_1980,au-chin_characteristic_1983,sun_unitarily_1984}
for restoring compatibility, as pursued in \cite{lackner_propagating_2015,lackner_high-harmonic_2017}
for electronic systems.
Thereby, we obtain a closure approximation which conserves the compatibility
of the RDMs as well as energy, respects symmetries such as parity or translational invariance
if existent and is unitarily invariant, i.e.\ gauge invariant with respect to the choice
of the constraint operator of the MCTDHB theory. In contrast to the cluster expansion for fermionic
systems, however, the employed cluster-expansion lacks size-extensitivity, i.e.\
testifies that two independent ideal Bose-Einstein condensates feature few-particle correlations stemming
solely from the bosonic particle-exchange symmetry.
Such correlations are physical, of course, but 
should not manifest themselves in the correlation {\it definition} on which a cluster expansion is build.
Otherwise, truncating the expansion by neglecting clusters may imply neglecting correlations
stemming from the bosonic symmetry.
We show that this flaw can in principle be cured by 
minimal modifications of the cluster expansion, at the price, however, of losing the unitary invariance
of the thereby defined correlations. For this reason, we do not apply these modifications
to the truncation approximation in our numerical investigations.

Using appropriate super-operators and our bosonic number-state based framework,
we derive two computational rules by means of which the clusters and thereby the
closure approximation can be calculated highly efficiently for high orders in a recursive manner.
This computational strategy allows us to go to truncation orders as high as $\bar o=12$,
meaning that up to $12$-particle correlations are taken into account.

We have applied the above methodological framework
to two scenarios, namely the tunneling dynamics in a double well and the interaction-quench
induced breathing dynamics in a harmonic trap, in order to investigate the accuracy and stability of the
numerical results in dependence on the truncation order. 
In both applications, we have found that the short-time dynamics can be highly accurately
described by the truncated BBGKY approach, where the accuracy of the results systematically
improves with increasing truncation order $\bar o$.
At longer times, the BBGKY gives also excellent 
results with controllable accuracy in the interaction-quench scenario for not too high
excitation energies. However, severe deviations from the corresponding MCTDHB simulations occur at longer times
in the tunneling scenario as well as for stronger interaction quenches.
In these cases, the accuracy does not monotonously improve
with increasing truncation order anymore and the truncated BBGKY EOM start to
suffer from exponential instabilities, which lead to unphysical states.
By inspecting the exact numerical results for the tunneling scenario,
we find that few-particle correlations on all orders quickly
play a significant role and eventually $N$-particle correlations dominate 
because the total system evolves into a NOON-state. This finding indicates
that the long-time physics of this scenario prevents to use a truncation approximation
which is based on neglecting $(\bar o+1)$-particle correlations.

Nevertheless, it is important to clearly separate the stability properties of the 
truncated BBGKY EOM from accuracy issues because (i) it is not
desirable to have a highly accurate theory which is exponentially unstable under slight e.g.\
numerical perturbations and (ii) also a not highly accurate truncation approximation
may give useful, sufficiently accurate results for low-dimensional observables such as the 
density if the EOM are sufficiently stable. Thus we have analyzed these instabilities
for the two scenarios in depth. Thereby, we have found that the instability sets in at the truncation 
order $\bar o$ and then propagates down to lower orders meaning that $o$-particle RDMs lack 
to be positive semi-definite in decreasing sequence with respect to the order $o$.
The time until 
which the highest-order RDM, $\hat \rho_{\bar{o}}$, stays positive 
semi-definite only gradually increases with the truncation order $\bar o$, while the
time when the lowest-order RDMs start lacking positivity decreases with increasing
truncation order in some cases. This may be explained by the enhanced non-linearity
of the closure approximation for higher truncation orders. Moreover, we have observed that
the instabilities go often hand in hand with avoided crossings of RDM eigenvalues close
to zero.

In order to stabilize the EOM, we have developed two novel minimal
invasive and energy conserving correction algorithms: In our first attempt, we extend
the dynamical purification algorithm 
\cite{mazziotti_purification_2002,lackner_propagating_2015,lackner_high-harmonic_2017}.
Yet being based on a first-order perturbation theory argument, this algorithm cannot properly 
cope with the avoided crossings of the RDM eigenvalues
in the vicinity of zero and thus fails to prevent the instabilities in our simulations.
For this reason, we have developed a second, non-perturbative correction algorithm, which
slightly alters the truncated BBGKY EOM such that negative RDM eigenvalues are exponentially
damped to zero. 
We find that this approach indeed stabilizes the BBGKY EOM truncated at the second order 
and 
leads to reasonable long-time results for the interaction-quench scenario.

Besides these major methodological developments and their numerical evaluation,
we have also proposed an imaginary-time relaxation approach for calculating the lowest order ground-state
RDMs of some reference Hamiltonian such that they can be used as the initial state for the BBGKY hierarchy.
Moreover, we have analytically shown that certain coherences in the
contraction-free component of the $2$-RDM are responsible for dynamical quantum depletion 
and fragmentation of
Bose gases.

Thereby, this exploratory work constitutes a major step forward to the treatment of correlated
ultracold bosonic systems in terms of the truncated BBGKY hierarchy of EOM.
In addition to the developed truncation approximation, we have presented
numerous technical as well conceptual results, which are independent of
the applied truncation approximation and as such also valuable 
for future works on closure approximations. 
In these regards,
still open questions remain such as how to enforce size-extensitivity in the
cluster definition while keeping its unitary invariance. 
Although our numerical simulations reveal challenges for long-time propagations
in far-off equilibrium situations, our thorough analysis gives valuable hints
for future research, namely (i) extending the highly successful EOM correction
algorithm of Section \ref{sec:corr_eom} to higher orders without using the Mazziotti 
ansatz and (ii) research on closure approximations for bosonic system 
with a fixed number of particles, going beyond the paradigm of the cluster expansion.
Finding novel closure approximations by machine-learning techniques 
might be a promising first step.

\begin{acknowledgments}
The authors thank Iva B\u{r}ezinov\'a, Joachim Burgd\"orfer, Ignacio J.\ Cirac, Uwe Manthe, David A.\ Mazziotti,
Hans-Dieter Meyer, Angel Rubio, Jan Stockhofe and Johannes M.\ Schurer for fruitful discussions.
This work has been supported by the excellence cluster ``The Hamburg Centre for Ultrafast Imaging 
- Structure, Dynamics and Control of Matter at the Atomic Scale'' of the Deutsche Forschungsgemeinschaft.
\end{acknowledgments}
\appendix

\section{Super-operators acting on bosonic operators}\label{app:superoperators}
In this appendix, we define important super-operators acting on bosonic 
few-particle operators such as the $o$-RDM. These super-operators are
represented in the second-quantization picture such that they can be applied
efficiently to e.g.\ the RDMs being represented as outlined in Section \ref{sec:rdm_rep}.
While only the basic concepts are discussed here, important technical details 
are covered by Appendix \ref{app:1st2ndquant}. In the following, let 
$\mathcal{B}_o$ denote the set of all hermitian bosonic
$o$-body operators, meaning each $\hat B_o\in\mathcal{B}_o$ obeys 
$\hat P_\pi\hat B_o=\hat B_o\hat P_\pi=\hat B_o$ with the particle-permutation 
operator $\hat P_\pi$ corresponding to an arbitrary permutation $\pi\in S(o)$ of
the first $o$ integers. 

\subsection{Partial traces}\label{app:part_tr}
Having $\hat B_o$ expanded with respect to $o$-particle
number states and using a mixed first and second quantization representation 
as outlined in Appendix \ref{app:1st2ndquant}, the partial trace of $\hat B_o$ over
one particle can be expressed as\footnote{\label{foot:part_numb_in_tr}Although the number of particles, $o$, 
occurs on the right hand side of Eq.\ \eqref{eq:part_tr_1}, we do not incorporate it in
the symbol $\tr_{1}$ for the partial trace since one may replace 
the factor $1/o$ in Eq.\ \eqref{eq:part_tr_1} by the inverse 
of the operator $(\hat N+1)$ with the total particle-number operator
$\hat N=\sum_{r=1}^m\ad{r}\a{r}$.
}
\begin{equation}
 \label{eq:part_tr_1}
 \tr_1\big(\hat B_o\big) \equiv\frac{1}{o}\sum_{r=1}^m \a{r}\,\hat B_o\,\ad{r}.
\end{equation}
An explicit formula for the right hand side of \eqref{eq:part_tr_1} as well as for 
the corresponding generalization 
to the partial trace over $k$ particles, 
$\tr_k(\hat B_o)\equiv\tr_1\circ...\circ\,\tr_1(\hat B_o)$, for $k\leq o$
are provided in Appendix \ref{app:1st2ndquant}.

\subsection{Raising and joining operations}\label{app:raise_join}
In our formal framework, we also need - loosely speaking - the inverse of the partial
trace, meaning an operation which raises an operator $\hat B_o\in\mathcal{B}_o$
to an $(o+1)$-body operator by adding a particle in an undefined state, i.e.\
the unit-operator. This raising operation is accomplished 
by\footnote{\label{foot:part_numb_in_raise}Similarly to Eq.\ \eqref{eq:part_tr_1},
the right hand side of \eqref{eq:raise_1} does not explicitly depend on the number of
particles since we may replace the right hand side by $\hat N^{-1}
\sum_{r=1}^m \ad{r}\,\hat B_o\,\a{r}$ (see also footnote
\ref{foot:part_numb_in_tr}). This expression is well-defined 
because the inverse of the 
total particle-number operator acts only on states with at least one particle. }
\begin{equation}
 \label{eq:raise_1}
 \hat R_1\big(\hat B_o\big) \equiv\frac{1}{o+1}\sum_{r=1}^m \ad{r}\,\hat B_o\,\a{r}.
\end{equation}
In Appendix \ref{app:uid}, we comment on the precise relationship between the raising
and the partial-trace operation in terms of Eq.\ \eqref{eq:rel_part_trace_raising}.

For defining few-particle correlations in Section \ref{sec:truncation}, we furthermore 
need a super-operator which joins two operators $\hat A_{o_1}\in\mathcal{B}_{o_1}$, 
$\hat B_{o_2}\in\mathcal{B}_{o_2}$ to a bosonic $(o_1+o_2)$-body operator
\begin{align}
 \label{eq:join_op}
 &\hat J_{o_1}^{o_2}\big(\hat A_{o_1},\hat B_{o_2}\big) \equiv\sum_{\vec{a},\vec{b}|o_1}
 \sum_{\vec{c},\vec{d}|o_2}A^{o_1}_{\vec{a},\vec{b}}B^{o_2}_{\vec{c},\vec{d}}
 \\\nonumber
 &\phantom{\hat J_{o_1}^{o_2}}\left(\prod_{r=1}^m\,\binom{a_r+c_r}{c_r}
\binom{b_r+d_r}{d_r}\right)^{\frac{1}{2}}\,
|\vec{a}+\vec{c}\rangle\!\langle \vec{b}+\vec{d}|,
\end{align}
where $A^{o_1}_{\vec{a},\vec{b}}\equiv\langle\vec{a}|\hat A_{o_1}|\vec{b}\rangle$
and $B^{o_2}_{\vec{c},\vec{d}}\equiv\langle\vec{c}|\hat B_{o_2}|\vec{d}\rangle$.
Expanding the argument $\hat B_o$ in Eq.\ \eqref{eq:raise_1} in number states 
and comparing the result with Eq.\ \eqref{eq:join_op}, one can easily verify
that raising $\hat B_o$ effectively means joining with the one-body 
unit-operator of the subspace spanned by the instantaneous SPFs, 
i.e.\ $\hat R_1(\hat B_o)=\hat J_o^1(\hat B_o,\hat{\mathbb{P}})/(o+1)$
with 
$\hat{\mathbb{P}}\equiv\sum_{r=1}^m|\varphi_r\rangle\!\langle\varphi_r|$. Similarly, one can 
show that $k$-fold raising $\hat R_k(\hat B_o)\equiv\hat R_1\!\circ...\circ\hat R_1(\hat B_o)$
effectively means joining with the corresponding bosonic $k$-body unit-operator, i.e.\
$\hat R_k(\hat B_o)=\hat J_o^k(\hat B_o,\hat{\mathbb{P}}_k^+)/\binom{o+k}{k}$
with $\hat{\mathbb{P}}_k^+\equiv\sum_{\vec{n}|k}|\vec{n}\rangle\!\langle\vec{n}|$.
Finally, we note that this bosonic joining operator plays the same role the wedge
product for the cluster expansion for fermions (see e.g.\ \cite{maziotti_rdm_book07}).

\section{Mixed first and second quantization representation}\label{app:1st2ndquant}
While an efficient representation of RDMs requires working with bosonic number states,
which is a second-quantization concept, operations such as partial traces
are most conveniently performed if individual particles can be addressed by 
particle labels, i.e.\ in a first quantization framework. Here, we provide
formulas for bridging between these two perspectives, which finally allows for evaluating
all expressions in the second quantization picture. Our starting-point is the well-known 
relationship between an $o$-particle Hartree product $|\varphi_{j_1}...\varphi_{j_o}\rangle$,
in which $n_r$ particles reside in the SPF $|\varphi_r\rangle$,
and
the corresponding bosonic number state $|{\vec n}\rangle$ (normalized to unity)
with $\vec n=(n_1,...,n_m)$ encoding the respective occupation numbers
\begin{equation}\label{eq:ns}
 |{\vec n}\rangle = \sqrt{ \frac{o!}{\prod_{i} n_i!}}\,\hat S_o
|\varphi_{j_1}...\varphi_{j_o}\rangle.
\end{equation}
Here, $\hat S_o$ refers to the idempotent $o$-particle symmetrization operator 
$\hat S_o=\sum_{\pi\in S(o)}\hat P_\pi/o!$ with
the sum running over all particle permutations $\pi$ and $\hat P_\pi$ denoting
the corresponding particle-exchange operator.
After explicating the summation over all permutations of particle labels and renaming the 
summation index of the orbital in which the $o$th particle resides, we arrive at \cite{cao_multi-layer_2013}
\begin{align}\label{eq:ns_1part_sep}
  |{\vec n}\rangle &= \sum_{r=1}^m \sqrt{\frac{n_r}{o}}\,  |{\vec n}-\uvec{r}\rangle^{(1,...,o-1)}
  \otimes|\varphi_r\rangle^{(o)}\\\nonumber
  &=\sum_{r=1}^m \left(\frac{\a{r}}{\sqrt{o}}\,  |{\vec n}\rangle\right)^{\!(1,...,o-1)}
  \otimes|\varphi_r\rangle^{(o)},
\end{align}
where the super-indices of the ket-vectors indicate the particle-labels for the
corresponding state and $\uvec{r}$ is an occupation number vector having vanishing
elements except for the $r$th one being set to unity.
In passing, we note that the state $\a{r}|{\vec n}\rangle/\sqrt{o}$
coincides with the so-called single-hole function of $|{\vec n}\rangle$ with respect
to the $r$th SPF as used in e.g.\ \cite{MCTDH_BJMW2000,cao_multi-layer_2013}. 
The second identity in \eqref{eq:ns_1part_sep} can be used to prove and explicate the expression
\eqref{eq:part_tr_1} for the partial trace of 
$\hat B_o\in\mathcal{B}_o$ over one particle,
given its representation $\sum_{{\vec n},{\vec m}|o}B^o_{{\vec n},{\vec m}}
|{\vec n}\rangle\!\langle{\vec m}|$
\begin{align}
 &\tr_1\big(\hat B_o\big) = \sum_{\vec n, \vec m|o}\sum_{r=1}^m
 B^o_{{\vec n},{\vec m}}\,{}^{(o)}\!\langle \varphi_r|\vec n\rangle\!\langle
 \vec m|\varphi_r\rangle^{(o)}\\\nonumber
 &=\frac{1}{o}\sum_{\vec n, \vec m|o}\sum_{r=1}^m
 B^o_{{\vec n},{\vec m}}\,\a{r}|\vec n\rangle\!\langle
 \vec m|\ad{r}
 =\frac{1}{o}\sum_{r=1}^m \a{r}\,\hat B_o\,\ad{r}\\\nonumber
 &=\frac{1}{o}\sum_{\vec a, \vec b|o-1}\sum_{r=1}^m
 \sqrt{(a_r+1)(b_r+1)}\,B^o_{{\vec a}+\uvec{r},{\vec b}+\uvec{r}}\,
 |{\vec a}\rangle\!\langle{\vec b}| .
\end{align}
By
successively applying the steps leading to \eqref{eq:ns_1part_sep} to the respectively 
occurring number states and using the identity \eqref{eq:ns} for the resulting Hartree
products, we may decompose an $o$-particle number state into a sum over products of 
$(o-k)$-particle and $k$-particle number states ($k<o$) associated with the ``first'' $o-k$
and the ``last'' $k$ particles
\begin{align}\label{eq:ns_kpart_sep}
  |{\vec n}\rangle = \binom{o}{k}^{-\frac{1}{2}}\sum_{{\vec l}|k}&\Theta({\vec n}-{\vec l})
  \Big[\prod_{r=1}^m\,\binom{n_r}{l_r}\Big]^{\frac{1}{2}}\\\nonumber
  &|{\vec n}-{\vec l}\rangle^{(1,...,o-k)}
  \otimes|{\vec l}\rangle^{(o-k+1,...,o)},
\end{align}
where $\Theta({\vec n})\equiv\prod_{i=1}^m\Theta(n_i)$ with the Heavyside
function $\Theta$ defined by $\Theta(x)=1$ for $x\geq0$ and zero otherwise.
The relation \eqref{eq:ns_kpart_sep} is a central technical result, on which the
recursive formulation of the cluster expansion is founded (see Section \ref{sec:trunc_recursive_cluster}
and Appendix \ref{app:recursion_rules}).

Furthermore, this identity allows for efficiently evaluating the partial trace 
${\rm tr}_k(\hat B_o)$
of a bosonic $o$-body operator $\hat B_o$ over $k$ particles ($k<o$). 
Expanding $\hat B_o$ with respect to number states
$|{\vec n}\rangle$,  $|{\vec m}\rangle$
and inserting the decomposition 
\eqref{eq:ns_kpart_sep},
one directly obtains the following
expression
\begin{align}\label{eq:part_tr_k}
  {\rm tr}_k&\big(\hat B_o\big)\equiv\\\nonumber
  &\equiv
  \sum_{i_{1},...,i_k}
  {}^{(o-k+1,...,o)}\!\langle
  \varphi_{i_{1}}...\varphi_{i_k}|
  \hat B_o
  |\varphi_{i_{1}}...\varphi_{i_k}\rangle^{(o-k+1,...,o)}\\\nonumber
  &=\sum_{{\vec l}|k}{}^{(o-k+1,...,o)}\!\langle {\vec l}|
  \hat B_o|{\vec l}\rangle^{(o-k+1,...,o)}\\\nonumber
  &= \binom{o}{k}^{-1}\sum_{{\vec a},{\vec b}|o-k}\sum_{{\vec l}|k}
  \left(\prod_{r=1}^m\,\binom{a_r+l_r}{l_r}
\binom{b_r+l_r}{l_r}\right)^{\frac{1}{2}}\\\nonumber
&\phantom{\hat B_o\big)=\sum_{{\vec l}|k}{}^{(o-k+1,...,o)}\!\langle {\vec l}|}
B^o_{{\vec a}+{\vec l},{\vec b}+{\vec l}}\;
|{\vec a}\rangle\!\langle{\vec b}|.
\end{align}
We note that this expression is meaningful also for $k=o$, resulting in
${\rm tr}_o(\hat B_o)={\rm tr}(\hat B_o)\,|0\rangle\!\langle 0|$
with $|0\rangle$ denoting the vacuum state.
Besides, the above formula provides one way
to derive the expression \eqref{eq:rdm_rep} for the $o$-RDM 
by setting $o$ to the total number of atoms $N$, $k$ to $N-o$ and 
$\hat B_N$ to $|\Psi_t\rangle\!\langle\Psi_t|$ and using the expansion \eqref{eq:mctdhb_ansatz}.

\section{Finite temperatures}\label{app:finiteT}
Let us now show that and in which sense the EOM \eqref{eq:mctdhb_part_eom} together
with the possibly truncated RDM EOM \eqref{eq:BBGKY_arb_gauge} result
in an optimal SPF dynamics also for mixed initial $N$-particle states,
given that the $N$-particle dynamics is unitary. This line of argumentation
strongly resembles the considerations on an alternative to the
so-called $\rho$MCTDH type-2 method for simulating Lindblad master equations
for distinguishable degrees of freedom in the limit of purely
unitary dynamics, as discussed in \cite{raab_multiconfigurational_2000}.
Given the spectral representation of the initial state
\begin{equation}
 \label{eq:spectral_N_rdm}
 \hat\rho_N(0)=\sum_{r=1}^{C^N_m}\lambda^{(N)}_r(0)\,|\phi^N_r(0)\rangle\!\langle\phi^N_r(0)|,
\end{equation}
a unitary dynamics governed by the Hamiltonian $\hat H$ leaves the probabilities 
$\lambda^{(N)}_r(t)$ invariant. Instead of solving the von-Neumann equation
for $\hat\rho_N$, one may purify the density operator to
\begin{equation}
 \label{eq:purif_rhoN}
 |\Phi_t\rangle = \sum_{r=1}^{C^N_m}\sqrt{\lambda^{(N)}_r}\,|\phi^N_r(t)\rangle\otimes
 |u_r\rangle,
\end{equation}
where $\{|u_r\rangle\}$ denotes some fixed time-independent orthonormal basis of a
$C^N_m$-dimensional auxiliary Hilbert space. This pure state is propagated
according to $i\partial_t |\Phi_t\rangle=\hat H\otimes\mathds{1}|\Phi_t\rangle$
and one exactly recovers the solution of the von-Neumann equation
by taking the partial trace over the auxiliary space, i.e.\
$\hat\rho_N(t)=\tr_{\rm aux}(|\Phi_t\rangle\!\langle \Phi_t|)$.
Now one can expand each $|\phi^N_r(t)\rangle$ in the MCTDHB manner 
$\sum_{\vec n|N}A^r_\vec n\,|\vec n\rangle$, where all states share the same set of 
SPFs, and minimize the action functional with the Lagrangian $\langle\Phi_t|(i\partial_t
-\hat H\otimes\mathds{1})|\Phi_t\rangle$ under orthonormality constraints
on both the SPFs and the $N$-particle states $|\phi^N_r(t)\rangle$. Thereby,
one finds that the coefficients $A^r_\vec n$ for fixed $r$ obey Eq.\
\eqref{eq:mctdhb_top_eom}. Differentiating $\rho^N_{\vec n,\vec m}=\sum_r\lambda_r^{(N)}\,
(A^r_\vec n)^*A^r_\vec m$ with respect to time, we directly obtain the RDM EOM \eqref{eq:BBGKY_arb_gauge}
at order $o=N$. Varying the action with respect to the SPFs exactly results 
in the SPF EOM \eqref{eq:mctdhb_part_eom} where the elements of the 1- and 2-RDM entering
the equations are the convex sum over the corresponding matrices \eqref{eq:rdm_rep}
for the state  $|\phi^N_r\rangle$ weighted with the probability $\lambda^{(N)}_r$.
The latter means that the 1- and 2-RDM in the EOM for the SPFs are exactly
the $(N-1)$- and $(N-2)$-fold partial trace of $\hat \rho_N$, respectively. 

Thereby, it is
shown that the EOM \eqref{eq:mctdhb_part_eom}, \eqref{eq:BBGKY_arb_gauge} applied
to a mixed, e.g.\ thermal initial $N$-particle state also result in a well-defined
variationally optimal dynamics, where the dynamical adaption of the SPFs 
is a compromise between an optimal representation of the various
eigenvectors $|\phi^N_r(t)\rangle$ of $\hat\rho_N(t)$. Here, eigenstates of higher
probability $\lambda_r^{(N)}$ have a stronger impact on the SPF dynamics than weakly
occupied eigenstates. 
\section{Propagation in negative imaginary time}\label{app:imag_time}

The purpose of this appendix is to show how one can derive EOM for the RDMs in
imaginary time which contract an initial guess to the ground-state RDMs. While the 
case of the $N$-RDM, i.e.\ the state of the full system, has already been
addressed in Section \ref{sec:init_state}, we illustrate the derivation
by exemplarily inspecting the case of the 1-RDM.

Fixing the constraint operator to $g_{ij}=0$ here, we may write the MCTDHB
EOM in imaginary time in the following compact form
\begin{align}\label{eq:mctdhb_imag}
 \partial_\tau|\Psi_\tau\rangle&=(E_\tau-\hat{\tilde H}_0)|\Psi_\tau\rangle\\\nonumber
 &\phantom{=}+
 \sum_{i=m+1}^\infty\sum_{j=1}^m\langle\varphi_i|(\partial_\tau|\varphi_j\rangle)\,
 \ad{i}\a{j}|\Psi_\tau\rangle.
\end{align}
Here, $\hat{\tilde H}_0$ is defined via Eq.\ \eqref{eq:H_tild} with the constraint-operator $g_{ij}$ set to zero
and with
the Hamiltonian $\hat H$ being replaced by the reference Hamiltonian $\hat H_0$, whose
ground-state RDMs shall be calculated. 
The term proportional to the energy expectation value $E_\tau\equiv\langle\Psi_\tau|\hat H_0|\Psi_\tau\rangle=
\langle\Psi_\tau|\hat{\tilde H}_0|\Psi_\tau\rangle$ ensures that the norm of $|\Psi_\tau\rangle$
does not contract to zero but stays unity. In Eq.\ \eqref{eq:mctdhb_imag}, we have furthermore
expanded the SPF notation: $|\phi_r\rangle$ still refers to the dynamically adapted MCTDHB SPFs
for $r=1,...,m$. For larger $r$, $|\phi_r\rangle$ refers to an unoccupied, i.e.\ virtual orbital outside of the
space spanned by the instantaneous SPFs. For the sake of concreteness, we assume the single-particle 
Hilbert space to be infinite dimensional, while in a numerical implementation $r$ would be bounded from above by the 
number of time-independent single-particle states used to represent the MCTDHB SPFs. Accordingly, we extend 
the notation of creation and annihilation operators also to the space of virtual orbitals. Finally,
$\partial_\tau|\varphi_j\rangle$ coincides with the negative of the right hand side of Eq.\ \eqref{eq:mctdhb_part_eom}
with $\hat g$ set to zero.

Now we can derive the corresponding EOM for the 1-RDM where we use the representation 
\eqref{eq:D_matrix} for simplicity. When differentiating $D^1_{i,j}=\langle\Psi_\tau|
\ad{j}\a{i}|\Psi_\tau\rangle$ for $i, j=1,...,m$ with respect to $\tau$, we can make use for the fact
$\partial_\tau\a{i}^{(\dagger)}=\sum_{r=m+1}^\infty\langle\partial_\tau\varphi_i|\varphi_r\rangle^{\!(*)}\,
\a{r}^{(\dagger)}$ for $i=1,...,m$, which is a consequence of $\hat g=0$. Thereby,
one obtains
\begin{align}
\label{eq:imag_time_1_RDM_EOM_1}
 \partial_\tau \langle\Psi_\tau|
\ad{j}\a{i}|\Psi_\tau\rangle =
\langle\Psi_\tau|\,\{E_\tau-\hat{\tilde H}_0,\ad{j}\a{i}\}\,
|\Psi_\tau\rangle.
\end{align}
Inserting $\hat{\tilde H}_0$ into the right hand side and normal ordering under consideration
of the permutation symmetry of both the occurring RDMs and $v_{ijqp}$ results in
\begin{align}
\label{eq:imag_time_1_RDM_EOM_2}
  \partial_\tau D^1_{i,j}&=2E_\tau D^1_{i,j}-2\Big(\sum_{r=1}^m\big(h_{rj}D^1_{i,r}+h_{ir}D^1_{r,j}\big)\\\nonumber
  &\phantom{=}+
  \sum_{q,p=1}^m h_{qp}D^2_{(ip),(jq)}\\\nonumber
  &\phantom{=}+
  \sum_{q,p,r=1}^m
  \big(v_{irqp}D^2_{(qp),(jr)}+v_{qpjr}D^2_{(ir),(qp)}\big)\\\nonumber
  &\phantom{=}+
  \sum_{q,p,r,s=1}^mv_{qprs}D^3_{(irs),(jqp)}\Big)
\end{align}
with
\begin{equation}
\label{eq:energy_functional_of_D}
 E_\tau = \sum_{q,p=1}^mh_{qp} D^1_{p,q}+\frac{1}{2}\sum_{q,p,r,s=1}^mv_{qprs}D^2_{(rs),(qp)}.
\end{equation}
Therefore, as in the case of a contracted Schr\"odinger equation \cite{valdemoroII,mazziotti_contracted_1998,many_electr_and_red_damats,maziotti_rdm_book07}, the dynamics of the 1-RDM couples 
to both the 2- and the 3-RDM. The EOM \eqref{eq:imag_time_1_RDM_EOM_2} can be easily 
translated to the more compact representation of RDMs \eqref{eq:rdm_rep}, of course.
This derivation can be applied also for the EOM of higher-order RDMs, with the result that
$\partial_\tau D^o$ couples to both the $(o+1)$- and the $(o+2)$-RDM as well as
to the 1- and 2-RDM by virtue of Eq.\ \eqref{eq:energy_functional_of_D}.
The latter coupling, however, can be transformed away by expressing $E_\tau$ solely as a functional
of $D^o$, which is always possible for $o\geq2$ when using a truncation approximation which
conserves the compatibility of the RDMs. As a starting-point for the truncation of the resulting
hierarchy of EOM, one can try the various closure approximations derived for
contracted Schr\"odinger equations \cite{valdemoroII,mazziotti_contracted_1998,many_electr_and_red_damats,maziotti_rdm_book07}.
\section{An alternative cluster expansion for bosons}
\label{app:trunc_alternative_cluster}
In this part of the appendix, we outline how to construct an alternative cluster expansion which is termwise
compatible and has multi-orbital mean-field states as (approximately) correlation-free
reference states. The latter is a desirable
property that implies (approximate) size-extensitivity in the sense specified
in Section \ref{sec:trunc_sym_cluster}. First, we discuss how to appropriately modify the terms
$\otimes_{\kappa=1}^o\hat\rho_1^{(\kappa)}\hat S_o$
of the expansion
\eqref{eq:cluster_symm}. Then we 
exemplarily describe the modification of terms such as $[\hat c_2^{(1,2)}\hat\rho_1^{(3)}+...]\hat S_3$
and finally a crucial mathematical subtlety is discussed.

For constructing this alternative cluster expansion, we first inspect the structure of the $o$-RDM 
given that the total system is in a multi-orbital mean-field state $|\Psi_{\rm MMF}\rangle
=|\vec{k}\rangle$. Tracing out $(N-o)$ atoms by means of \eqref{eq:part_tr_k},
we obtain
\begin{equation}
\label{eq:MMF_o_RDM}
 \hat \rho_o^{\rm MMF}=\binom{N}{o}^{-1}\,\sum_{{\bf 
n}|o}\Theta(\vec k-\vec n)\Big[\prod_{r=1}^m\binom{k_r}{n_r}\Big]\,|{\bf n}\rangle\!\langle{\bf n}|,
\end{equation}
which at first order boils down to $\hat \rho_1^{\rm MMF}=\sum_r\frac{k_r}{N}|\hat r\rangle\!\langle\hat r|$.
The latter means that the NOs $|\phi^1_r\rangle$ coincide with the single-particle states
underlying the permanent $|\vec{k}\rangle$ and that the NPs read $\lambda_r^{(1)}\equiv k_r/N$. Now let us inspect why the cluster 
expansion \eqref{eq:cluster_symm} diagnoses few-particle correlations for $|\Psi_{\rm MMF}\rangle$. This means
inspecting why the first term in the expansion \eqref{eq:cluster_symm}  at order $o$ deviates from the analytical
result \eqref{eq:MMF_o_RDM}. Using the spectral decomposition of $\hat \rho_1^{\rm MMF}$ and the relationship between
symmetrized Hartree products and permanents \eqref{eq:ns}, such 
a symmetrized product of 1-RDMs can be calculated as
\begin{align}
\label{eq:sym_cluster_for_MMF}
\bigotimes_{\kappa=1}^o\hat\rho_1^{(\kappa)}\hat S_o &=\hat S_o\bigotimes_{\kappa=1}^o\hat\rho_1^{(\kappa)}\hat S_o\\\nonumber
&=\sum_{i_1,...,i_o=1}^m\Big[\prod_{\kappa=1}^o\lambda_{i_\kappa}^{(1)}\Big]\,
\hat S_o|\phi^1_{i_1}...\phi^1_{i_o}\rangle\!\langle \phi^1_{i_1}...\phi^1_{i_o}|\hat S_o
\\\nonumber
 &= \sum_{{\bf n}|o}\Big[\prod_{r=1}^m\big(\lambda_r^{(1)}\big)^{n_r}\Big]\,|{\bf n}\rangle\!\langle{\bf n}|,
\end{align}
where we have abbreviated $\hat\rho_1\equiv\hat \rho_1^{\rm MMF}$. Straightforward
calculations show that \eqref{eq:sym_cluster_for_MMF} is neither termwise compatible in
the sense $\tr_1(\otimes_{\kappa=1}^{o+1}\hat\rho_1^{(\kappa)}\hat S_{o+1})\neq
\otimes_{\kappa=1}^o\hat\rho_1^{(\kappa)}\hat S_o$ nor does it
serve as a good approximation for \eqref{eq:MMF_o_RDM}. These flaws can actually
be linked to the fact that the norm of a symmetrized Hartree product 
$\hat S_o|\phi^1_{i_1}...\phi^1_{i_o}\rangle$, as it occurs in the second identity of \eqref{eq:sym_cluster_for_MMF}, depends on the corresponding
occupation numbers (see Eq.\ \eqref{eq:ns}). The latter, however, can be compensated by introducing
a modified symmetrization operator
\begin{equation}
 \hat{\mathcal{S}}_o\equiv\Big(\frac{o!}{\bigotimes_{r=1}^m\hat n_r!}\Big)^\frac{1}{2}\hat S_o
\end{equation}
where $\bigotimes_{r=1}^m\hat n_r!$
is defined as
$\sum_\vec{n}(\prod_r n_r!)|\vec n\rangle\!\langle\vec n|$ with the NOs
$|\phi^1_r\rangle$ as the underlying single-particle basis. Thereby, we ensure $\hat{\mathcal{S}}_o|\phi^1_{i_1}...\phi^1_{i_o}\rangle
=|\vec{n}\rangle$. Now we are equipped to replace the 1-RDM product terms
\eqref{eq:sym_cluster_for_MMF} of the cluster expansion \eqref{eq:cluster_symm}
by the following expression
\begin{align}
\label{eq:alt_cluster_for_MMF}
 \hat{\mathcal{S}}_o\bigotimes_{\kappa=1}^o\hat\rho_1^{(\kappa)} \hat{\mathcal{S}}_o=
 o!\sum_{{\bf n}|o}\Big[\prod_{r=1}^m\frac{\big(\lambda_r^{(1)}\big)^{n_r}}{n_r!}\Big]\,|{\bf n}\rangle\!\langle{\bf n}|.
\end{align}
It is easy to see that the partial trace of \eqref{eq:alt_cluster_for_MMF}
equals \eqref{eq:alt_cluster_for_MMF} with $o$ replaced by $(o-1)$ implying that this
class of terms in the alternative cluster expansion is termwise compatible indeed.
Inserting $\lambda_r^{(1)}\equiv k_r/N$ into \eqref{eq:alt_cluster_for_MMF}, 
we see that \eqref{eq:alt_cluster_for_MMF}
coincides with \eqref{eq:MMF_o_RDM} up to $1/N$ corrections
if $o\ll k_r$ for all finite $k_r$, i.e.\ in the case of a multi-orbital mean-field
state with only macroscopically occupied orbitals. Thus, size-extensitivity is
approximately ensured. In passing, we note that the relationship between
anti-symmetrized Hartree products and fermionic number states does not
involve a state-dependent normalization factor, which might be the reasons
why the anti-symmetrization of the cluster expansion \eqref{eq:cluster_indist_spinless}
leads to size-extensitivity. 

We further illustrate how to construct a termwise compatible cluster expansion
by inspecting how the terms $[\hat c_\sigma^{(1,...,\sigma)}\hat\rho_1^{(\sigma+1)}+...]\hat S_{\sigma+1}$
have to be modified. Let us abbreviate the modified version
of this term by $\hat G_{1,\sigma}^{1,1}$ (in analogy to the notation \eqref{eq:cluster_symbol})
and assume that termwise compatibility has already been
ensured up to order $\sigma$ implying  $\tr_1(\hat c_\sigma)=0$. For the term $\hat G_{1,\sigma}^{1,1}$,
we then make the ansatz
\begin{equation}
 \hat G_{1,\sigma}^{1,1}=\sum_{\vec{n},\vec{m}|\sigma+1}|\vec{n}\rangle\!\langle\vec{m}|
 \sum_{\substack{r=1\\
 n_rm_r>0}}^mc^\sigma_{\vec{n}-\uvec{r},\vec{m}-\uvec{r}}\,\lambda_r^{(1)}\,
 f_{\vec{n},\vec{m},r},
 \end{equation}
where again the NOs $|\phi^1_r\rangle$ serve as the underlying single-particle basis.
Motivated by the factorial factors of Eq.\ \eqref{eq:alt_cluster_for_MMF} compared to
\eqref{eq:sym_cluster_for_MMF}, we have introduced
a real-valued occupation-number factor $f_{\vec{n},\vec{m},r}$ to be 
determined by the termwise compatibility requirement $\tr_1(\hat G_{1,\sigma}^{1,1})=
\hat c_\sigma$. 
When evaluating the left hand side of this requirement, one realizes that one can only
benefit from $\tr_1(\hat c_\sigma)=0$  if $f$ depends only on $n_r$ and $m_r$, i.e.\
$f_{\vec{n},\vec{m},r}=g(n_r,m_r)$. Then the requirement $\tr_1(\hat G_{1,\sigma}^{1,1})=
\hat c_\sigma$ becomes equivalent to $g$ solving
\begin{equation}
 \sqrt{(l+1)(k+1)}\,g(l+1,k+1)-\sqrt{lk}\,g(l,k)=\sigma+1,
\end{equation}
for all $l,k=0,...,\sigma$. This set of linear equations possess the unique
solution $g(l,k)=(\sigma+1)[\delta_{lk}+(1-\delta_{lk})\,
\sqrt{\min\{l/k,k/l\}}]$. By making a similar ansatz involving occupation-number factors,
other classes of terms in the cluster expansion \eqref{eq:cluster_symm}
can be modified to obey termwise compatibility.

Although termwise compatibility and (approximate) size-extensitivity 
are desirable properties which the cluster expansion \eqref{eq:cluster_symm}
lacks, we do not employ this alternative approach for truncating the BBGKY hierarchy
because the thereby defined clusters fail to be unitarily invariant in the case of NP
degeneracies. Suppose that there are only two SPFs ($m=2$) and that the NPs
are degenerate, i.e.\ $\lambda^{(1)}_1=\lambda^{(1)}_2=0.5$. Then one can analytically show
that the alternative definition for the two-particle cluster $\hat c_2=\hat \rho_2-
\hat{\mathcal{S}}_2\,\hat\rho_1^{(1)}\hat\rho_1^{(2)}\hat{\mathcal{S}}_2$ does depend
on the concrete choice for the degenerate NOs. This ambiguity stems from the fact that
the operator $\bigotimes_{r=1}^m\hat n_r!$ is not form-invariant under unitary 
transformations of the NOs. Due to this finding, we do not develop this alternative
cluster expansion further but use the unitarily invariant decomposition, being
described in the following section, for making the symmetrized cluster expansion
\eqref{eq:cluster_symm} compatible.

\section{Unitarily invariant decomposition of bosonic operators}\label{app:uid}

According to the UID \cite{coleman1974,coleman_reduced_1980,au-chin_characteristic_1983,sun_unitarily_1984}, 
any given hermitian bosonic $o$-body
operator $\hat B_o\in\mathcal{B}_o$ can be uniquely decomposed into
$\hat B_o=\oplus_{k=0}^o\hat B_{o;k}$ with respect to all irreducible representations
of the unitary group $U(m)$ (the unitary transformations within the SPF space) 
on $\mathcal{B}_o$. This decomposition has the property that
the $l$-fold partial trace of $\hat B_o$ is fully determined by the first $(o-l+1)$
addends:  $\tr_l(\hat B_o)=\tr_l(\oplus_{k=0}^{o-l}\hat B_{o;k})$. Explicit
formulas for the components can be obtained
by making use of the fact that for each\footnote{We note that the space of hermitian bosonic $0$-body
operators is given by $\mathcal{B}_0=\{\alpha\,|0\rangle\!\langle0|,\;\alpha\in\mathbb{R}\}$.} 
$l=0,...,o-1$ there is a unique $\hat T_l\in\mathcal{B}_l$
such that $\oplus_{k=0}^l\hat B_{o;k}=\hat R_{o-l}\big(\hat T_l\big)$
and determining the $\hat T_l$'s recursively by employing
the relation between the raising operation \eqref{eq:raise_1} 
and the partial trace \eqref{eq:part_tr_1} \cite{kummer_nrepresentability_1967,sun_unitarily_1984}
\begin{align}
 \label{eq:rel_part_trace_raising}
 \tr_1\Big(\hat R_1\big(\hat B_o\big)\Big)&=\frac{2o+m}{(o+1)^2}\hat B_o\\\nonumber
 &+\Big(\frac{o}{o+1}\Big)^2\,\hat R_1\Big(\tr_1\big(\hat B_o\big)\Big).
\end{align}
We, however, are not interested in the individual terms of the complete UID 
but only in separating the contraction-free 
component $\hat B_o^{\rm irr}\equiv\hat B_{o;o}$ from the rest, 
i.e.\ in the decomposition\footnote{We note that the superscripts ``red'' and ``irr'' do not refer to (irr)reducible 
representations of the unitary group $U(m)$ but encode whether or not the component contains genuine $o$-particle correlations (see Section
\ref{sec:trunc_recursive_cluster}).}
${\hat B_o=
\hat B_o^{\rm red}\oplus\hat B_o^{\rm irr}}$. Using the results  of \cite{sun_unitarily_1984} for $\hat B_{o;k}$
and a computer algebra program \cite{mathematica} for summing over $k$,
we find
\begin{align}
\label{eq:red_comp_UID}
 \hat B_o^{\rm red} =-\sum_{k=0}^{o-1}(-1)^{o+k}\frac{\binom{o}{k}^2}{\binom{2o+m-2}{o-k}}
 \hat R_{o-k}\Big(\tr_{o-k}\big(\hat B_o\big)\Big)
\end{align}
and $\hat B_o^{\rm irr}=\hat B_o-\hat B_o^{\rm red}$. Eq.\ \eqref{eq:red_comp_UID} implies
that the reducible component $\hat B_o^{\rm red}$ depends linearly on all partial
partial traces of $\hat B_o$. We stress here that all considerations regarding
the UID rely on having a finite-dimensional single-particle Hilbert space, i.e.\ on
$m$ being finite \cite{coleman1974,coleman_reduced_1980,au-chin_characteristic_1983,sun_unitarily_1984}. 
For the actual evaluation of \eqref{eq:red_comp_UID}, we 
use the relationship between the $k$-fold raising operation and the joining operation
as stated in Sect.\ \ref{app:raise_join} together with Eq.\ \eqref{eq:join_op}.
\section{Proof of the recursive formulation of the bosonic cluster expansion}\label{app:recursion_rules}
The proofs of the recursion relations \eqref{eq:cluster_rule1}, \eqref{eq:cluster_rule2}
rely on (i) the property $\hat S_o|\vec{n}\rangle
=|\vec{n}\rangle$ for any $o$-particle number state $|\vec{n}\rangle$ and (ii) the 
number-state decomposition \eqref{eq:ns_kpart_sep}. To prove \eqref{eq:cluster_rule1},
we first inspect the matrix element of $\hat F_\sigma^n$ with respect to
any two $(n\sigma)$-particle number states
\begin{align}
 \nonumber
 \langle\vec{a}|\hat F_\sigma^n|\vec{b}\rangle&\overset{(i)}{=}
 \langle\vec{a}|[\hat c_\sigma^{(1,...,\sigma)}...\hat c_\sigma^{([n-1]\sigma+1,...,n\sigma)}
 +\text{dist.\ perm.}]|\vec{b}\rangle\\\label{eq:cluster_rule1_proof1}
 &=\frac{(n\sigma)!}{n!(\sigma!)^n}
 \langle\vec{a}|\hat c_\sigma^{(1,...,\sigma)}...\hat c_\sigma^{([n-1]\sigma+1,...,n\sigma)}|\vec{b}\rangle
\end{align}
where we have also used the invariance of number states under particle permutations in the last identity.
The resulting prefactor denotes the number 
of distinct sequences of $n$ pairwise distinct $\sigma$-tuple if the order of
the sequence does not matter. Next we apply \eqref{eq:ns_kpart_sep} to both
$|\vec a\rangle$ and $|\vec b\rangle$
\begin{widetext}
\begin{align}
\label{eq:cluster_rule1_proof2}
 \langle\vec{a}|\hat F_\sigma^n|\vec{b}\rangle=
 \frac{(n\sigma)!}{n!(\sigma!)^n}\binom{n\sigma}{\sigma}^{\!-1}\sum_{\vec{r},\vec{s}|\sigma}
 \Theta(\vec{a}-\vec{r}) \Theta(\vec{b}-\vec{s})
 &\Big[\prod_{i=1}^m\binom{a_i}{r_i}\binom{b_i}{s_i}\Big]^\frac{1}{2}\;\times\\\nonumber
 &\times\;\langle\vec{a}-\vec{r}|\hat c_\sigma^{(1,...,\sigma)}...\hat c_\sigma^{([n-2]\sigma+1,...,[n-1]\sigma)}|\vec{b}-\vec{s}\rangle\,
 \langle\vec{r}|\hat c_{\sigma}|\vec{s}\rangle.
\end{align}
\end{widetext}
By re-using the result \eqref{eq:cluster_rule1_proof1}, we may express the
second last factor of \eqref{eq:cluster_rule1_proof2} as 
\begin{align}
 \label{eq:cluster_rule1_proof3}
 \langle\vec{a}-\vec{r}|\hat c_\sigma^{(1,...,\sigma)}\;...\;\hat c_\sigma^{([n-2]\sigma+1,...,[n-1]\sigma)}&|\vec{b}-\vec{s}\rangle=\\\nonumber
 \frac{(n-1)!(\sigma!)^{n-1}}{[(n-1)\sigma]!}\langle\vec{a}-\vec{r}|\hat F_\sigma^{n-1}&|\vec{b}-\vec{s}\rangle.
\end{align}
Inserting \eqref{eq:cluster_rule1_proof2}, \eqref{eq:cluster_rule1_proof3}
into the expansion $\hat F_\sigma^n=\sum_{\vec{a},\vec{b}|n\sigma}
\langle\vec{a}|\hat F_\sigma^n|\vec{b}\rangle\,|\vec{a}\rangle\!\langle
\vec{b}|$ and substituting the sum over $\vec{a}$ ($\vec{b}$) by 
a sum over $\vec{a}'\equiv\vec{a}-\vec{r}$ ($\vec{b}'\equiv\vec{b}-\vec{s}$) we
finally find $\hat F_{\,\sigma}^n =\hat J_{(n-1)\sigma}^{\sigma}(\hat F_{\,\sigma}^{n-1},\hat c_\sigma)/n$.

In order to prove the second relation \eqref{eq:cluster_rule2}, we abbreviate
$o'=\sum_{r=1}^{K-1} n_r\sigma_r$ as well as $o=o'+n_K\sigma_K$ and  decompose 
the distinguishable permutations over particle labels in \eqref{eq:cluster_symbol}
as follows
\begin{widetext}
 \begin{align}
 \label{eq:cluster_rule2_proof1}
  \langle \vec{a}|\hat F_{\,\sigma_1,...,\sigma_K}^{n_1,...,n_K}|\vec{b}\rangle=
 \langle \vec{a}|\Big[\,&\big[\,\hat c_{\sigma_1}^{(1,...,\sigma_1)}\;...\;
 \hat c_{\sigma_{K-1}}^{(o'-\sigma_{K-1}+1,...,o')}\;+\;\text{dist.\ perm.}\,\big]\,\times\\\nonumber
 \times&\big[\,\hat c_{\sigma_K}^{(o'+1,...,o'+\sigma_K)}\;...\;
 \hat c_{\sigma_{K}}^{(o-\sigma_{K}+1,...,o)}\;+\;\text{dist.\ perm.}\,\big]\;+
 \;\text{dist.\ perm.}\,\Big]|\vec{b}\rangle\\\nonumber
 \overset{(i)}{=}
 \langle \vec{a}|\Big[\,&\big[\,\hat c_{\sigma_1}^{(1,...,\sigma_1)}\;...\;
 \hat c_{\sigma_{K-1}}^{(o'-\sigma_{K-1}+1,...,o')}\;+\;\text{dist.\ perm.}\,\big]\hat S_{o'}^{(1,...,o')}\,\times\\\nonumber
 \times&\big[\,\hat c_{\sigma_K}^{(o'+1,...,o'+\sigma_K)}\;...\;
 \hat c_{\sigma_{K}}^{(o-\sigma_{K}+1,...,o)}\;+\;\text{dist.\ perm.}\,\big]
 \hat S_{n_K\sigma_K}^{(o'+1,...,o)}
 \;+
 \;\text{dist.\ perm.}\,\Big]|\vec{b}\rangle\\\nonumber
  =
 \langle \vec{a}|\Big[\,&\big[\,\hat F_{\,\sigma_1,...,\sigma_{K-1}}^{n_1,...,n_{K-1}}\,\big]^{(1,...,o')}\,
 \big[\,\hat F_{\,\sigma_K}^{n_K}\,\big]^{(o'+1,...,o)}\;+
 \;\text{dist.\ perm.}\,\Big]|\vec{b}\rangle\\\nonumber
 =\langle \vec{a}|\phantom{\Big[}\,&\big[\,\hat F_{\,\sigma_1,...,\sigma_{K-1}}^{n_1,...,n_{K-1}}\,\big]^{(1,...,o')}\,
 \big[\,\hat F_{\,\sigma_K}^{n_K}\,\big]^{(o'+1,...,o)}\,|\vec{b}\rangle\;\binom{o}{o'}\\\nonumber
 \end{align}
\end{widetext}
where we have again employed the bosonic symmetry of the number states for the
last identity. Using \eqref{eq:ns_kpart_sep}, the binomial factor in \eqref{eq:cluster_rule2_proof1} is canceled and eventually the 
same number-state substitutions as before lead to
$\hat F_{\,\sigma_1,...,\sigma_K}^{n_1,...,n_K}=
\hat J_{\,o-n_K\sigma_K}^{n_K\sigma_K}(\hat F_{\,\sigma_1,...,\sigma_{K-1}}^{n_1,...,n_{K-1}},\hat F_{\,\sigma_K}^{n_K})$.
\section{Minimal invasive purification of the RDM}\label{app:corr_rdm}

In this appendix, we show how one can translate the minimal-invasive purification
scheme as outlined in Section \ref{sec:purif_sol} into a linear (quadratic)
program when using the $1$-norm ($2$-norm). The hermiticity of the correction
operator $\hat{\mathcal{C}}_2$ is incorporated in the following formalism by
decomposing $\mathcal{C}^2_{\vec n,\vec m}=\mathcal{C}^{2,\Re}_{\vec n,\vec m}
+i\,\mathcal{C}^{2,\Im}_{\vec n,\vec m}$ with $\mathcal{C}^{2,\Re}_{\vec n,\vec m}$
($\mathcal{C}^{2,\Im}_{\vec n,\vec m}$)
denoting a real-valued symmetric (anti-symmetric) matrix and mapping
the upper triangles of these matrices to real-valued vectors $\vec{c}^\Re$
and  $\vec{c}^\Im$, respectively, which are stacked to the vector
$\vec c=(\vec{c}^\Re, \vec{c}^\Im)^T$ containing $[C^2_m]^2$ elements where $C^2_m=m(m+1)/2$
denotes the number of distinct bosonic two-body configurations.
Having determined $\vec c$ as a solution of an optimization problem, we use it in order to
reconstruct the hermitian matrix $\mathcal{C}^2_{\vec n,\vec m}$. 

The aim of this appendix is to formulate 
our correction scheme as the problem of
minimizing\footnote{We note that this cost functional differs slightly
from \eqref{eq:p_norm} because pairs of off-diagonal elements $\mathcal{C}^2_{\vec n,\vec m}$,
$\mathcal{C}^2_{\vec m,\vec n}$
enter $|\vec c|_p$ only by a single representative off-diagonal element.
We, however, do not expect that differences between \eqref{eq:p_norm} and $|\vec c|_p$ 
have a severe impact on the purification scheme.
} $|\vec c|_p\equiv\sum_r|c_r|^p$ under the linear constraints 
$\vec A\,\vec c=\vec b$. In the following, we construct the matrix $\vec A$ and 
the vector $\vec b$ of this underdetermined system of linear equations.
Here, the overall strategy is to formulate the $i$-th constraint as follows
$\sum_{{\bf n},{\bf m}|2} O_{{\bf m},{\bf n}}^i\,\mathcal{C}^2_{{\bf n},{\bf m}}=b_i$.
Mapping the number states $\vec n$, $\vec m$ to integer-valued indices $I$, $J$,
we may decompose the latter equation as
\begin{align}\label{eq:constr_strategy}
b_i&=\sum_{I\leq J} {\tilde{A}}_{I,J}^{i,\Re}\,\mathcal{C}^{2,\Re}_{I,J}
+i\sum_{I< J} {\tilde{A}}_{I,J}^{i,\Im}\,\mathcal{C}^{2,\Im}_{I,J}
\\\nonumber
&\equiv[\vec a^{i,\Re}]^T\cdot\vec c^\Re+[\vec a^{i,\Im}]^T\cdot\vec c^\Im
 =[\vec a^i]^T\cdot\vec c,
\end{align}
where ${\tilde{A}}_{I,J}^{i,\Re}=[O_{I,J}^i+O_{J,I}^i]/(1+\delta_{I,J})$
and ${\tilde{A}}_{I,J}^{i,\Im}=O_{J,I}^i-O_{I,J}^i$. In the second identity
of \eqref{eq:constr_strategy}, we have mapped the upper triangles of the 
matrices ${\tilde{A}}_{I,J}^{i,\Re}$
and $i\,\tilde{A}_{I,J}^{i,\Im}$ to the vectors $\vec a^{i,\Re}$ and $\vec a^{i,\Im}$, respectively, which turn out to be real-valued for hermitian $O_{I,J}^i$. In this case,
these vectors are stacked to $\vec a^i=(\vec a^{i,\Re},\vec a^{i,\Im})^T$,
which constitutes the $i$-th row of the matrix $\vec A$.

(i) Let us translate the constraints of  $\hat{\mathcal{C}}^2$ being contraction-free.
Due to hermiticity of $\langle\phi_i|\tr_1(\hat{\mathcal{C}}^2)|\phi_j\rangle$,
we obtain $m^2$ independent constraints. Using the expression \eqref{eq:part_tr_1} for the partial trace,
we have
\begin{equation}
\sum_{l=1}^m\sqrt{(1+\delta_{il})(1+\delta_{jl})}\,
 \mathcal{C}^2_{\uvec{i}+\uvec{l},\uvec{j}+\uvec{l}}=0,
\end{equation}
which can expressed as \eqref{eq:constr_strategy} with $b_r=0$ ($r$ shall label
the constraint corresponding to $i\leq j$) and 
\begin{equation}
 O^r_{\vec m,\vec n}=\sum_{l=1}^m\sqrt{(1+\delta_{il})(1+\delta_{jl})}
 \delta_{\vec n,\uvec i+\uvec l}\delta_{\vec m,\uvec j+\uvec l}
\end{equation}
where we introduced the Kronecker delta for number states as $\delta_{\vec a,\vec b}
=\prod_{k=1}^m\delta_{a_k,b_k}$. Now, we have to distinguish two cases.
If $i=j$, $O^r_{\vec m,\vec n}$ is a real-valued symmetric matrix such
that $\tilde{A}_{I,J}^{r,\Im}=0$. Thereby, we obtain $m$ constraints affecting only
the symmetric component $\mathcal{C}^{2,\Re}_{\vec n,\vec m}$. For $i<j$, however,
$O^r_{\vec m,\vec n}$ turns out to be a real-valued asymmetric matrix, resulting
in non-vanishing real-valued matrices ${\tilde{A}}_{I,J}^{r,\Re}$
and $\tilde{A}_{I,J}^{r,\Im}$. Thus, we obtain for each $i<j$ two independent constraints
affecting either the symmetric component $\mathcal{C}^{2,\Re}_{\vec n,\vec m}$
or the antisymmetric component $\mathcal{C}^{2,\Im}_{\vec n,\vec m}$ only. 
The corresponding rows of $\vec A$ are given by 
$\vec a^i=(\vec a^{i,\Re},\vec a^{i,\Im})^T$ with $\vec a^{i,\Re}$ 
covering the upper triangle of ${\tilde{A}}_{I,J}^{r,\Re}$
as well as $\vec a^{i,\Im}$ set to zero and 
$\vec a^{i,\Re}$ set to zero as well as $\vec a^{i,\Im}$ covering
the upper triangle of  ${\tilde{A}}_{I,J}^{r,\Im}$, respectively. Thereby, we obtain further
$m(m-1)$ constraints, which amounts to $m^2$ constraints related to the partial
trace of $\hat{\mathcal{C}}_2$ in total.

(ii) Energy conservation can be easily formulated as a linear constraint. 
By means of the relation $\langle\hat H\rangle_t
=N\,\tr(\hat k_2\,\hat\rho_2)$ with
the auxiliary $2$-particle Hamiltonian 
$\hat k_2=[\hat h_1+\hat h_2+(N-1)\hat v_{12}]/2$ \cite{bopp59}, we have to require
$\tr(\hat k_2\,\hat{\mathcal{C}}_2)=0$. The latter boils down to 
$\tr(\hat v_{12}\,\hat{\mathcal{C}}_2)=0$ as $\hat{\mathcal{C}}_2$
shall be contraction-free. Apparently, we have $b_r=0$ for this
constraint. Using the cyclic invariance of the trace and $\hat{\mathcal{C}}_2$ being bosonic,
we obtain $\tr(\hat S_2\hat v_{12}\hat S_2\,\hat{\mathcal{C}}_2)=0$, which allows
for identifying 
$O^r_{\vec m,\vec n}$ with the matrix elements of $\hat S_2\hat v_{12}\hat S_2$
in the number-state basis, namely as
\begin{equation}
 O^r_{\uvec i+\uvec j,\uvec q+\uvec  p} =\frac{v_{ijqp}+v_{jiqp}}{\sqrt{(1+\delta_{ij})(1+\delta_{qp})}}.
\end{equation}

(iii) In the case of a symmetry such as invariance under parity or translation operations, 
we proceed as follows. We remind that the SPFs stay invariant under the corresponding
symmetry operation, $\hat \pi_1|\phi_j\rangle=\exp(i\theta_j)|\phi_j\rangle$
with some real phase $\theta_j$ (see Section \ref{sec:trunc_cons_laws}). Correspondingly,
an $o$-particle number-state transforms as $\hat\Pi_o|\vec n\rangle=\exp[i\theta(\vec n)]|\vec n\rangle$
with $\theta(\vec n)=\sum_{j=1}^mn_j\theta_j$. Now we introduce the following
equivalence relation between number states $\vec n\sim\vec m$ if
$\theta(\vec n)\!\!\mod 2\pi=\theta(\vec m)\!\!\mod 2\pi$. Then the correction
respects the symmetry if $\mathcal{C}^2_{\vec n,\vec m}=0$ for all $\vec n
\nsim\vec m$. In order to construct the corresponding rows in $\vec A$, 
we loop over all pairs of inequivalent number states $\vec f
\nsim\vec g$ whose corresponding integer labels $I$ (for $\vec f$) and $J$
(for $\vec g$) obey $I<J$ in order to avoid redundant constraints. For 
each such $\vec f$, $\vec g$, we set $b_r=0$ and $ O^r_{\vec m,\vec n}=
\delta_{\vec m,\vec g}\delta_{\vec n,\vec f}$.
The latter constitutes a real-valued asymmetric matrix such 
that for each $\vec f
\nsim\vec g$ two independent constraints arise, affecting either $\mathcal{C}^{2,\Re}_{{\bf n},{\bf m}}$
or  $\mathcal{C}^{2,\Im}_{{\bf n},{\bf m}}$ only (see constraints (i)).

(iv) The constraint that each negative NP below a certain threshold
$\lambda^{(2)}_r<\epsilon$ is raised to zero in first order perturbation theory
with respect to $\hat{\mathcal{C}}_2$ can be expressed as 
$\tr(|\phi^2_r\rangle\!\langle\phi^2_r|\,\hat{\mathcal{C}}_2)=-\lambda^{(2)}_r$.
From this, we may directly read-off $b_r=-\lambda^{(2)}_r$ as well 
as $O^r_{\vec m,\vec n}=[\phi^2_{r;\vec n}]^*\phi^2_{r;\vec m}$ where
$\phi^2_{r;\vec m}=\langle \vec m|\phi^2_r\rangle$.

(v) In order to formulate the constraints related to negative $\hat K_2$ eigenvalues,
 we first need to clarify the relation between $\hat K_2$ and $\hat \rho_1$, $\hat \rho_2$.
 Explicating \eqref{eq:K_matrix}, we find
\begin{align}\nonumber
 K^2_{(i_1,j_1),(i_2,j_2)}\,\mathcal{N}_K
 &=N(N-1)\,f_{i_2j_1} f_{i_1j_2}\,\rho^2_{\uvec{i_2}+\uvec{j_1},\uvec{i_1}+\uvec{j_2}}
 \\\label{eq:K_rho_relation}
 &+\delta_{j_1,j_2} N\;\rho^1_{i_2i_1}-N^2\;\rho^1_{j_1,i_1}\rho^1_{i_2,j_2},
 \end{align}
where we have again used the abbreviation $f_{ij}=\sqrt{(1+\delta_{i,j})/2}$.
Updating $\hat\rho_2$ by the contraction-free operator $\hat{\mathcal{C}}_2$
apparently leaves $\hat\rho_1$ and thus also $\mathcal{N}_K=N(N+m-1)-N^2\,\tr(
[\hat\rho_1]^2)$ invariant. Thereby, the update of $\hat\rho_2$ implies the update
$\hat K_2+\hat\Delta_2$ with
\begin{align}\label{eq:Delta2}
 \Delta^2_{(i_1,j_1),(i_2,j_2)}&=\\\nonumber
&\frac{N(N-1)}{\mathcal{N}_K}\,f_{i_2j_1} f_{i_1j_2}\,\mathcal{C}^2_{\uvec{i_2}+\uvec{j_1},\uvec{i_1}+\uvec{j_2}}.
\end{align}
Now the constraint that a negative $\hat K_2$ eigenvalue $\xi_r$ below the
threshold $\epsilon$ shall be raised to zero in first-order perturbation theory
with respect to $\hat\Delta_2$, i.e.\ $\tr(|\Xi_r\rangle\!\langle\Xi_r|\,
\hat \Delta_2)=-\xi_r$, can be rephrased in terms of the correction $\hat{\mathcal{C}}_2$.
The result is setting $b_r=-\xi_r$ and 
\begin{align}
 O^r_{\uvec{q}+\uvec{p},\uvec{i}+\uvec{j}}&\equiv\frac{N(N-1)}{4\mathcal{N}_Kf_{ij}f_{qp}}
 \Big(M^r_{(i,p),(q,j)}\\\nonumber
 &\phantom{=}+M^r_{(j,p),(q,i)}+M^r_{(i,q),(p,j)}+M^r_{(j,q),(p,i)}\Big)
\end{align}
where $M^r_{(i,p),(q,j)}\equiv\langle\varphi_i\varphi_p|\Xi_r\rangle\,
\langle\Xi_r|\varphi_q\varphi_j\rangle$. 

We remark that the inhomogeneity
$\vec b$ of the system of linear equations features only extremely small or vanishing
numerical values.
For increasing the numerical stability when solving the optimization problem,
we replace $\vec b$ by $\vec b/\mathcal{N}_b$
with $\mathcal{N}_b = (\sum_{i|\lambda^{(2)}_i<\epsilon}|\lambda^{(2)}_i|
+\sum_{i|\xi_i<\epsilon}|\xi_i|)/(d+d')$ where
$d$ ($d'$) denotes the number of $\hat \rho_2$ ($\hat K_2$) eigenvalues below
the threshold. Thereafter, the solution $\vec c$ is rescaled as $\vec c\,\mathcal{N}_b$.

Finally, let us investigate how underdetermined the correction operator $\hat{\mathcal{C}}_2$ is, which can
be parametrized by $[C^2_m]^2=m^2(m+1)^2/4$ independent real numbers. If no
symmetry has to be incorporated, there are $m^2+d+d'+1$ independent constraints
(note that the energy-conservation constraint has to be imposed also in
cases where the total Hamiltonian is explicitly time-dependent, otherwise
$\frac{{\rm d}}{{\rm d}t}\langle\hat H \rangle=\langle\partial_t\hat H \rangle$
would be violated). If there is parity symmetric and half of the initial SPFs are of even
and half of them are of odd parity ($m$ shall be even), then $m^4/8+m^3/4$ additional
constraints have to be taken into account.
\section{Minimal invasive correction of the EOM}\label{app:corr_eom}

The technical implementation of the minimal-invasive correction scheme for the
$2$-RDM EOM is very much alike the steps discussed in Appendix \ref{app:corr_rdm}.
Therefore, we only work out differences here.

The constraints on the correction $\hat{\mathcal{C}}_2$ to be
(i) contraction-free, (ii) energy conserving and (iii) respecting symmetries if existent
can be exactly implemented as discussed in Appendix \ref{app:corr_rdm}.
For enforcing negative NPs $\lambda^{(2)}_r$ below the threshold $\epsilon$
to be exponentially damped to zero, we may use the same $O^r_{\vec m,\vec n}$
as described in Appendix \ref{app:corr_rdm} (vi), where one 
has to replace, however, $b_r=-\lambda^{(2)}_r$ by
$b_r=-\eta\lambda^{(2)}_r+i\langle\phi^2_r|\hat I_2(\hat \chi_3)|\phi^2_r\rangle$
with $\hat \chi_3=\hat\rho_3^{\rm appr}$ for $\bar o=2$
and $\hat \chi_3=\hat\rho_3$ for $\bar o>2$.

For the requirement that also negative $\hat K_2$ eigenvalues are damped to zero, we first have to 
express the EOM  for $\hat K_2$ in terms of $\hat\rho_1$, $\hat \rho_2$, $\hat R_1$
and $\hat R_2$ by differentiating
\eqref{eq:K_rho_relation} with respect to time 
\begin{widetext}
\begin{align}
 \partial_tK^2_{(i_1,j_1),(i_2,j_2)}\equiv\langle\varphi_{i_1}\varphi_{j_1}|\hat T_2|\varphi_{i_2}\varphi_{j_2}\rangle&=\frac{1}{\mathcal{N}_K}\Big(
 2N^2\,\tr\big(\hat R_1\hat\rho_1\big)\,K^2_{(i_1,j_1),(i_2,j_2)}
 +
 N(N-1)\,f_{i_2j_1}f_{i_1j_2}\,R^2_{\uvec{i_2}+\uvec{j_1},\uvec{i_1}+\uvec{j_2}}
 \\\nonumber
  &\phantom{=}
  +\delta_{j_1,j_2}\,N\,R^1_{i_2,i_1}-N^2\,R^1_{j_1,i_1}\,\rho^1_{i_2,j_2}
  -N^2\,R^1_{i_2,j_2}\,\rho^1_{j_1,i_1}\Big).
\end{align}
\end{widetext}
Since $\hat{\mathcal{C}}_2$ is enforced to 
be contraction-free, $\hat R_1$ is left invariant under the correction of
$\hat R_2$. Thus, we induce the correction $\hat T_2\rightarrow\hat T_2+\hat \Delta_2$
with $\hat\Delta_2$ given by \eqref{eq:Delta2}. Thereby, we may use the 
same 
$O^r_{\vec m,\vec n}$
as in Appendix \ref{app:corr_rdm} (v) but need to substitute $b_r=-\xi_r$ 
by $b_r=-\eta\xi_r-\langle\Xi_r|\hat T_2|\Xi_r\rangle$.
\section{Unitarily invariant decomposition of the collision integral}\label{app:UID_coll_int}
While we have so far applied the UID only to the RDM, the purpose of this
appendix is to gain insights into the unitarily invariant components of the collision
integral. Let $\hat A_{o+1}\in\mathcal{B}_{o+1}$. By means of \eqref{eq:part_tr_1},
we may evaluate the partial trace of the collision integral
\begin{widetext}
\begin{align}
 \tr_1\big(\hat I_o(\hat A_{o+1})\big)
 &=\frac{N-o}{o(o+1)}\sum_{r,i,j,q,p=1}
  v_{qjpi}\,\a{r}\big[\ad{q} \a{p}, \a{i}\,\hat A_{o+1}\,\ad{j}\big]\ad{r}
  =\frac{N-o}{o(o+1)}\sum_{r,i,j,q,p=1}
  v_{qjpi}\,\big[\ad{q} \a{p}, \a{i}\a{r}\,\hat A_{o+1}\,\ad{r}\ad{j}\big]
  \\\nonumber
  &=\hat I_o\big(\tr_1(\hat A_{o+1})\big).
\end{align}
\end{widetext}
From this identity, we may conclude that the reducible component of the 
collision integral depends solely on the reducible component of its 
argument, $[\hat I_o(\hat A_{o+1})]^{\rm red}=[\hat I_o(\hat A_{o+1}^{\rm red})]^{\rm red}$.
The irreducible component of the collision integral $[\hat I_o(\hat A_{o+1})]^{\rm irr}$, however, depends
on both reducible and the irreducible component of $\hat A_{o+1}$ in general, which we
have confirmed by an explicit calculation for the cases $o=1,2$. Thus,
the collision integral with a contraction-free argument is itself contraction-free.
\section{Numerical integration of the truncated BBGKY EOM}\label{app:eom_intr}
In both scenarios of Section \ref{sec:appl}, we employ the variable-coefficient ordinary differential equation
solver ZVODE \cite{zvode} for integrating the EOM \eqref{eq:mctdhb_part_eom}, \eqref{eq:BBGKY_arb_gauge}. 
The conservation of hermiticity of the RDMs
is numerically ensured by only propagating the lower triangle of the 
matrix-valued EOM \eqref{eq:BBGKY_arb_gauge}, which at the same time reduces
the number of variables to be integrated. Since the applied truncation approximation
conserves the compatibility of the RDMs, we propagate only the BBGKY EOM 
\eqref{eq:BBGKY_arb_gauge} at the truncation order $\bar o$ and obtain the
RDMs of lower order by partial tracing. Moreover, we operate in the single-particle
Hamiltonian gauge, $g_{ij}=h_{ij}$ (see Section \ref{sec:BBGKY_EOM_rep}).
\bibliography{bbgky_lit}
\bibliographystyle{unsrt}

\end{document}